\documentclass[11pt,a4paper]{article}                           

\usepackage{amsmath}
\usepackage{amssymb}
\usepackage[usenames]{color} 
\usepackage{multirow}
\usepackage{graphicx}
\usepackage{epstopdf}
\usepackage{array}
\usepackage{hhline}
\usepackage{cite}
\usepackage{microtype,colortbl}
\usepackage[bf]{caption}
\usepackage{color}
\usepackage[usenames,dvipsnames]{xcolor}
\usepackage{pstool}
\usepackage{tikz}
\usepackage{ytableau}
\usepackage{latexsym,stmaryrd}
\usepackage[bookmarks=false]{hyperref}
 \usepackage{lscape}
 \usepackage{subfig}

\usepackage{enumitem}

\usepackage{ifpdf}
\def\hybrid{\topmargin -20pt    \oddsidemargin 0pt
        \headheight 0pt \headsep 0pt
        \textwidth 6.25in       
        \textheight 9 in       
        \marginparwidth .875in
        \parskip 5pt plus 1pt 
          \jot = 1.5ex
   }
\hybrid
\numberwithin{equation}{section}
\numberwithin{table}{section}

\newcommand{\beq}{\begin{equation}}  \newcommand{\eeq}{\end{equation}}
\newcommand{\bal}{\begin{aligned}}   \newcommand{\eal}{\end{aligned}}
\newcommand{\bea}{\begin{eqnarray}}  \newcommand{\eea}{\end{eqnarray}}

\newcommand{\bmat}{\left(\begin{array}}
\newcommand{\emat}{\end{array}\right)}

\newcommand{\bbC}{\mathbb{C}}

\newcommand{\nn}{\nonumber}

\newcommand{\cO}{\mathcal{O}}

\newcommand{\cK}{\mathcal{K}}
\newcommand{\cN}{\mathcal{N}}

\newcommand{\cF}{\mathcal{F}}

\newcommand{\cV}{\mathcal{V}}

\newcommand{\cM}{\mathcal M}

\renewcommand{\Im}{\mathrm{Im}\,}

\usepackage{cancel}

\newcommand{\be}{\begin{equation}}
\newcommand{\ee}{\end{equation}}

\DeclareMathOperator{\img}{img}

\definecolor{Gray}{gray}{0.95}

\begin{document}
\baselineskip=14pt
\parskip 5pt plus 1pt 

\vspace*{2cm}
\begin{center}
{\Large \bfseries Modeling General Asymptotic
Calabi-Yau Periods}\\[.3cm]

\vspace{1cm}
{\bf Brice Bastian}\footnote{b.bastian@uu.nl},
{\bf Thomas W.~Grimm}\footnote{t.w.grimm@uu.nl},
{\bf Damian van de Heisteeg}\footnote{d.t.e.vandeheisteeg@uu.nl},

{\small
\vspace*{.5cm}
Institute for Theoretical Physics, Utrecht University\\ Princetonplein 5, 3584 CC Utrecht, The Netherlands\\[3mm]
}
\end{center}
\vspace{1cm}
\begin{abstract} \noindent
In the quests to uncovering the fundamental structures that underlie some of the asymptotic Swampland conjectures we initiate the 
general study of asymptotic period vectors of Calabi-Yau manifolds. Our strategy is to exploit the constraints imposed by 
completeness, symmetry, and positivity, which are formalized in asymptotic Hodge theory. 
We use these general principles to study the periods near any boundary in complex structure moduli space and explain that 
near most boundaries leading exponentially suppressed corrections must be present for consistency. The only exception are period vectors near the well-studied large complex structure point. Together with the classification of possible boundaries, our procedure makes it possible to construct general models for these asymptotic periods. 
The starting point for this construction is the $sl(2)$-data classifying the boundary, which we use 
to construct the asymptotic Hodge decomposition known as the nilpotent orbit. We then use the latter to determine the asymptotic period vector. 
We explicitly carry out this program for all possible one- and two-moduli boundaries in Calabi-Yau threefolds and write down general models for their asymptotic periods.
\end{abstract}

\newpage

\setcounter{footnote}{0}

\tableofcontents
\newpage
\section{Introduction}
One of the main goals of the swampland programme is to characterize low energy effective theories that can be consistently coupled to quantum gravity. Owing to the richness in variety of such models, one is led to postulate general properties that these effective theories should possess.  In turn, these conjectures are then put under scrutiny by studying explicit examples realized in string theory constructions, with the purpose of providing evidence for their validity. Many of these proposals turn out to be intimately connected to each other, giving rise to an intricate and expanding web (for reviews, see \cite{Brennan:2017rbf,Palti:2019pca,vanBeest:2021lhn}). In recent years, the emerging interconnected nature of these conjectures hints at the existence of some deeper underlying structures that 
are at play in the background. It is therefore crucial to examine classes of compactifications that are as general possible to avoid drawing too strong conclusions from 
simplified models.  

A promising arena for this endeavor is provided by studying effective theories evaluated in the asymptotic regions of the moduli space associated to a string compactification.
Several swampland conjectures suggest that for such regions in moduli space a universal asymptotic behaviour sets in and asymptotic structures can be used to describe general 
properties of the resulting lower-dimensional theories. To exemplify this, it was recently argued in \cite{Grimm:2018ohb,Blumenhagen:2018nts,Lee:2018urn,Lee:2018spm,Grimm:2018cpv,Corvilain:2018lgw,Lee:2019tst,Font:2019cxq,Marchesano:2019ifh,Lee:2019xtm,Grimm:2019wtx,Kehagias:2019akr,Lee:2019wij,Grimm:2019ixq,Baume:2019sry,Enriquez-Rojo:2020pqm,Andriot:2020lea,Gendler:2020dfp,Lanza:2020qmt,Heidenreich:2020ptx,Bastian:2020egp,Klaewer:2020lfg,Calderon-Infante:2020dhm,Grimm:2020ouv,Cota:2020zse,Brodie:2021ain,Lanza:2021qsu,Castellano:2021yye} 
that some of the swampland conjectures can be tested rather generally 
in Type II string compactifications on Calabi-Yau threefolds, when restricting to the asymptotic regions of the complex structure or K\"ahler structure moduli space of these 
geometries.  
In many of these tests the framework of asymptotic Hodge theory turned out to be an indispensable tool to unearth part of the deep structures rooted in some of these conjectures.\footnote{This is complemented by the observation of \cite{Cecotti:2020rjq} that Hodge theory plays a crucial role in the $\cN=2$ swampland program, and it has been speculated that this approach can even be extended to non-supersymmetric settings \cite{Cecotti:2021cvv}.}
As we will also see in this work, it provides a powerful tool to make general statements about the behavior of important physical quantities in the asymptotic regimes of moduli space. 

In this work our main focus will be the study of the complex structure moduli space of Calabi-Yau threefolds and the derivation 
of the so-called period vectors that are associated with these compact geometries. These period vectors are defined to be the 
integrals of the, up to rescaling, unique $(3,0)$-form over a basis of three-cycles. They are crucial in determining the effective theories arising 
in string compactifications on Calabi-Yau threefolds. In particular, they determine the $\cN=2$ vector sector of Type IIB string compactifications \cite{Andrianopoli:1996cm,Craps:1997gp} and 
encode part of the $\cN=1$ K\"ahler potential and flux superpotential in Type IIB orientifold \cite{Grana:2005jc,Douglas:2006es}.
A detailed understanding of the asymptotic form of the periods therefore allows one to test swampland conjectures in these general classes of compactifications. 
It is well-known that the period vectors can be computed by using Picard-Fuchs equations \cite{Hosono:1993qy,Hosono:1994ax,CoxKatz}\footnote{See also \cite{ruddat2019period} for recently developed methods that take a different approach, which in particular does not require one to embed the Calabi-Yau manifold in some ambient space.} whose solutions are in general 
complicated functions of the complex structure moduli. While this strategy to study the periods has been employed successfully for many Calabi-Yau examples in the past, 
it turns out to be convenient to take a different route when addressing general questions about the form of all possible periods that can occur.

In this work we employ the mathematical machinery of asymptotic Hodge theory that was originally developed in the works of Schmid \cite{Schmid} and Cattani, Kaplan and Schmid \cite{CKS}. These works describe in detail how the Hodge decomposition behaves near the boundaries of moduli space where the Calabi-Yau manifold $Y_3$ develops a singularity. 
As a first mathematical fact we recall that this behaviour can be encoded by a set of complex vector spaces known as 
the nilpotent orbit \cite{Schmid}. This orbit depends in a simple way on some data associated to the boundary under consideration, namely the monodromy transformation matrices and some vector spaces defined on the boundary.  Nilpotent orbits can be systematically classified, leading to a list of possible boundaries that can occur in complex structure moduli space. In order to perform 
this classification one uses another deep result in asymptotic Hodge theory  \cite{Schmid,CKS}, namely, that one can associate to each boundary of complex co-dimension $n$ an $sl(2)^n$ algebra and decompose the middle cohomology $H^3(Y_3,\bbC)$ into representations of this algebra. Importantly, the $sl(2)^n$  can also be used to construct 
a minimal normal form of the boundary data. The strategy employed in this work is to start with the minimal normal form of the $sl(2)^n$ boundary data 
and work backwards to construct the nilpotent orbit and eventually the period vector. Note that this is exactly in the spirit of the holographic perspective recently put 
forward in \cite{Grimm:2020cda,Grimm:2021ikg}.\footnote{Note that in \cite{Grimm:2020cda,Grimm:2021ikg} it was not assumed that a nilpotent orbit exists and therefore the constructions in 
this works are more involved and essentially mimic the proof of the Sl(2)-orbit theorem. Here we will use the Sl(2)-orbit theorem directly. A complementary perspective was given in \cite{Cecotti:2020uek}, where also an auxiliary bulk action principle was proposed which differs in the coupling to gravity. }

While it is 
straightforward to use the nilpotent orbit to determine the leading polynomial part of the period vector,
we show that this polynomial approximation only suffices for boundaries of large complex structure type. For all other boundaries we find that non-perturbative corrections to the period vector have to be included. This generalizes the observation already made in \cite{Grimm:2020cda} and nicely complements \cite{Palti:2020qlc,Cecotti:2020rjq}, in that we now note the importance of exponentially suppressed corrections near boundaries away from large complex structure. One way to see the necessity of these non-perturbative terms is that the K\"ahler metric derived from the polynomial periods can be singular. More generally, these issues can be traced back to a completeness principle. It is a fundamental result for Calabi-Yau threefolds $Y_3$ that it is possible to span the entire three-form cohomology $H^3(Y_3,\mathbb{C})$ from the period vector and its derivatives. The necessity of exponentially suppressed terms is expected from finite distance boundaries such as the conifold point, but we show using asymptotic Hodge theory that it extends to infinite distance boundaries as well. More precisely, we are able to systematically characterize which correction terms are needed depending on the boundary type. Due to their exponential dependence in the complex structure moduli we refer to them throughout this work as non-perturbative or instanton terms, although the interpretation as world-sheet instantons can only be made more precise at large complex structure.

Our construction of the periods uses the techniques laid out in the mathematical works \cite{BrosnanPearlsteinRobles,KaplanPearlstein,BrosnanPearlstein,CattaniFernandez2000,CattaniFernandez2008}. We begin by constructing the most general nilpotent orbit compatible with the  $sl(2)^n$ boundary data, following the approach of \cite{BrosnanPearlsteinRobles,KaplanPearlstein,BrosnanPearlstein}. In particular, this procedure captures how to get the monodromy transformations from a given set of $sl(2)^n$-data. We then turn to \cite{CattaniFernandez2000,CattaniFernandez2008} where the holomorphic expansion of the periods was encoded in terms of an analytic matrix-valued function $\Gamma$. This function generates the exponential corrections to the polynomial periods, hence we dub it the instanton map. The form of this instanton map is constrained by imposing consistency with the boundary Hodge decomposition. Furthermore, its coefficients must obey certain differential equations following from orthogonality relations between the periods and its derivatives. We then derive a rank condition on this instanton map which ensures that the entire three-form cohomology $H^3(Y_3, \mathbb{C})$ can be spanned by the period vector and its derivatives. This allows us to determine precisely which coefficients in $\Gamma$ are needed, consequently indicating the essential instanton terms required in the periods.  By bringing these building blocks together, asymptotic Hodge theory thus makes it possible through principles of completeness, symmetry and positivity to construct general models for period vectors including the relevant corrections near any boundary.

We explicitly carry out this programme for all possible one- and two-moduli boundaries in complex structure moduli space. In the one-modulus case boundaries are straighforwardly classified by their singularity type, while for two moduli one has to deal with intersections of divisors where it can enhance. Nevertheless the two-moduli case can also be completely classified, as was worked out in \cite{Kerr2017}. At the one-modulus level we find two classes of boundaries that require instantons, which cover for instance the conifold point of the (mirror) quintic \cite{Candelas:1990rm} and the so-called Tyurin degeneration \cite{Tyurin:2003}.   For two moduli we encounter three classes, among them the recently studied coni-LCS boundaries \cite{Demirtas:2020ffz,Blumenhagen:2020ire}. As a more involved example, we show how our results cover a degeneration for the Calabi-Yau threefold in $\mathbb{P}^{1,1,2,2,6}$ which played an important role in geometrically engineering of Seiberg-Witten models \cite{Kachru:1995fv}.

An interesting application of our work is that the resulting periods can be used in the study of four-dimensional supergravity theories, without explicit knowledge of the mathematical machinery going into their construction. We illustrate this by computing the K\"ahler potential, flux superpotential and leading polynomial part of the scalar potential that characterize these theories. For the K\"ahler potential we find the crucial instanton terms that remedy the singular behavior of the metric. Furthermore, we observe the emergence of a continuous axion shift symmetry near all boundaries we considered, only broken by terms subleading compared to these instantons. For the flux superpotential we find that instanton terms can be essential for coupling all the fluxes, both for finite and even some infinite distance boundaries. In contrast, for the corresponding scalar potential all fluxes already appear at leading polynomial order. Let us stress that we have control over more instanton terms than just those required for the leading polynomial physical quantities. These subleading terms correct for instance the polynomial scalar potential at exponentially suppressed order.

The paper is organized as follows. In section \ref{sec:Basics} we introduce some basic facts about period vectors of Calabi-Yau threefolds and summarize some results from asymptotic Hodge theory. Our focus will be on introducing the nilpotent orbit and explaining how it serves as the central tool to encode the asymptotic behavior of the Hodge decomposition. We also review the classification of boundaries in complex structure moduli space. In section \ref{sec:instantonperiods} we first argue that exponentially suppressed terms are required away from large complex structure, and give a lower bound on the required number based on the boundary data. We then lay out the procedure to construct the periods using the instanton map. In section \ref{sec:models} we provide general models for period vectors for all possible one- and two-moduli boundaries. In sections \ref{app:one-modulus} and \ref{sec:two-moduli} we go through the explicit construction for the one- and two-moduli models respectively. In appendix \ref{app:boundarydata} we construct the nilpotent orbit data for the two-moduli periods, and in appendix \ref{app:embedding} we embed some geometrical examples from the literature into our models for the periods.

\section{Basics on Calabi-Yau periods and asymptotic Hodge theory}
\label{sec:Basics}

In this section we first provide some background material on period vectors for Calabi-Yau threefolds and briefly describe  their relevance 
in string compactifications. It will be necessary to introduce these objects  starting with a brief abstract discussion of  the variation of Hodge structures
in section \ref{VHS-section}. We then have a closer look at the asymptotic form of the period vector near boundaries of the complex structure 
moduli space. In section \ref{ssec:asympperiods} we recall some basics of asymptotic Hodge theory and introduce one of its central objects, the so-called nilpotent orbit. This orbit encodes the asymptotic form of the periods and we give a first indication how this information is recovered from the orbit. In section \ref{ssec:classification} we review the classification of the possible boundaries, which is based on the data provided by 
the nilpotent orbit. Finally, in section \ref{sec:sl2splitting} we explain how in strict asymptotic regimes an $sl(2)^n$-structure emerges from the nilpotent orbit.  For a mathematical review on the subject we refer to \cite{CKAsterisque}, and to \cite{Schmid, CKS} for the original articles.

\subsection{Variation of Hodge structure and physics of string compactifications} \label{VHS-section}

The geometric input for our physical theory is provided by the compactification manifold which in our case is a compact Calabi-Yau threefold $Y_3$. As a complex manifold, $Y_3$ requires the choice of a complex structure. The latter is far from being unique and there is a whole moduli space $\mathcal{M}_{\rm cs}(Y_3)$ parametrizing the complex structure whose coordinates are identified as the scalars in the vector multiplet of the supergravity theory. On the geometric side, we can consider the middle cohomology group $H^3(Y_3,\mathbb{Z})$ which admits a Hodge decomposition over the complex numbers
\begin{align}
H^3(Y_3,\mathbb{C})=H^3(Y_3,\mathbb{Z}) \otimes \mathbb{C}=\bigoplus_{k=0}^3 H^{p-k,k}\, , \quad \text{where } \bar{H}^{p,q}= H^{q,p} \,. \label{eq:HodgeDecomp}
\end{align}
We denote the dimensions of the subspaces as $h^{p,q}=\dim_{\mathbb{C}} (H^{p,q})$. From the definition, it is clear that this decomposition depends on the complex structure. Changing the latter by moving through moduli space $\mathcal{M}_{\rm cs}(Y_3)$ amounts to changing what we call holomorphic and anti-holomorphic. This means that while the total space $H^{3}(Y_3,\mathbb{C})$ remains unchanged the orientation of the subspaces $H^{p,q}$ inside $H^3(Y_3,\mathbb{C})$ varies, e.g. due to the Calabi-Yau condition $H^{3,0}$ is simply a complex line passing through the origin spanned by the nowhere vanishing holomorphic three-form $\Omega$ on $Y_3$. Alternatively, we can describe the space $H^3(Y_3,\mathbb{C})$ in terms of a decreasing filtration 
\begin{align}
0 \subset F^3 \subset F^2 \subset F^1 \subset F^0 = H^3 (Y_3,\mathbb{C})\, ,
\end{align}
where the relation to the previous description \eqref{eq:HodgeDecomp} is given by 
\begin{align}
F^p= \bigoplus_{k=p}^{3} H^{k,3-k} \, , \qquad F^p \cap \bar{F}^q = H^{p,q}\, ,
\end{align}
for $p+q=3$.  One generally refers to the $F^p$ as a Hodge filtration. The usefulness of this description comes from the fact that the $F^p$ vary holomorphically with the complex structure, i.e.
\begin{align}
\frac{\partial F^p}{\partial z} \subset F^{p-1} \, , \qquad \frac{\partial F^p}{\partial \bar{z}} \subset F^p \,. \label{eq:Transversality}
\end{align}
where $z$ and $\bar{z}$ are a holomorphic respectively anti-holomorphic complex structure moduli. We thus see that by taking derivatives with respect to the holomorphic complex structure moduli we move down the filtration but only by one degree at a time. The latter property is generally referred to as \textit{horizontality}. In a Calabi-Yau threefold we can recover the whole $H^3(Y_3,\mathbb{C})$ by taking holomorphic derivatives of $\Omega$ with respect to the complex structure moduli. Furthermore, the Hodge structure is said to be polarized if there exists an anti-symmetric bilinear pairing $\langle \, \cdot \, , \, \cdot \, \rangle$ such that the following holds 
\begin{equation}
\begin{aligned}
\langle F^p, \bar{F}^{4-p} \rangle &=0 \, ,\\
i^{p-q} \langle \omega,\bar{\omega} \rangle &> 0 \, ,\quad \text{ for } \omega \in H^{p,q} \text{ and } \omega \neq 0 \,. \label{eq:HRrelations}
\end{aligned}
\end{equation}
These two conditions are generally referred to as the Hodge-Riemann bilinear relations. Given such a bilinear pairing $\langle \, \cdot \, , \, \cdot \, \rangle$, a polarized variation of Hodge structure is defined to be a variation of Hodge structure such that the Hodge-Riemann bilinear relations hold for each point in complex structure moduli space $\mathcal{M}_{\rm cs}(Y_3)$. One of the deep insights of asymptotic Hodge theory is that this variation cannot be arbitrary, especially in asymptotic regions of $\mathcal{M}_{\rm cs}(Y_3)$ for which there exist powerful theorems that imply severe constraints thus making the problem far more tractable. 

There is of course nothing directly physical about an abstract variation of a Hodge structure, so we should say a few words about how it relates to the
characteristic functions that appear in effective theories obtained by compactifying string theory on $Y_3$. One conveniently formulates the 
effective theories in terms of the so-called period vector $\mathbf{\Pi}$, which is given in terms of the holomorphic $(3,0)$-form $\Omega$ by
\begin{align}
\mathbf{\Pi}= \begin{pmatrix}
\int_{A^I} \Omega \\
\int_{B_I} \Omega
\end{pmatrix}\, ,
\end{align}
where $A^I , B_I \in H_3(Y_3,\mathbb{Z})$ denote a suitable symplectic basis of three-cycles in $Y_3$ that satisfies the following intersection rules
\begin{align}
A^I \cap B_J = \delta_I^J \, , \qquad A^I \cap A^J = B_I \cap B_J =0 \,.
\end{align}
The anti-symmetric bilinear pairing $\langle \, \cdot \, , \, \cdot \, \rangle$ is naturally provided by the integration of differential forms over the Calabi-Yau threefold
\begin{align}
\langle u,v \rangle =\int_{Y_3} u \wedge v = \textbf{u}^T \eta \textbf{v} \,,
\end{align}
where $\textbf{u}, \textbf{v}$ are the vectors of coefficients of the corresponding differential forms in a given basis of $H^3(Y_3,\mathbb{C})$ and $\eta$ is the matrix representation of the bilinear pairing $\langle \, \cdot \, , \, \cdot \, \rangle$. By appropriately choosing our cohomology basis we can bring the bilinear pairing into the following standard form 
\begin{align}
\eta = \begin{pmatrix}
0 & \mathbb{I} \\
- \mathbb{I} & 0
\end{pmatrix}\, , \label{eq:PairingMatrix}
\end{align}
with $\mathbb{I}$ denoting the $(h^{2,1}+1)$-dimensional identity matrix. For the rest of this work it will be understood that we will assume a choice of basis such that $\eta$ is of the form given in \eqref{eq:PairingMatrix}. The horizontality property in \eqref{eq:Transversality} then takes the form
\begin{align}\label{eq:horizontality}
\langle \Omega , \partial_i \Omega \rangle = \mathbf{\Pi}^T \eta \partial_i \mathbf{\Pi}=0 \, , \qquad  \langle \Omega , \partial_i \partial_j \Omega \rangle = \mathbf{\Pi}^T \eta \partial_i \partial_j \mathbf{\Pi}=0 \,.
\end{align}
The pairing with the third order derivative does not vanish in general. 

To see how the period vector relates to the data of the effective theory, let us first recall that Type IIB string theory on $Y_3$ 
yields a $\cN=2$ supergravity effective action in four dimensions. This theory has $h^{2,1}(Y_3)$ vector multiplets and one vector 
in the gravity multiplet. The kinetic terms for these vectors can be encoded by a complex matrix $\cN_{IJ}$ that itself can be 
derived from the periods $\mathbf{\Pi}$. The explicit formulas can be found, for example, in refs.~\cite{Andrianopoli:1996cm,Craps:1997gp}. 
The kinetic terms for the complex scalars $z^i$ in the vector multiplets are encoded by a K\"ahler metric derived from the 
K\"ahler potential 
\beq \label{Kpot_Omega}
   K(z,\bar z) = - \log i \int_{Y_3} \bar \Omega \wedge \Omega  = - \log i \mathbf{\bar \Pi}^T \eta \mathbf{\Pi}\ .
\eeq 
The period vector also allows one to give an explicit expression for the central charge of a BPS particle of charge $\mathbf{q}$ by 
writing $Z = e^{K/2} \mathbf{q}^T \eta \mathbf{\Pi}$. This highlights that an explicit knowledge of $\mathbf{\Pi}$ is crucial if one wants 
to evaluate the characteristic functions in the $\cN=2$ vector sector. 

The importance of deriving $\mathbf{\Pi}$ also extends to certain $\cN=1$ effective actions. For example, the K\"ahler potential \eqref{Kpot_Omega} also appears in $\cN=1$ orientifold settings with O3/O7-planes, in which Type IIB string theory is compactified on the quotient $Y_3/\sigma$. Here $\sigma$ is the orientifold involution which might freeze, or project out, some of 
the $h^{2,1}(Y_3)$ complex structure moduli. In this setting the periods $\mathbf{\Pi}$ furthermore specify the superpotential \cite{Gukov:1999ya} when we turn on R-R or NS-NS fluxes $F_3$ and $H_3$, and it reads
\begin{equation}\label{eq:superpotential}
W = \int_{Y_3} G_3 \wedge \Omega = \langle G_3 \, , \Pi \rangle \, ,
\end{equation} 
where $G_3=F_3- \tau H_3$ and  $\tau $ denotes the axio-dilaton. Using this superpotential one can determine the scalar potential using the standard $\cN=1$ identities. Considering only the classical K\"ahler potential for the K\"ahler moduli, see e.g.~\cite{Grimm:2004uq} for details, we then find
\begin{equation}\label{eq:potential}
V = \frac{1}{\cV^2 \, \Im \tau}e^K  K^{I \bar{J}} D_I W D_{\bar{J}} \bar{W} = \frac{1}{4 \cV^2\, \Im \tau}  ( \langle \bar{G}_3\, , \ \ast G_3 \rangle - i \langle  \bar{G}_3\ , \, G_3  \rangle )\, ,
\end{equation} 
where the sum over $I,J$ runs over the complex structure moduli $t^i$ and the axio-dilaton $\tau$. The volume factor $\cV$ depends on the K\"ahler moduli of $Y_3$, which are not relevant to our work. For illustration we will compute these quantities \eqref{Kpot_Omega} \eqref{eq:superpotential} and \eqref{eq:potential} for each of the one- and two-moduli period vectors we construct in section \ref{sec:models}. Let us already point out that exponentially suppressed terms in the superpotential can conspire in such a way with the inverted K\"ahler metric such that they give contributions to the leading polynomial part of the scalar potential $V_{\rm lead}$ for finite and some infinite distance boundaries.

In summary, we realize that the period vector $\mathbf{\Pi}$ encodes all information about the Hodge structure of $Y_3$ and provides convenient 
way to encode the information of certain effective theories. However, it is crucial to realize that  $\mathbf{\Pi}$ generally is a very complicated 
function of the complex structure moduli. While it can be computed for explicit Calabi--Yau examples, it is hard to draw general conclusions about its structure. This is where asymptotic Hodge theory comes in and tells us that as we move to an asymptotic region in moduli space the period vector consists essentially of two pieces of data. The first is of algebraic nature and is given by the so-called \textit{nilpotent orbit}, while the second piece is analytic and given by a holomorphic, Lie algebra-valued, map that we will refer to as \textit{instanton map}. Facing our task to develop general asymptotic models for the period vector, we are in the fortunate situation that the  algebraic data captured by 
the nilpotent orbit can be classified \cite{Robles_2015, Kerr2017} as we will review in section \ref{ssec:classification}.  
Furthermore, it was shown in \cite{CattaniFernandez2000,CattaniFernandez2008} that the instanton map can subsequently be obtained by solving differential constraints that have the linear algebraic data as input. We will discuss this procedure in detail in section \ref{sec:instanton_map}.

\subsection{Asymptotic expansion and nilpotent orbits}\label{ssec:asympperiods}

Let us now give a more thorough description of the asymptotic regions in the complex structure moduli space $\mathcal{M}_{\rm cs}(Y_3)$ associated with the Calabi-Yau threefold $Y_3$. The moduli space has dimension $h^{2,1}$ and provides us with a continuous family of Calabi-Yau manifolds that give rise to, in general, different effective actions. The asymptotic regions in $\mathcal{M}_{\rm cs}(Y_3)$ are reached when approaching a boundary point in moduli space at which the manifold $Y_3$ develops singularities. 
A single modulus limit is given by sending one field, say $z^k$, to a boundary of  $\mathcal{M}_{\rm cs}(Y_3)$ that can be described as the codimension-one locus $z^k=0$. When $n$ moduli are involved, the boundary can be given as a normal intersection of $n$ loci of the form $z^k=0$ and thus, after appropriate reordering of the coordinates, can be locally described as $z^1=\dots=z^n=0$. Throughout this work, we refer to such an intersection locus as co-dimension $n$ boundary component, where it is understood that we exclude loci corresponding to further intersections that would thus correspond to higher co-dimension boundaries. Alternatively, we can use the coordinates\footnote{This amounts to working on the universal cover of the considered near boundary region in $\cM_{\rm cs}(Y_3)$.}
\begin{align}
t^i = x^i + i y^i= \frac{1}{2 \pi i } \log[z^i] \, ,\label{eq:Coordinates}
\end{align}
which are more directly related to physics as $x^i$ and $y^i$ have the interpretation as axions and saxions respectively in the supergravity framework. The relation \eqref{eq:Coordinates} should be kept in mind as we will use the coordinates interchangeably throughout this work. As follows from their definition, the limit towards the boundary in the coordinates $t^i$ corresponds to 
\begin{align}
t^i \to i \infty \,.
\end{align}
Of course, we are not obliged to take all the $h^{2,1}$ coordinates to the boundary. Sending only $n$ moduli to the boundary we refer to the remaining coordinates as spectator moduli and denote them as $\zeta^i$ with $i=n+1, \dots, h^{2,1}$.

Having set the stage we are now ready to discuss what happens in asymptotic regions of complex structure moduli space $\cM_{\rm cs}(Y_3)$. First, we give the key results in the general setting of the Hodge filtration $F^p$. Subsequently we rephrase everything in terms of the quantity that is of central importance in supergravity, namely the period vector $\mathbf{\Pi}$. We remind the reader that this is only a representative of the $F^3$ part of the Hodge filtration, so in physics we are generally in a situation where we profit from the horizontality property \eqref{eq:Transversality} to recover the rest of the information about the filtration. Circling the boundary divisor $z^k=0$ by $z^k \to e^{2\pi i} z^k$ corresponds to sending $t^k \to t^k + 1$ and induces a monodromy transformation on elements of the Hodge filtration\footnote{The matrix action on a form is understood as an action on the vector of coefficients when expanded in a given three-form basis.}
\begin{align}
\omega^p(t^k + 1) = T_k\, \omega^p (t^k) \,, \qquad \omega^p \in F^p\, ,
\end{align}
where the $T_k$ are the matrix generators of the monodromy and are elements of Sp$(2(h^{2,1}+1), \mathbb{R})$, which is also the duality group of the associated $\cN=2$ supergravity theories. It follows from \cite{Landman} that these generators are unipotent,\footnote{In general, we could have quasi-unipotent matrices, i.e. $(T^q- \mathbb{I})^{m+1}=0$ for some integers $q,m$. However, we can always make them unipotent, i.e. setting $q=1$, by coordinates redefinitions of the form $z \to z^q$.} i.e. $(T-\mathbb{I})^m=0$ for some positive integer $0 \leq m \leq \dim_{\mathbb{C}}Y_3=3$. From these generators we can define the so-called log-monodromy matrices $N_i=\log(T_i)$ that satisfy 
\begin{align}
[N_i,N_j]=0\, , \qquad N_i^T \eta + \eta N_i =0\, , \quad \quad \forall i,j\, ,
\end{align}
where $\eta$ was defined in \eqref{eq:PairingMatrix}. Given the unipotency of the monodromy generators $T_i$ one checks that the log-monodromy matrices are nilpotent of degree $0 \leq m \leq 3$. 

It follows from the nilpotent orbit theorem of asymptotic Hodge theory \cite{Schmid} that close to the boundary defined by $t^1,\dots,t^n \to i \infty$, we can write down an asymptotic expansion summarized by the vector space identity
\begin{align} \label{eq:AsymptoticFiltration}
F^p(t) = e^{t^i N_i} e^{\Gamma(z)} F^p_{0}\ ,
\end{align}
where $\Gamma(z)$ is a holomorphic $\mathfrak{sp}(2h^{2,1}+2)$-valued map that satisfies $\Gamma(0)=0$. The complex vector spaces $F^p_0$ define the so-called limiting filtration and do not depend anymore on the coordinates that have been taken to the boundary. Together, the limiting filtration $F^p_0$ and the log-monodromy matrices $N_i$ define a so-called limiting mixed Hodge structure that arises due to the degeneration of the pure Hodge structure on $H^{3}(Y_3,\mathbb{C})$ when the manifold develops singularities. Important properties of the limiting mixed Hodge structures will be picked up along the way when needed and a more detailed definition can be found in section \ref{ssec:classification}. Based on the asymptotic expansion \eqref{eq:AsymptoticFiltration}, the following limit is well-defined 
\begin{align}
\lim_{t^i \to i \infty} e^{-t^i N_i} F^p(t) = F_0^p \,. \label{eq:Limiting filtration}
\end{align}
This allows us to extract the limiting filtration $F^p_0$ for a given $F^p$ and shows that $F^p_0$ can still depend on spectator moduli. The nilpotent orbit theorem further implies that the full Hodge filtration $F^p$ can be successfully approximated by its nilpotent orbit
\begin{align} \label{Fnil_orbit}
F_{\rm nil}^p= e^{t^i N_i} F^p_{0} \,.
\end{align}
An estimate on how good the approximation of $F^p$ by $F_{\rm nil}^p$ is, can be found in \cite{Schmid}. In particular, note that $\Gamma(z)$ in \eqref{eq:AsymptoticFiltration} parametrizes the difference between the two sets of vector spaces, and it indeed vanishes at the boundary since $\Gamma(0)=0$. The crucial point is that $F^p_{\rm nil}$ also satisfies the Hodge-Riemann relations \eqref{eq:HRrelations} and thus still defines a proper polarized pure Hodge structure in the near-boundary region.  

One of the tasks in this work is to translate the asymptotic behavior of the Hodge filtration $F^p$ into an equivalent statements about the period vector $\mathbf{\Pi}$, which is the quantity we prefer for physical calculations. In other words we are searching for a concrete representative of $F^3$, 
which captures the information about the asymptotic Hodge filtration $F_{\rm nil}^p$. 
To begin with, we note that it is straightforward that \eqref{eq:AsyExpPeriod} implies that the period vector as a representative of $F^3$ also enjoys the asymptotic expansion
\beq \label{eq:AsyExpPeriod}
\mathbf{\Pi}(t)=e^{t^iN_i}e^{\Gamma(z)} \mathbf{a}_0= e^{t^iN_i}(\mathbf{a}_0 + \mathbf{a}_{1,r} e^{2 \pi i t^r} + \mathbf{a}_{2,rs} e^{2 \pi i (t^r + t^s)}+ \dots) \,,
\eeq
where $\mathbf{a}_0$ denotes the representative of $F^3_0$. 
For the second equality, we expanded the holomorphic map $\Gamma(z)$ in $z=e^{2\pi i t}$ and defined the vectors $\mathbf{a}_{k,r_1 \dots r_k}$ that capture the action of $\exp(\Gamma)$ on $\mathbf{a}_0$ at each order in the exponentially suppressed terms. Note that all $\mathbf{a}_{k,r_1 \dots r_k}$, with $k=0,...,\infty$, are independent of the coordinates $t^i$ that are close to the boundary. 
It also follows naturally that there is a corresponding leading polynomial part of the period vector given by  
\begin{align} 
\mathbf{\Pi}_{\rm pol}(t) = e^{t^i N_i} \mathbf{a}_0 \label{eq:nilpOrbit} \,,
\end{align}
which approximates thus $\mathbf{\Pi}$ as a representative of $F^3_{\rm nil}$. By comparing \eqref{eq:nilpOrbit} to the full period vector \eqref{eq:AsyExpPeriod}, we see that the terms $\mathbf{a}_{k,r_1 \dots r_k}$ with $k>0$ precisely correspond to exponential corrections
to a leading polynomial term obtained from $\mathbf{a}_0$. Employing a terminology familiar from the large complex structure point, we will 
refer to the exponentially suppressed terms loosely as \textit{non-perturbative corrections} or \textit{instanton terms}. This nomenclature stems from the mirror 
symmetric setting, which is well understood for the near-boundary region of the large complex structure point. In fact, in this mirror setting the exponential corrections actually stem from world-sheet instantons that wrap curves in the mirror Calabi-Yau threefold. In the following we will 
use this terminology also for the terms in a general period vector near any boundary.\footnote{Note that in Type IIB on a Calabi-Yau threefold the vector sector of the resulting effective theory is actually not corrected by world-sheet instantons, which couple to the K\"ahler moduli.}

It is crucial to stress the difference in using the full nilpotent orbit $F^p_{\rm nil}$, which gives a the complete information about the split of $H^3(Y_3,\mathbb{C})$,
compared with only looking at the polynomial $\mathbf{\Pi}_{\rm pol}$ given in \eqref{eq:nilpOrbit} parameterizing the leading polynomial part of an element of $F^3$.
The important point is that when using only $\mathbf{\Pi}_{\rm pol}$ it is not necessarily true that the information about the full filtration can be recovered from it by taking holomorphic derivatives. This is most easily understood by explicitly writing down the action of holomorphic derivatives on \eqref{eq:nilpOrbit} to find
\begin{align}
\partial_{t^i} \mathbf{\Pi}_{\rm pol}&= e^{t^m N_m} N_i \mathbf{a}_0\ , \nonumber \\
\partial_{t^i} \partial_{t^j} \mathbf{\Pi}_{\rm pol}&= e^{t^m N_m} N_i N_j  \mathbf{a}_0 \,, \quad \quad i,j,k=1,\dots,n  \, , \label{eq:DerivativesNilpOrbit}  \\ 
\partial_{t^i} \partial_{t^j} \partial_{t^k} \mathbf{\Pi}_{\rm pol}&= e^{t^m N_m} N_i N_j N_k \mathbf{a}_0 \, . \nonumber 
\end{align}
There is no need to consider higher order derivatives as in a Calabi-Yau threefold every element in $H^{3}(Y_3,\mathbb{C})$ can be expressed as linear combination of up to three derivative terms.
From \eqref{eq:DerivativesNilpOrbit} we can easily see that we only get something non-zero if the given combination of the log-monodromy matrices does not annihilate the vector $\mathbf{a}_0$. We want to emphasize here that this can already happen at the first derivative level, i.e.~for a single $N_i$ acting on it, which then implies that the polynomial part $\mathbf{\Pi}_{\rm pol}$ in this case contains information about $F^3_{\rm nil}$ but misses information about elements further down in the filtration. This data has been truncated when going from the asymptotic expansion \eqref{eq:AsyExpPeriod} of $\mathbf{\Pi}$  to the polynomial expression $\mathbf{\Pi}_{\rm pol}$. The prototypical example of such a case is the conifold point \cite{PhysRevLett.62.1956,Candelas:1989js,Strominger_1995}. However, it is also important to note that not all the higher order terms in the $z^i$ expansion would be necessary to recover the Hodge filtration $F^p$ but only a finite number as we will argue more precisely in section \ref{sec:instantonsnes}. 

Let us point out that there is also the special situation for which the expressions \eqref{eq:DerivativesNilpOrbit} are non-zero for all values of $i,j,k$. This happens for the well-studied large complex structure regime in which all $h^{2,1}$ moduli of $Y_3$ are taken to approach a boundary. The expressions \eqref{eq:DerivativesNilpOrbit} then imply that all the information in the asymptotic Hodge filtration $F^p_{\rm nil}$ can be recovered from $\mathbf{\Pi}_{\rm pol}$. This does not mean that there is no additional information in the higher order terms in $\Gamma(z)$. In fact, it is well-know in the context of mirror symmetry that the higher order terms generically appear in 
the period vector and correspond to actual world-sheet instanton corrections in string compactifications on the mirror Calabi-Yau threefold. 
In most cases, we have an intermediate situation where part of the information about the Hodge filtration is contained in the polynomial part
$\mathbf{\Pi}_{\rm pol}$, but additional information is required to reconstruct $F^p_{\rm nil}$ from the $(3,0)$-form periods. If we want to recover the residual information about the Hodge filtration, we have to rely on the $\Gamma(z)$ map, which extends the approximation at the boundary into the bulk of moduli space. 
To get a better handle on how much of the information about $F^p_{\rm nil}$ is captured by $\mathbf{\Pi}_{\rm pol}$, we next briefly review 
a classification of the different boundaries that can arise in the moduli space of any $Y_3$. This will also allow us to introduce the facts 
from asymptotic Hodge theory that are useful in our construction of the general models for the asymptotic periods.

\subsection{Classification of boundaries in complex structure moduli space}\label{ssec:classification}

Asymptotic Hodge theory can be used to systematically classify the possible boundaries that can occur
in the complex structure moduli space \cite{Robles_2015, Kerr2017}. This yields to a classification of possible nilpotent orbits $F^p_{\rm nil}$ introduced in \eqref{Fnil_orbit}. The main idea is to encode characteristic features of the boundary and associated nilpotent orbit by a finer splitting of the space of three-forms  $H^{3}(Y_{3},\mathbb{C})$ by vector spaces $I^{p,q}$. We begin with a brief review of these so-called 
Deligne splittings, and how they are used to classify boundaries in complex structure moduli space. 
Finally we also discuss briefly how this framework constrains intersections of singular divisors in moduli space. In particular, we summarize the recent classification of all possible intersections in a two-moduli setting by \cite{Kerr2017}, which will serve as our starting point for constructing general models of two-moduli periods later in sections \ref{sec:two-moduli}.

Let us begin by introducing the Deligne splitting $I^{p,q}$ of the space of three-forms $H^{3}(Y_{3},\mathbb{C})$. In this refined splitting 
one does not require $p+q=3$ as is the case for a pure Hodge structure, but rather only requires $0 \leq p,q \leq 3$. 
In fact, the Deligne splitting can be used to define a \textit{mixed Hodge structure} instead of a pure Hodge structure. The input needed to determine the spaces $I^{p,q}$ is the limiting filtration $F^p_0$ defined in \eqref{eq:Limiting filtration}, together the log-monodromy matrices $N_i$. Given the $N_i$ one first determines the so-called monodromy weight filtration $W_{l}(N)$, where $N=c_1 N_1+\ldots c_k N_k$ is any linear combination with coefficients $c_k >0$. We can compute these vector spaces $W_l(N)$ from the kernels and images of powers of this nilpotent element $N$ as \cite{SteenbrinkZucker}
\begin{equation}\label{eq:W}
W_{l}(N)= \sum_{j \geq \max(-1,l-3)} \ker N^{j+1} \cap \img N^{j-l+3}\, .
\end{equation}
One then computes the vector spaces $I^{p,q}$ from the vector spaces $F^{p}_{0}, W_{l}$ as
\begin{equation}\label{eq:Ipq}
I^{p,q} = F_{0}^{p} \cap W_{p+q} \cap \bigg( \bar{F}_{0}^{q}\cap W_{p+q}+\sum_{j\geq 1} \bar{F}_{0}^{q-j}\cap W_{p+q-j-1} \bigg)\, .
\end{equation}
It turns out that the resulting splitting is independent of the choice of $c_k$, so we typically pick $N\equiv N_{(k)}=N_1+\ldots + N_k$. 
The Deligne splitting then decomposes the vector spaces $F^{p}_{0}$ and $W_{l}$, given by
\begin{equation}\label{eq:FWsplit}
F^p_{0} = \sum_{r \geq p} \sum_{s} I^{r,s}\, , \qquad W_{l} = \sum_{p+q \leq l}I^{p,q}\, .
\end{equation} 
Let us now discuss how this Deligne splitting into $I^{p,q}$ can be used to classify boundaries in complex structure moduli space. This classification is based on the dimensions of these vector spaces, which can be combined into a Hodge-Deligne diamond as
\begin{equation}
 \begin{array}{ccccccccc} & & & i^{3,3}& &\\ 
& & i^{3,2}\hspace*{-.15cm}& & \hspace*{-.15cm} i^{2,3} \\ 
& i^{3,1}\hspace*{-.15cm} & & \hspace*{-.15cm} i^{2,2}\hspace*{-.15cm} & & \hspace*{-.15cm} i^{1,3}\\
 i^{3,0} \hspace*{-.15cm} & & i^{2,1} \hspace*{-.15cm} & & \hspace*{-.15cm} i^{1,2} & &\hspace*{-.15cm} i^{0,3}\\
&i^{2,0}\hspace*{-.15cm} & & \hspace*{-.15cm} i^{1,1}\hspace*{-.15cm} & & \hspace*{-.15cm} i^{0,2}\\
& &i^{1,0}\hspace*{-.15cm} & &  \hspace*{-.15cm} i^{0,1} \\
& & &i^{0,0}& & \\  
\end{array}\ , \label{eq:HDDiamond}
\end{equation}
where we denoted the dimensions by $i^{p, q} = \dim_\bbC I^{p, q}$. The numbers in this Hodge-Deligne diamond then admit various symmetries
\begin{equation}\label{eq:symmetries}
i^{p,q}=i^{q,p} = i^{3-q,3-p}\, ,
\end{equation}
while they also satisfy the inequality
\begin{equation}
i^{p-1,q-1} \leq i^{p,q} \, , \qquad p+q \leq 3\, .
\end{equation}
The dimensions $i^{p,q}$ can be related to the Hodge numbers $h^{p,q} = \dim_\mathbb{C} H^{p,q}$ of the underlying pure Hodge structure by
\begin{equation}\label{eq:hformula}
h^{p,3-p} = \sum_{q=0}^3 i^{p,q}\, .
\end{equation}
An interesting feature to point out is that one can move downwards in the Deligne splitting by acting on elements with the log-monodromy matrices $N_i$  (for $1 \leq i \leq n$). Namely, the $N_i$ are $(-1,-1)$-maps with respect to the Deligne splitting $I^{p,q}$, meaning these are elements of
\begin{equation}\label{eq:Nmin1min1}
N_i \in \Lambda_{-1,-1}\, ,
\end{equation}
where we defined spaces of operators acting on the $I^{p,q}$ by 
\begin{equation}\label{eq:lambdapq}
\cO_{p,q} \in \Lambda_{p,q} : \qquad  \cO_{p,q} I^{r,s} \subseteq I^{r+p,s+q}\, .
\end{equation}
In other words, application of log-monodromy matrices lowers elements by two rows in the Hodge-Deligne diamond \eqref{eq:HDDiamond}. 

Another aspect we would like to point out is that the spaces $I^{p,q}$ into primitive and non-primitive pieces under the nilpotent operator $N$. The primitive components of the $I^{p,q}$ can be computed from
\begin{equation}
P^{p,q}(N)=I^{p,q}(N) \cap \ker N^{p+q-2}\, ,
\end{equation}
with $p+q \geq 3$. The $I^{p,q}$ can then be spanned by the primitive pieces as
\begin{equation}
I^{p,q}(N) = \oplus_k N^k P^{p+k,q+k}(N)\, .
\end{equation}
Bilinear operators of the form $\langle \cdot, \, N^\ell \cdot \rangle$ then satisfy certain constraints with respect to these primitive spaces $P^{p,q}$, known as \textit{polarization conditions}. For our purposes these provide us with positivity conditions given by
\begin{equation}\label{eq:pol}
i^{p-q} \langle N^{p+q-3} \bar{v}  \, , \  v \rangle >0 \, , 
\end{equation}
for $v \in P^{p,q}$. We choose to work with a fixed expression \eqref{eq:PairingMatrix} for the symplectic pairing, so these inequalities impose constraints on the log-monodromy matrices $N_i$. In practice these positivity conditions fix the signs of  coefficients in the log-monodromy matrices $N_i$, which will prove to be useful in the explicit construction of one- and two-moduli periods.

Having reviewed the most relevant aspects of the Deligne splitting, we now turn to the classification of boundaries in the complex structure moduli space. By specializing to Calabi-Yau threefolds we have to put $h^{3,0}=1$, which implies that only one of the numbers $i^{3,d}$ can be non-vanishing according to \eqref{eq:hformula}. Therefore we can make a separation of cases in this classification based on whether $i^{3,d}=1$ for $d=0,\, 1,\, 2, \, 3$. We label these cases by the principal types $\mathrm{I},\, \mathrm{II}, \, \mathrm{III},\, \mathrm{IV}$ respectively. For which value of $d$ we have $i^{3,d}=1$ turns out to have interesting implications for the behavior of the period vector. One finds that the leading order term $\mathbf{a}_0$ in the expansion \eqref{eq:AsyExpPeriod} spans the one-dimensional space $I^{3,d}$, since from \eqref{eq:Ipq} it can be deduced that $I^{3,d}=F^3_0$. The integer $d$ then captures the maximal number of times $N_{(k)}$ can be applied on $\mathbf{a}_0$ as
\begin{equation}\label{eq:ddef}
N_{(k)}^{d} \mathbf{a}_0 \neq 0\, , \qquad N_{(k)}^{d+1} \mathbf{a}_0 = 0\,.
\end{equation}
After exploiting the symmetries \eqref{eq:symmetries} one then finds that the remainder of the Hodge-Deligne diamond is made up by the middle components $i^{p,q}$ with $1\leq p,q \leq 2$. Invoking the sum of all $i^{p,q}$ to be equal to the total dimension $2h^{2,1}+2$, only $i^{2,2}$ remains as a free index for the singularity type. Attaching the dimension $i^{2,2}$ as a subscript to the principal types, we end up with $4h^{2,1}$ possible types of singularities. This classification has been summarized in table \ref{table:HDclass}. 

\begin{table}[h!]
\centering
\renewcommand*{\arraystretch}{2.0}
\begin{tabular}{| c| c | c | c | c |}
\hline singularity & $\mathrm{I}_a$ & $\mathrm{II}_b$ & $\mathrm{III}_c$ & $\mathrm{IV}_d$ \\ \hline \hline 
\begin{minipage}{0.15\textwidth}
\vspace{-1.3cm}
HD diamond
\vspace{1.3cm}
\end{minipage} &
\rule[-0.25cm]{.0cm}{3.5cm} \begin{tikzpicture}[scale=0.65,cm={cos(45),sin(45),-sin(45),cos(45),(15,0)}]
  \draw[step = 1, gray, ultra thin] (0, 0) grid (3, 3);

  \draw[fill] (0, 3) circle[radius=0.05];
  \draw[fill] (1, 2) circle[radius=0.05] node[above]{$a'$};
  \draw[fill] (2, 1) circle[radius=0.05] node[above]{$a'$};
  \draw[fill] (1, 1) circle[radius=0.05] node[above]{$a$};
  \draw[fill] (2, 2) circle[radius=0.05] node[above]{$a$};
  \draw[fill] (3, 0) circle[radius=0.05];
\end{tikzpicture} &
\begin{tikzpicture}[scale=0.65,cm={cos(45),sin(45),-sin(45),cos(45),(15,0)}]
  \draw[step = 1, gray, ultra thin] (0, 0) grid (3, 3);

  \draw[fill] (0, 2) circle[radius=0.05];
  \draw[fill] (1, 3) circle[radius=0.05];
  \draw[fill] (1, 2) circle[radius=0.05] node[above]{$b'$};
  \draw[fill] (1, 1) circle[radius=0.05] node[above]{$b$};
  \draw[fill] (2, 1) circle[radius=0.05] node[above]{$b'$};
  \draw[fill] (2, 2) circle[radius=0.05] node[above]{$b$};
  \draw[fill] (2, 0) circle[radius=0.05];
  \draw[fill] (3, 1) circle[radius=0.05];
\end{tikzpicture} &
\begin{tikzpicture}[scale=0.65,cm={cos(45),sin(45),-sin(45),cos(45),(15,0)}]
  \draw[step = 1, gray, ultra thin] (0, 0) grid (3, 3);
  \draw[fill] (0, 1) circle[radius=0.05];
  \draw[fill] (1, 0) circle[radius=0.05];
  \draw[fill] (1, 2) circle[radius=0.05] node[above]{$c'$};
  \draw[fill] (2, 1) circle[radius=0.05] node[above]{$c'$};
  \draw[fill] (2, 3) circle[radius=0.05];
  \draw[fill] (1, 1) circle[radius=0.05] node[above]{$c$};
  \draw[fill] (2, 2) circle[radius=0.05] node[above]{$c$};
  \draw[fill] (3, 2) circle[radius=0.05];
\end{tikzpicture} &
\begin{tikzpicture}[scale=0.65,cm={cos(45),sin(45),-sin(45),cos(45),(15,0)}]
  \draw[step = 1, gray, ultra thin] (0, 0) grid (3, 3);

  \draw[fill] (0, 0) circle[radius=0.05];
  \draw[fill] (1, 1) circle[radius=0.05] node[above]{$d$};
  \draw[fill] (1, 2) circle[radius=0.05] node[above]{$d'$};
  \draw[fill] (2, 1) circle[radius=0.05] node[above]{$d'$};
  \draw[fill] (2, 2) circle[radius=0.05] node[above]{$d$};
  \draw[fill] (3, 3) circle[radius=0.05];
\end{tikzpicture} \\ \hline
index &  \begin{minipage}{.15\textwidth}\centering\vspace*{-0.cm} \begin{equation*}\begin{aligned} a+a'&=h^{2,1} \\ 0\leq a &\leq h^{2,1} \end{aligned}\end{equation*}  \vspace*{-0.cm} \end{minipage} & \begin{minipage}{.18\textwidth}\centering\vspace*{-0.cm} \begin{equation*}\begin{aligned}b+b'&=h^{2,1}-1 \\ 0\leq b &\leq h^{2,1}-1 \end{aligned}\end{equation*}  \vspace*{-0.cm} \end{minipage}& \begin{minipage}{.18\textwidth}\centering\vspace*{-0.cm} \begin{equation*}\begin{aligned} c+c'&=h^{2,1}-1 \\ 0 \leq c &\leq h^{2,1}-2 \end{aligned}\end{equation*}  \vspace*{-0.cm}\end{minipage}&\begin{minipage}{.15\textwidth}\centering\vspace*{-0.cm} \begin{equation*}\begin{aligned} d+d'&=h^{2,1} \\ 1 \leq d &\leq h^{2,1} \end{aligned}\end{equation*}  \vspace*{-0.cm} \end{minipage}\\ \hline
\begin{minipage}{0.156\textwidth}
\vspace{-1.1cm} 
\ \ \ \  (signed) \\
Young diagram 
\end{minipage}&   \begin{minipage}{0.156\textwidth}
\vspace{-1.1cm} \begin{tikzpicture}[scale=0.4]
    \draw (0, 0) rectangle (1, -1);
    \draw (.5, -.5) node {$+$};
    \draw (1, 0) rectangle (2, -1);
    \draw (1.5, -.5) node {$-$};
    \draw (2, -.5) node[right] {\small $\otimes \, a$};
    \draw (0, -1) rectangle (1, -2);
    \draw (2, -1.5) node[right] {\small $\otimes \, 2a' + 2$};
  \end{tikzpicture}
  \end{minipage} & 
   \rule[-0.4cm]{.0cm}{2.1cm}   \begin{tikzpicture}[scale=0.4]
    \draw (0, 0) rectangle (1, -1);
    \draw (.5, -.5) node {$+$};
    \draw (1, 0) rectangle (2, -1);
    \draw (1.5, -.5) node {$-$};
    \draw (2, -.5) node[right] {\small $\otimes \, b$};
    \draw (0, -1) rectangle (1, -2);
    \draw (.5, -1.5) node {$-$};
    \draw (1, -1) rectangle (2, -2);
    \draw (1.5, -1.5) node {$+$};
    \draw (2, -1.5) node[right] {\small $\otimes \, 2$};
    \draw (0, -2) rectangle (1, -3);
    \draw (2, -2.5) node[right] {\small $\otimes \, 2b'$};
  \end{tikzpicture} &
    \begin{tikzpicture}[scale=0.4]
    \draw[step = 1] (0, 0) grid (3, -1);
    \draw (3, -.5) node[right] {\small $\otimes \, 2$};
    \draw (0, -1) rectangle (1, -2);
    \draw (.5, -1.5) node {$+$};
    \draw (1, -1) rectangle (2, -2);
    \draw (1.5, -1.5) node {$-$};
    \draw (3, -1.5) node[right] {\small $\otimes \, c$};
    \draw (0, -2) rectangle (1, -3);
    \draw (3, -2.5) node[right] {\small $\otimes \, 2c' - 2$};
  \end{tikzpicture} &
    \begin{tikzpicture}[scale=0.4]
    \draw[step = 1] (0, 0) grid (4, -1);
    \draw (.5, -.5) node {$-$};
    \draw (1.5, -.5) node {$+$};
    \draw (2.5, -.5) node {$-$};
    \draw (3.5, -.5) node {$+$};
    \draw (4, -.5) node[right] {\small $\otimes \, 1$};
    \draw (0, -1) rectangle (1, -2);
    \draw (.5, -1.5) node {$+$};
    \draw (1, -1) rectangle (2, -2);
    \draw (1.5, -1.5) node {$-$};
    \draw (4, -1.5) node[right] { \small $\otimes \, d - 1$};
    \draw (0, -2) rectangle (1, -3);
    \draw (4, -2.5) node[right] {\small $\otimes \, 2d'$};
  \end{tikzpicture} \\ \hline
$\text{rk}(N,N^2,N^3)$ & $(a,\, 0,\, 0)$ & $(2+b,\, 0,\, 0)$ & $(4+c,\, 0,\, 0)$ & $(2+d,\, 2,\, 1)$ \\ \hline
eigvals $\eta N$ & $a$ negative & \begin{minipage}{0.15\textwidth}\centering\vspace*{0.2cm}
$b$ negative \\
2 positive  \vspace*{0.15cm}
\end{minipage} & not needed & not needed \\ \hline
\end{tabular}
\caption{\label{table:HDclass} Classification of singularity types in complex structure moduli space based on the $4h^{2,1}$ possible different Hodge-Deligne diamonds. In each Hodge-Deligne diamond we indicated non-vanishing $i^{p,q}$ by a dot on the roster, where the dimension has been given explicitly when $i^{p,q} >1$. In the last two rows we listed the characteristic properties of the log-monodromy matrix $N$ and the symplectic pairing $\eta$ that are sufficient to make a distinction between the types.}
\end{table}

For the construction of periods later it is useful to note that these Deligne splittings also characterize the allowed forms for the nilpotent element $N$. This is achieved most straightforwardly by mapping the classification of Hodge-Deligne diamonds to signed Young diagrams. These signed Young diagrams characterize the form of a nilpotent element $N$ up to basis transformations that preserve the symplectic pairing \eqref{eq:PairingMatrix}. To be more precise, signed Young diagrams classify the conjugacy classes of nilpotent elements under the adjoint action of Sp$(2h^{2,1}+2, \mathbb{R})$, i.e.~$N \to g N g^{-1}$ for $g \in \text{Sp}(2h^{2,1}+2,\mathbb{R})$. In table \ref{table:HDclass} we listed the signed Young diagram for each of the Hodge-Deligne diamonds. While we do not review the details of this correspondence here, let us simply point out that there is a minimal set of building blocks that can be used to assemble these nilpotent elements for the given signed Young diagrams, which are included in table \ref{table:buildingblocks} for completeness.

In one-dimensional moduli spaces it suffices to characterize boundaries by a single type, but in higher-dimensional moduli spaces we can extract more data. Instead of considering a limit where all coordinates are sent to the boundary at the same rate, we can take ordered limits in the sense that $y^1 \gg y^2 \gg \ldots \gg y^n$, which we will refer to as \textit{strict asymptotic regimes}. In other words, we first take $y^1 \to \infty$, thereafter $y^2 \to \infty$, up to $y^n \to \infty$. At each step one then classifies the limit involving $y^1  , \, \ldots , \, y^k$ by the type associated with the sum of log-monodromy matrices $N_{(k)}$. Sending additional coordinates to their limit enhances the type, so we can combine the types of an ordered limit into an \textit{enhancement chain} as
\begin{equation}\label{eq:enhancementchain}
\mathrm{I}_0\xrightarrow{\ y^{1} \rightarrow \infty\ }\  {{\sf Type\ A}_{(1)}}\ \xrightarrow{\ y^{2} \rightarrow \infty\ }\  {\sf Type\ A}_{(2)} \ \xrightarrow{\ y^{3} \rightarrow \infty\ }\  
\ldots \ \xrightarrow{\ y^{n} \rightarrow \infty\ }\  {\sf Type\ A}_{(n)} \, .
\end{equation}
We always start in the non-degenerate case $\text{I}_0$  in the interior of the moduli space, and the types ${{\sf  A}_{(k)}}$ in the subsequent steps indicate the type associated with taking the limit $y^1  , \, \ldots , \, y^k \to \infty$. By considering all possible orderings of the coordinates $y^1,\ldots,y^{h^{2,1}}$ we can then extract a more refined set of invariants for singularities. In \cite{Grimm:2019bey} this approach was used to classify, using mirror symmetry, Calabi-Yau threefolds based on the limit pattern of its K\"ahler moduli space.

In order to make the above story more explicit, we turn our attention to classifications of one- and two-moduli setups. The only possible one-modulus types compatible with the bounds in table \ref{table:HDclass} are $\mathrm{I}_{1}$, $\mathrm{II}_{0}$ and $\mathrm{IV}_{1}$. The type $\mathrm{IV}_{1}$ can be viewed as a one-modulus large complex structure point so it does not require any instanton terms in $\mathbf{\Pi}$ to recover the nilpotent orbit $F_{\rm nil}$, but the other two types provide interesting examples for our study. The type $\mathrm{I}_{1}$ is realized as a conifold point in e.g.~the moduli space of the quintic \cite{Candelas:1990rm}. The type $\mathrm{II}_{0}$ is less well-known and arises from a so-called Tyurin degeneration \cite{Tyurin:2003}, and has also been studied later in \cite{Doran:2005gu,doran2016mirror, Joshi:2019nzi}. We construct general models for the periods near these two types of boundaries in section \ref{app:one-modulus}, and summarize the obtained results in section \ref{sec:onemodels}.

\begin{figure}[h!]
\begin{center}
\includegraphics[width=7.5cm]{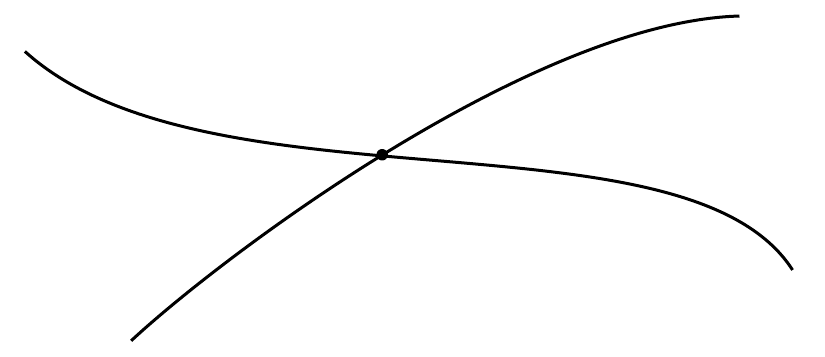}
\end{center}
\begin{picture}(0,0)\vspace*{-1.2cm}
\put(131,104){\small $\mathrm{I}_1$}
\put(294,118){\small $\mathrm{IV}_1$}
\put(207,84){\small $\mathrm{IV}_2$}
\end{picture}\vspace*{-1.0cm}
\caption{\label{fig:2cube}Example of the 2-cube $\langle \mathrm{I}_1 | \mathrm{IV}_2 | \mathrm{IV}_1 \rangle$ that characterizes certain two-moduli coni-LCS boundaries. A $\mathrm{I}_1$ and $\mathrm{IV}_1$ divisor in complex structure moduli space intersect, where they enhance to a $\mathrm{IV}_2$ singularity type.}
\end{figure}
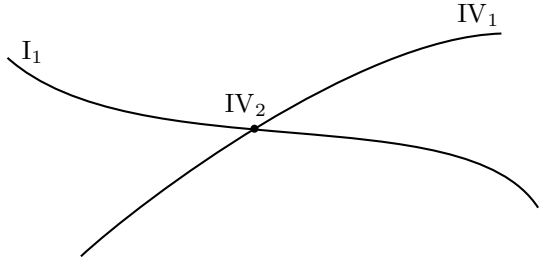

Studying two-moduli setups is more intricate, since we have to consider different orderings of the limits towards the singularity as discussed below \eqref{eq:enhancementchain}. This results in a richer picture for its classification by Hodge-Deligne diamonds. In addition to the type arising at the intersection $y^1=y^2 = \infty$, we also have to consider the types obtained from sending just one coordinate to the boundary, i.e.~$y^{1} =  \infty$ or $y^{2} =  \infty$.  To this end it is interesting to point out the recent mathematical work \cite{Kerr2017}, where precisely these two-moduli boundaries have been classified. The limit types were combined into a so-called \textit{2-cube} as $\langle \mathrm{A}_1 | \mathrm{A}_{(2)} | \mathrm{A}_2 \rangle$, where $A_i$ denotes the limit type for $y^i \to \infty$, and $A_{(2)}$ denotes the limit type for $y^1,y^2 \to \infty$. This 2-cube then characterizes the intersection of two boundary divisors in a two-dimensional moduli space as depicted in figure \ref{fig:2cube}. We do not review the details of this classification \cite{Kerr2017}, but will simply use the exhaustive set of 2-cubes that was obtained. We present expressions for the periods near each of these boundaries in section \ref{sec:twomodels}, and refer to section \ref{sec:two-moduli} for their construction. 

\subsection{Strict asymptotic regimes and the $sl(2)^n$-splitting}\label{sec:sl2splitting}
Asymptotic Hodge theory and more precisely, the $sl(2)$-orbit theorem \cite{Schmid, CKS}, allow for a further approximation to the nilpotent orbit. This condenses the boundary data into a minimal form, which can be effectively formulated in terms of $sl(2)^n$ representation theory. The systematic classification reduces to a few characteristic building blocks that are derived from simple principles, providing the perfect starting point for our program of reverse engineering the period vector in section \ref{sec:instantonperiods}. The goal of this subsection is not to give the $sl(2)$-orbit construction in all of its details but merely to give the reader a rough overview on how the nilpotent orbit data gets truncated into this simpler form. This will hopefully result in a better feeling for the procedure of constructing periods following in section \ref{sec:instantonperiods}.

Going from the nilpotent orbit data associated with a given codimension $n$ boundary to the $sl(2)^n$-data, requires us to specify how we approach the boundary by fixing the relative scaling of the involved complex structure moduli. This was introduced as strict asymptotic regime in the previous subsection and determines for us an enhancement chain. There is of course more than one way of approaching a boundary and thus naturally multiple possible enhancements chains. Consequently, the exact form of the $sl(2)^n$-data depends on this choice. However, when running the process in reverse as we do later, the result does not depend on what starting data we have used, as it should. Once the strict asymptotic regime is specified, the $sl(2)$-orbit theorem guarantees us that there is a iterative algorithm that allows us to encode the nilpotent orbit data, which consists out of the filtration $F_0^p$ and the log-monodromy matrices $N_i$, in terms of a set of $n$ mutually commuting $sl(2)$-triples $(N_i^-,N_i^+,N_i^0)$, one for each $N_i$, and a filtration $\tilde{F}_0^p$. From $\tilde{F}_0^p$ and the $N_i^-$ we can construct an associated Deligne splitting $\tilde{I}^{p,q}_{(n)}$, which is a direct sum of finite dimensional $sl(2)^n$ representations under these triples. Roughly, one should  think of each column (without multiplicity) of the Hodge-Deligne diamond as an irreducible representation with the $N_i^-$ and $N_i^+$ acting as standard lowering respectively raising operators, while the $N_i^0$ are representing the weight operators. The presentations of the data in terms of the triples and either the filtration $\tilde{F}^p_0$ or the space $\tilde{I}^{p,q}_{(n)}$ are equivalent and will be used interchangeably from here on. One starts the algorithm at the final singularity type of the enhancement chain \eqref{eq:enhancementchain} associated with the co-dimension $n$ boundary and then goes down step-by-step until reaching the $I_0$ boundary.

The idea is that we associate a distinguished $sl(2)^n$-split limiting filtration $\tilde{F}^p_{0,k}$ to each step $k$ in the enhancement chain. We can obtain these $sl(2)^n$-split filtrations iteratively from the limiting filtration $F_0^p$ that defines the nilpotent orbit of the codimension $n$ boundary under consideration. It was shown in \cite{CKS} that there exist two uniquely defined rotation operators that relate a limiting filtration to its $sl(2)^n$-split counterpart. We denote the operators that relate $F^p_{0}$ to its $sl(2)$-split $\tilde{F}^{p}_{0}=\tilde{F}^p_{0,n}$ by $\delta_n$ and $\zeta_n$, giving us the $sl(2)^n$-split filtration at the last step of the enhancement chain. We can then move one step down in the enhancement chain, and construct a limiting filtration $F^p_{0,n-1}$ at step $n-1$ from $\tilde{F}^p_{0,n}$. Generally this limiting filtration is not $sl(2)^n$-split, so one has to perform another rotation into $\tilde{F}^p_{0,n-1}$ by computing new operators $\delta_{n-1}$ and $\zeta_{n-1}$ One can repeat this process until reaching the beginning of the enhancement chain corresponding to the $\text{I}_0$ boundary. At this moment, we take the existence of the rotation operators for granted. More details are given in section \ref{sec:nilpotent_orbit}, and we also refer to \cite{Grimm:2018cpv} where this algorithm has been applied to explicit examples. The rest of the construction is neatly summarized by the two relations
\begin{align}\label{eq:sl2recursion}
 \tilde{F}^p_{0,k}= e^{ \zeta_k} e^{-i \delta_k}  F^p_{0,k}\, , \qquad F_{0,k-1}^p=\exp(i N_k) \tilde{F}_{0,k}^p \, ,
 \end{align}
where $k=1,\dots,n $ refers to the enhancement step. Using the $\tilde{F}^p_{0 , k}$ together with the corresponding weight filtration $W(N_{(k)})$, one defines the $sl(2)^n$-splittings $\tilde{I}^{p,q}_{(k)}$ for each enhancement step according to the definition \eqref{eq:Ipq}. These have the nice property that under complex conjugation
\begin{align}
\overline{\tilde{I}^{p,q}_{(k)}} = \tilde{I}^{q,p}_{(k)} \,.
\end{align}
Second, we construct the mutually commuting $sl(2)$-triples $(N^-_i, N^+_i,N^0_i)$. These satisfy the standard algebraic relations
\begin{equation}
[N_i^+, N_i^-]=2 N_i^0\, , \qquad [N_i^{\pm}, N_i^0] = \pm N_i^\pm \, , \label{eq:sl2Relations}
\end{equation}
supplemented by the fact that they are infinitesimal isometries of the bilinear pairing defined in \eqref{eq:PairingMatrix}, i.e. $\langle N^{\bullet}_i \, \cdot \, , \, \cdot \, \rangle = - \langle \, \cdot \, , \, N^{\bullet}_i \, \cdot \, \rangle $. We start by determining the weight operators $N^0_i$ for the $sl(2)$-triples. This requires an intermediate step where we construct a weight operator $N^0_{(k)}$ for each $\tilde{I}^{p,q}_{(k)}$ space in the enhancement chain with the natural action
\begin{equation}
N^0_{(k)}\, \omega= (p+q-3) \, \omega\, , \qquad \omega \in \tilde{I}^{p,q}_{(k)} \, . \label{eq:IntermediateWeightOp}
\end{equation}
The weight operators that appear in the $sl(2)$-triples are then defined as $N_{i}^0=N_{(i)}^0-N_{(i-1)}^0$ with the convention $N_{(0)}^0=0$, reflecting the fact that the $I_0$ singularity at the beginning of the enhancement chain corresponds to a pure Hodge structure of weight three, making all the eigenvalues identically zero, i.e. $p+q=3$. In the next step, we determine the lowering operators $N_i^-$ from the log monodromy matrices $N_i$. For that we can write down the following decomposition
\begin{equation}
N_i = N_i^- + \sum_{\ell \geq 2} N_{i, -\ell}\, ,  
\end{equation}
where $\ell$ specifies the weight under $N_{(i-1)}^0$, i.e. $[N_{(i-1)}^0, N_{i,-\ell}]=- \ell N_{i,-\ell}$, making it clear that the operator we are looking for is nothing else but the part with eigenvalue zero under the adjoint action of $N_{(i-1)}^0$. It also follows immediately that we always have $N_1^-=N_1$ because of $N_{(0)}^0=0$. For the last step, one can solve the defining relations \eqref{eq:sl2Relations} to determine all the raising operators $N^+_i$. These will however not play a central role in our work, so we will typically only write down the weight and lowering operators $N_i^0,N_i^-$ when we refer to an $sl(2)$-triple.

\ytableausetup
 {mathmode, boxsize=1.2em}
\newcolumntype{A}{>{\centering\arraybackslash} m{.3\linewidth} }

\begin{table}[h!]
\begin{center}
{\small
\begin{tabular}{|c|c|c|c|}
\hline
    \rule[-.1cm]{0cm}{0.5cm} signed Young diagram &  $N_{(n)}^-$  & $N^0_{(n)}$ \\
   \hline
   \hline
  \begin{ytableau}
    \ \\
    \ 
  \end{ytableau}
  &    \rule[-.5cm]{0cm}{1.2cm} $\left( \begin{array}{cc} 0 & 0 \\ 0 & 0\end{array} \right)$  & $\left( \begin{array}{cc} 0 & 0 \\ 0 & 0\end{array} \right)$  
  \\
   \hline
  \begin{ytableau}
    + & -
  \end{ytableau}
  & \rule[-.5cm]{0cm}{1.2cm}  $\left( \begin{array}{cc} 0 & 0 \\ -1 & 0\end{array} \right)$  &$\left( \begin{array}{cc} 1 & 0 \\ 0 & -1\end{array} \right)$  
  \\
    \hline
  \begin{ytableau}
    - & +
  \end{ytableau}
  &  \rule[-.5cm]{0cm}{1.2cm}  $\left( \begin{array}{cc} 0 & 0 \\ 1 & 0\end{array} \right)$& $\left( \begin{array}{cc} 1 & 0 \\ 0 & -1\end{array} \right)$  
  \\    
       \hline
  \begin{ytableau}
    \ & \ & \ \\
    \ & \ & \
  \end{ytableau}
  &  \rule[-1.3cm]{0cm}{2.8cm}   $\left( \begin{array}{cccccc} 0 & 0 & 0 & 0 & 0 & 0 \\ \ 1 & 0 & 0 & 0 &0 & 0\\0 & \ 1 & 0 &0 & 0 & 0\\ 0 & 0 &0 & 0 & -1 & 0\\ 0 &0 & 0 & 0 & 0 & -1\\ 0 & 0 & 0 & 0 & 0 & 0 \end{array} \right)$
    & $\left( \begin{array}{cccccc} 2 & 0 & 0 & 0 & 0 & 0 \\ \ 0 & 0 & 0 & 0 &0 & 0\\0 &  0 & -2 & 0 & 0 & 0\\ 0 & 0 &0 & -2 & 0 & 0\\ 0 &0 & 0 & 0 & 0 & 0\\ 0 & 0 & 0 & 0 & 0 & 2 \end{array} \right)$ 
    \\
    \hline
  \begin{ytableau}
    - & + & - & +
  \end{ytableau}
    & \rule[-0.9cm]{0cm}{2.0cm}  $\left( \begin{array}{cccc} 0 & 0 & 0 & 0  \\ \ 1 & 0 & 0 & 0 \\0 & 0 & 0 &-1 \\ 0 & 1 &0 & 0 
     \end{array} \right)$
    &  $\left( \begin{array}{cccc}    3 & 0 & 0 &0  \\   0 & 1 & 0 & 0 \\  0 & 0 & -3 & 0  \\   0 & 0 & 0 & -1 \end{array} \right)$    \\    
    \hline
\end{tabular}
}
\caption{Building blocks for the lowering and weight operators $N^-_{(n)},N^0_{(n)}$ for all relevant signed Young diagrams, where the pairing matrix always takes the standard form \eqref{eq:PairingMatrix}. We can obtain simple normal forms for these matrices by combining the building blocks into the complete signed Young diagrams given in table \ref{table:HDclass}.} \label{table:buildingblocks}
\end{center}
\end{table}

We conclude this section by presenting the building blocks that come out of the classification from the $sl(2)^n$-data. In section \ref{ssec:classification} it was briefly explained how singularity types can be efficiently classified with the help of signed Young diagrams and the correspondence was given in table \ref{table:HDclass}. We summarize the relevant building blocks for the $sl(2)^n$-data in table \ref{table:buildingblocks}. Assembling them appropriately such that they are compatible with the bilinear pairing \eqref{eq:PairingMatrix}, we can obtain the most general form of the operators $N^-_{(n)}$ and $N^0_{(n)}$. In turn, from these all the triples and in particular the lowering operators $N^-_i$  can be deduced. Together with the  filtration $\tilde{F}^{p}_0$ , or equivalently the $sl(2)^n$-splitting $\tilde{I}^{p,q}_{(n)}$, these form the starting point for the analysis of section \ref{sec:instantonperiods} where we explain how to build up general expressions for periods and how to integrate the crucial Hodge-theoretic information into the period vector.

\section{Instanton expansion of the periods}\label{sec:instantonperiods}
In this section we elucidate the structure behind instanton terms in the periods. We begin by explaining why these instanton terms are expected to be present from the perspective of asymptotic Hodge theory. To be concrete, we obtain in section \ref{sec:instantonsnes} a criterion \eqref{eq:instantonpresence} for the presence of instanton terms and a lower bound \eqref{eq:columncount} on the number of instanton terms required. We then turn to the techniques used to construct asymptotic expressions for the periods. In \ref{sec:nilpotent_orbit} we explain how to write down the most general nilpotent orbit compatible with a given set of $sl(2)^n$-data, i.e.~how to construct the log-monodromy matrices $N_i$ and the filtration $F^p_0$. In \ref{sec:instanton_map} we describe the instanton map $\Gamma(z)$ that encodes the instanton expansion the period vector \eqref{eq:AsyExpPeriod}. In particular, we discuss a rank condition \eqref{eq:rankGamma} for $\Gamma(z)$ indicating the essential instanton terms necessary for recovering the entire filtration $F^p_0$ from just the $(3,0)$-form periods. Finally, let us note that we focus on boundary components of codimension $h^{2,1}$ in this section, i.e.~we set $n=h^{2,1}$ in the following.

\tikzstyle{block} = [draw,, rectangle, 
    minimum height=3em, minimum width=2em, align=center]
   
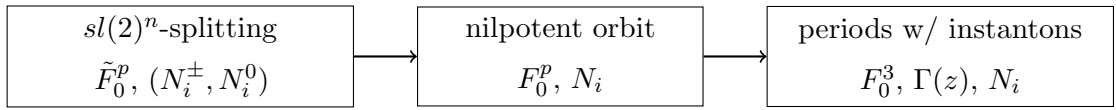
\begin{figure}[h!]
\centering
\begin{tikzpicture}
\node[block, text width=4.3cm] (a) at (0,0) {$sl(2)^n$-splitting\\ \vspace{0.2cm}
$\tilde{F}^p_{0}$, $(N_i^{\pm}, N_i^0)$ };
\node[block, text width=3.5cm] (b) at (5,0) {nilpotent orbit \\ \vspace{0.2cm}
$F^p_0$, $N_i$ };
\node[block, text width=4.3cm] (c) at (10,0) {periods w/ instantons \\ \vspace{0.2cm}
$F^3_0$, $\Gamma(z)$, $N_i$ };

\draw [ thick, ->] (a) -- (b);
\draw [ thick, ->] (b) -- (c);
\end{tikzpicture}
\caption{Flowchart illustrating the steps in constructing the periods. We start by writing down the data of the $sl(2)^n$-splitting that characterizes a strict asymptotic regime near the boundary. We then extend from this strict asymptotic regime to other regions near the boundary by constructing the most general nilpotent orbit compatible with this data. 
As second step we lift the data encoded in the nilpotent orbit $F^p_0$ into the leading terms of an instanton map $\Gamma(z)$ acting only on $F^3_0$, resulting in exponential corrections to the periods.}\label{fig:flowchart}
\end{figure}

\subsection{Presence of instanton terms} \label{sec:instantonsnes}

Based on the classification of boundary types by using the dimensions of Deligne splittings discussed in section \ref{ssec:classification}, we can already infer non-trivial information about the instanton terms $\mathbf{a}_i$ in the period vector expansion \eqref{eq:AsyExpPeriod}. We find that only one type of boundary region does not require the presence of instanton terms, while all other types need these terms in order to be able to recover the full mixed Hodge structure from the period vector.  Here we consider the boundary components that are of co-dimension $h^{2,1}$ in the 
complex structure moduli space, i.e.~that arise from sending all moduli to a limit. We can state this result as
\begin{equation}\label{eq:instantonpresence}
\fbox{\rule[-0.5cm]{.0cm}{1.1cm} \
\text{\parbox{.85\textwidth}{\centering Every period vector $\mathbf{\Pi}$ near a co-dimension $h^{2,1}$ boundary component that is \textit{not} of type $\mathrm{IV}_{h^{2,1}}$ \textit{must} contain 
instanton terms $\mathbf{a}_{i}$ in its expansion \eqref{eq:AsyExpPeriod}.  }} \ }
\end{equation}
Before we argue for this result, let us try to put it into a broader perspective. First, let us stress that such a simple statement cannot generally be formulated 
for periods near lower co-dimension boundary components. While we can apply a similar strategy near boundary components of lower co-dimensions, the necessity of 
instantons will depend on more details of the local period vector that are not captured by simply stating the boundary type. Second, we recall that 
a well-known class of type $\mathrm{IV}_{h^{2,1}}$ boundaries are the large complex structure points. These points evade the above statement, and indeed do not require the presence of instanton terms for consistency. Mathematically this follows from the fact that the vector $\mathbf{a}_0$ and its descendants obtained by applying $N_i$ suffice to span the complete vector space $H^3(Y_3,\mathbb{C})$. Interestingly, there is another closely-related class of $\mathrm{IV}_{h^{2,1}}$ boundaries, which we dub coni-LCS points. These boundary components 
can be obtained, in certain examples, by considering a large complex structure point and then sending one modulus away to a conifold locus. 
In terms of the associated mixed Hodge structure one then finds that this one-modulus limit results in a $\mathrm{I}_a$ boundary component, whereas additionally sending the remaining moduli to the large complex structure regime can still yield a $\mathrm{IV}_{h^{2,1}}$ point at the intersection. While these coni-LCS points evade the above theorem as well, the type $\mathrm{I}_a$ mixed Hodge structure does require us to consider instanton terms for the conifold modulus. In fact, we will study these points later explicitly in section \ref{sec:coniLCS} as the intersection of $\mathrm{I}_1$ and $\mathrm{IV}_{1,2}$ divisors in two-dimensional moduli spaces. 

Let us now argue for the statement \eqref{eq:instantonpresence}. We know that the vector $\mathbf{a}_0$ spans the vector space $I_{(n)}^{3,d}$ in the Deligne splitting. Application of the log-monodromy matrices $N_i$ then lowers us within the same column according to \eqref{eq:Nmin1min1}. This implies that the dimension of the vector space spanned by $\mathbf{a}_0$ and its descendants is bounded from above by
\begin{equation}
\dim_\mathbb{C} \big(\text{span}_{\mathbb{C}}( N_{i_1} \cdots N_{i_k} \mathbf{a}_0 ) \big) \leq \sum_{k=0}^d i^{3-k,d-k}\, ,
\end{equation}
where the span runs over all values $k=0,1,2,3$ and $0 \leq i_1,i_2,i_3 \leq n$. In other words, the vector $\mathbf{a}_0$ and its descendants span at most the column of $I_{(n)}^{3,d}$ in the Deligne splitting. Looking at table \ref{table:HDclass}, this means that we can only generate the vector space $H^3(Y_3,\mathbb{C})$ in its entirety via $\mathbf{a}_0$ for type $\mathrm{IV}_{h^{2,1}}$ singularities. In order to span the other columns of the Deligne splitting, we need other elements to enter in the input $F^p_0$. As pointed out below \eqref{eq:DerivativesNilpOrbit}, these elements enter through the instanton terms $\mathbf{a}_i$ in the period vector expansion, completing our vector space $H^3(Y_3,\mathbb{C})$. Thus we find that we must require the presence of instanton terms whenever a singularity is not of type  $\mathrm{IV}_{h^{2,1}}$. 

From \eqref{eq:instantonpresence} we do know whether instanton terms $\mathbf{a}_i$ must be present for a given boundary type, but let us now try to make the minimal number required more precise. We need additional elements in the boundary filtration $F^p_0$ in order to span the other columns of the Deligne splitting $I_{(n)}^{p,q}$, besides the column of $I_{(n)}^{3,d}$ corresponding to $\mathbf{a}_0$. Roughly speaking each instanton term $\mathbf{a}_i$ can only be identified with one column, since descendants via application of $N_i$ end up in the same column according to \eqref{eq:Nmin1min1}. Therefore we only need to count the number of columns in order to get a lower bound on the number of instanton terms required. Looking at table \ref{table:HDclass} we find that
\begin{equation}\label{eq:columncount}
\mathrm{I}_a : \ 2h^{2,1}-a+1\, , \qquad \mathrm{II}_b :\ 2h^{2,1}-b-1\, , \qquad \mathrm{III}_c : \ c+1\, , \qquad \mathrm{IV}_d :\ 2(h^{2,1}-d)\, .
\end{equation}
Note that in this counting scheme we interpret $i^{p,q}>1$ at the top of a column as having $i^{p,q}$ columns. Namely, in this case we need at least $i^{p,q}$ instanton terms $\mathbf{a}_i$ in order to span the $i^{p,q}$-dimensional space $I^{p,q}$.

\subsection{Reconstructing the periods I: nilpotent orbit}\label{sec:nilpotent_orbit} 
In the following two subsections we lay out how to reverse engineer asymptotic periods that include these essential instanton terms. In this subsection we describe how to work out the first step in figure \ref{fig:flowchart}, i.e.~construct the nilpotent orbit. We already covered in section \ref{sec:sl2splitting} how the nilpotent orbit can be approximated by an $sl(2)$-splitting when moving to a strict asymptotic regime. Here we will turn the story around, and show how to write down the most general nilpotent orbit compatible with a given $sl(2)$-splitting. In other words, we extend from a particular strict asymptotic regime characterized by this $sl(2)$-splitting into other regions near this boundary.

Our construction starts from a given $sl(2)$-splitting. Recall from section \ref{sec:sl2splitting} that this splitting is encoded in a set of commuting $sl(2)$-triples $(N_i^\pm, N_i^0)$ and an $sl(2)$-split Deligne splitting $\tilde{I}^{p,q}_{(n)}$. This data characterizes the boundary in a strict asymptotic regime, which we take to be $y_1 \gg \ldots \gg y_n$ without loss of generality. Crucially for us, the possible $sl(2)$-splittings that can arise are classified through the limiting mixed Hodge structures given in table \ref{table:HDclass}. Furthermore, there is a systematic procedure to write down simple expressions for the defining data of these $sl(2)$-splittings. This procedure translates the classifying Deligne splittings into signed Young diagrams as discussed in section \ref{sec:sl2splitting}. For our purposes the details in this correspondence are not important, but we simply note that there exists a set of simple building blocks given in table \ref{table:buildingblocks} that can be used to assemble the defining elements of the $sl(2)$-splitting. 

The task of writing down the most general nilpotent orbit compatible with the $sl(2)$-splitting is then twofold. Firstly, we want to construct the most general log-monodromy matrices $N_i$ that match with the lowering operators $N_i^-$ in the strict asymptotic regime. Secondly, we want to consider the most general rotation away from the $sl(2)$-split $\tilde{I}^{p,q}_{(n)}$ for the Deligne splitting. In the remainder of this subsection we describe how to carry out both of these tasks.

Let us begin by describing how to construct the most general log-monodromy matrices $N_i$ out of the lowering operators $N_i^-$. Our procedure follows the approach taken in the study of Deligne systems in the mathematics literature, see for instance example 6.61 in \cite{BrosnanPearlsteinRobles} and also \cite{KaplanPearlstein,BrosnanPearlstein}. These systems formalize the structure behind the log-monodromy matrices $N_i$ and $sl(2)$-triples $(N_i^{\pm}, N_i^0)$ into a purely linear algebraic setup, without any reference to an underlying geometrical origin. Following section \ref{sec:sl2splitting}, we want to reverse engineer the decomposition of the log-monodromy matrices
\begin{equation}\label{eq:NtoNmin}
N_k = N_k^- + \sum_{\ell \geq 2}  N_{k,-\ell} \, ,
\end{equation}
where the components $N_{k,-\ell}$ have as weights
\begin{equation}
\begin{aligned}
 \ [N^0_{(k-1)}, \, N_{k,-\ell}] &=-\ell N_{k,-\ell}\, , \\
[N^0_{k}, \, N_{k,-\ell}] &=(\ell -2) N_{k,-\ell}\, , \qquad  [ N^+_k, \, N_{k,-\ell}]=0\, ,
\end{aligned}
\end{equation} 
where the last commutator tells us that $N_{k,-\ell}$ is a highest weight $\ell-2$ state under the $sl(2)$-triple $(N_k^\pm, N_k^0)$. For consistency we must supplement these commutators by some additional constraints. For instance we require the resulting log-monodromy matrices to be infinitesimal isometries of the symplectic pairing and to commute with each other, i.e.~$(N_{k,-\ell})^T\eta+\eta N_{k,-\ell}=0$ and $[N_k,N_r]=0$. Additionally, we have to impose that $N_k$ is a $(-1,-1)$-map with respect to the $sl(2)$-split Deligne splitting $\tilde{I}^{p,q}_{(k)}$ as described by \eqref{eq:Nmin1min1}. Altogether this provides us with a set of linear algebraic constraints that have to be satisfied by the components $N_{k,-\ell}$. In practice, we can therefore systematically solve these equations to obtain the most general expressions for the log-monodromy matrices $N_i$ compatible with the given $sl(2)$-splitting. Finally, let us note that the resulting log-monodromy matrices still have to be constrained by the polarization conditions \eqref{eq:pol}, typically yielding positivity conditions on the free parameters in the components $N_{k,-\ell}$.

Next we discuss how to rotate the Deligne splitting away from the $sl(2)$-split $\tilde{I}^{p,q}_{(n)}$ characterizing the intersection. We parametrize this rotation by two real matrices $\delta,\zeta$, and use that $\zeta$ is fixed componentwise by phase operator $\delta$ as shown by \cite{CKS}. The most convenient way to describe this rotation is by acting with these matrices on the filtration $\tilde{F}^{p}_{0}$ of the $sl(2)$-splitting as described by \eqref{eq:sl2recursion}.  For completeness let us record the reverse of this identity here, which states that the rotation away from the $sl(2)$-splitting is given by
\begin{equation}
F^p_0 = e^{i \delta} e^{-\zeta} \tilde{F}^p_0\, .
\end{equation}
Finding the most general rotation away from the $sl(2)$-splitting thus amounts to writing down the most general phase operator $\delta$. Following up on the discussion in section \ref{sec:sl2splitting}, we now show how to constrain $\delta$. To begin with, it is an infinitesimal isometry of the symplectic pairing similar to the log-monodromy matrices, i.e.~$\delta^T \eta+\eta \delta =0$. It can be decomposed with respect to the $sl(2)$-splitting $\tilde{I}^{p,q}_{(n)}$ as
\begin{equation}\label{eq:delta}
\delta = \sum_{p,q \geq 1} \delta_{-p,-q}\, , \qquad \delta_{-p,-q} \in \Lambda_{-p,-q}\, ,
\end{equation}
where $ \Lambda_{p,q}$ is defined in  \eqref{eq:lambdapq}. Furthermore, we must require that $\delta$ also commutes with the log-monodromy matrices
\begin{equation}\label{eq:deltacom}
[N_i \, , \, \delta ] = 0\, .
\end{equation}
These conditions together can then be solved in order to determine the most general phase operator $\delta$ allowed for a given $sl(2)$-splitting. The componentwise relations for $\zeta$ can be written out as \cite{Kato}
\begin{equation}\label{eq:zeta}
\begin{aligned}
\zeta_{-1,-1} &= \zeta_{-2,-2} = 0  \, , \qquad \zeta_{-1,-2} = -\frac{i}{2} \delta_{-1,-2}\, , \qquad &\zeta_{-1,-3} &= -\frac{3i}{4} \delta_{-1,-3}\, , \\
\zeta_{-2,-3} &= -\frac{3i}{8} \delta_{-2,-3} - \frac{1}{8}\big[ \delta_{-1,-1}, \, \delta_{-1,-2} \big] \, , \quad &\zeta_{-3,-3} &= -\frac{1}{8} \big[ \delta_{-1,-1} , \, \delta_{-2,-2} \big] \, ,
\end{aligned}
\end{equation}
and the other components of $\zeta$ follow by complex conjugation. More concretely, let us note that this rotation captures additional model-dependent parameters in our expressions for the periods, since for instance it rotates the leading term $a_0 \in F^3_0$ in the expansion of the period vector \eqref{eq:AsyExpPeriod}. A well-known example of this sort is the $\alpha'$-correction that arises in the mirror Calabi-Yau threefold and corrects the period vector near the LCS point with a term proportional to the Euler characteristic.

\subsection{Reconstructing the periods II: instanton map}\label{sec:instanton_map}

We now turn to the instanton map $\Gamma(z)$. We use this map to describe the expansion in instanton terms $z=e^{2\pi i t}$ for the period vector. This map has originally been studied in great detail in \cite{CattaniFernandez2000,CattaniFernandez2008}, and we review the relevant aspects of their work here. Let us first state how we can recover the Hodge filtration $F^p$ from the boundary structure by using $\Gamma(z)$. We can write it in terms of the limiting filtration $\tilde{F}^p_0$ of the $sl(2)$-splitting $\tilde{I}^{p,q}_{(n)}$ as\footnote{In comparison to \cite{CattaniFernandez2000} we chose to expand $F^p_0 = e^{i\delta} e^{-\zeta} \tilde{F}^p_0$, and rewrite in terms of the filtration $\tilde{F}^p_0$ of the $sl(2)$-split mixed Hodge structure instead of $F^p_0$. Furthermore we commuted the exponentials involving $\delta$ and $\zeta$ to the left, which means that the instanton maps are related by $\Gamma |_{\rm here} =e^{\zeta} e^{-i\delta} \Gamma |_{\rm there}  e^{i\delta} e^{-\zeta}  $.  \label{fn:footnoteGamma}}
\begin{equation}\label{eq:FpGamma}
F^p = e^{i\delta} e^{-\zeta} e^{t^i N_i} e^{\Gamma(z)} \, \tilde{F}^p_0\, ,
\end{equation}
where $\Gamma(z)$ is a matrix-valued function holomorphic in $z=e^{2\pi it}$ with $\Gamma(0)=0$. Vanishing at $z=0$ ensures that the nilpotent orbit $e^{t^{i}N_{i}}F^{p}_{0}$ provides a good approximation for $F^{p}$ for $y^{i} \gg 1$. To be more precise, $\Gamma(z)$ is a map valued in the Lie algebra $\mathfrak{sp}(2h^{2,1}+2)$, located in 
\begin{equation}\label{eq:locGamma}
\Gamma(z) \in \Lambda_{-} = \bigoplus_{p<0}\bigoplus_q \Lambda_{p,q}\, ,
\end{equation}
where we consider the operator spaces $\Lambda_{p,q}$ with respect to the $sl(2)$-split Deligne splitting $\tilde{I}^{p,q}_{(n)}$. From a practical perspective this means one needs to determine a basis for elements of $\Lambda_{-}$ that lie in the Lie algebra $\mathfrak{sp}(2h^{2,1}+2)$. One can then write out $\Gamma(z)$ by expanding in terms of this basis, where holomorphic functions vanishing at $z=0$ are taken as coefficients. Later we find that these holomorphic coefficients can be constrained by differential equations obtained from \eqref{eq:gammamin1} and \eqref{eq:recursion}. 

For the purposes of this work we want to translate the vector space relation \eqref{eq:FpGamma} into an expression for the period vector. By taking a representative $\mathbf{\tilde{a}}_0$ of $\tilde{F}^{3}_{0}$, we find that we can write the period vector $\mathbf{\Pi}$ as
\begin{equation}\label{eq:PiGamma}
\boxed{
\rule[-.25cm]{.0cm}{0.8cm} \quad
\mathbf{\Pi}(t) = e^{i\delta} e^{-\zeta} e^{t^i N_i} e^{\Gamma(z)} \, \mathbf{\tilde{a}}_0\, . \quad}
\end{equation}
where we wrote again $z=e^{2\pi i t}$ for convenience. When comparing the asymptotic expansion \eqref{eq:PiGamma} to \eqref{eq:AsyExpPeriod}, one should keep in mind that although we used the same notation for the $\Gamma(z)$ map to make things simpler, they are related as explained in footnote \ref{fn:footnoteGamma}. In the context of the asymptotic expansion \eqref{eq:PiGamma} the vanishing condition $\Gamma(0)=0$ can be understood as the statement that the nilpotent orbit approximation \eqref{eq:nilpOrbit} provides a good estimate for the period vector for $y^i \gg 1$. 

In order to constrain the instanton map $\Gamma(z)$, we can now use the horizontality property of the Hodge filtration as described by \eqref{eq:Transversality}. The idea is that besides \eqref{eq:locGamma} the instanton map should satisfy certain differential conditions to produce a consistent period vector. To obtain these conditions, it is convenient to first combine the exponential maps acting on  $\tilde{F}^p_0$ in \eqref{eq:FpGamma} (respectively on $\mathbf{\tilde a}_{0}$ in \eqref{eq:PiGamma}) into a single Sp$(2h^{2,1}+2,\mathbb{C})$-valued matrix. We define this matrix as
\begin{equation}\label{eq:Edef}
E(t) =   \exp[X(t)]  \equiv e^{i\delta} e^{-\zeta} e^{t^i N_i} e^{\Gamma(z)}  \, ,
\end{equation}
where $X(t)$ is valued in $\mathfrak{sp}(2h^{2,1}+2,\mathbb{C})$ and $\Lambda_{-}$, since $\delta,\zeta,N_i,\Gamma(z)$ are all valued in these operator subspaces. For later reference let us write out the component in $\Lambda_{-1}=\bigoplus_{q} \Lambda_{-1,q}$ explicitly as
\begin{equation}\label{eq:Xmin1}
X_{-1}(t) = i\delta_{-1}-\zeta_{-1}+t^{i}N_{i}+\Gamma_{-1}(z)\, ,
\end{equation}
which follows simply from expanding the exponentials in \eqref{eq:Edef}. The horizontality property \eqref{eq:Transversality} can now be recast into a condition on $E(t)$ by rewriting the Hodge filtration $F^p$ with \eqref{eq:FpGamma}. This leads to a vector space relation that reads
\begin{equation}\label{eq:EdEsubset}
\big( E^{-1}\partial_{i} E \big) \tilde{F}^{p}_{0} \subseteq \tilde{F}^{p-1}_{0}\, .
\end{equation}
From writing out the exponentials in \eqref{eq:Edef} it already follows that $E^{-1}\partial_{i} E \in \Lambda_-$. However, we also know that the $\tilde{F}^p_0$ can be split into $\tilde{I}_{(n)}^{p,q}$ according to \eqref{eq:FWsplit}, so $E^{-1}\partial_{i} E$ can only be valued in the operator subspaces $\Lambda_{-1,q}$. Therefore we must impose
\begin{equation}\label{eq:EdE}
E^{-1}\partial_{i} E \in \Lambda_{-1}\, .
\end{equation}
By expanding the exponentials in \eqref{eq:Edef} we find that this implies
\begin{equation}\label{eq:EdEdX}
E^{-1}\partial_{i} E = \partial_i X_{-1} \, ,
\end{equation} 
since higher order terms are valued in the operator subspaces $\Lambda_{-2,q}$ or lower. Note in particular from \eqref{eq:Edef} and \eqref{eq:Xmin1} that the operators $\delta$ and $\zeta$ drop out of this relation, which can be seen immediately on the right-hand side since they are constant, while on the left-hand side they can be moved past the partial derivative.

The differential constraint \eqref{eq:EdEdX} on the instanton map $\Gamma(z)$ ensures that we can integrate the boundary data into a consistent period vector. However, imposing \eqref{eq:EdEdX} directly is not the most practical way to constrain this instanton map. In \cite{CattaniFernandez2000} a convenient approach was given to reduce \eqref{eq:EdEdX}. The idea is to first derive a necessary and sufficient condition \eqref{eq:gammamin1} on the component $\Gamma_{-1}(z)$ of the instanton map. Subsequently the lower-charged components $\Gamma_{-p}(z)$ with $p\geq 2$ can be fixed recursively through \eqref{eq:recursion}. The differential condition on $\Gamma_{-1}(z)$ is obtained by taking another derivative $\partial_{j}$ of \eqref{eq:EdEdX} and antisymmetrizing in $i,j$, which yields\footnote{This condition is more naturally obtained by introducing an exterior derivative $d=\partial_{i} dt^{i}$ on the moduli space.}
\begin{equation}\label{eq:dXdX}
\partial_{[i} X_{-1} \partial_{j]}X_{-1} = 0\, .
\end{equation}
By using \eqref{eq:Xmin1} we can formulate this as a differential constraint on $\Gamma_{-1}(z)$ as
\begin{equation}\label{eq:gammamin1}
[N_{i}, \partial_{j} \Gamma_{-1}(z) ] + [\partial_i \Gamma_{-1}(z), N_j ] +[\partial_i \Gamma_{-1}(z), \partial_j \Gamma_{-1}(z) ] = 0 \, ,
\end{equation}
where we used that $[N_i, N_j] =0$ since the log-monodromy matrices commute. Next we need to obtain constraints on the lower-charged components $\Gamma_{-q}(z)$ with $q>2$ of the instanton map. First we write out \eqref{eq:EdEdX} by multiplying from the left with $\exp[\Gamma(z)]$ as
\begin{equation}
\partial_{i} \exp[\Gamma(z)] = [\exp[\Gamma(z)], N_{i}]+  \exp[\Gamma(z)]  \partial_{i} \Gamma_{-1}(z)\, .
\end{equation}
This condition can be translated into a constraint on the components in the subspaces $\Lambda_{-p} = \bigoplus_q \Lambda_{-p,q}$ as
\begin{equation}\label{eq:recursion}
\partial_{i} \exp[\Gamma(z)]_{-p} = [\exp[\Gamma(z)]_{-p+1}, N_{i}]+  \exp[\Gamma(z)]_{-p+1}  \partial_{i} \Gamma_{-1}(z)\, .
\end{equation}
From  the left-hand side we obtain the term $\Gamma_{-p}(z)$ by expanding the exponential, while the other terms that appear in the equation are of charge $\Gamma_{-p+1}(z)$ or lower. This means we can fix $\Gamma_{-p}(z)$ uniquely in terms of the lower-charged components $\Gamma_{-1}(z),\ldots, \Gamma_{-p+1}(z)$. By induction we thus find that the entire map $\Gamma(z)$ is uniquely determined by its piece $\Gamma_{-1}(z)$, provided this piece solves the consistency requirement \eqref{eq:gammamin1}.

It is worthwhile to check how coordinate redefinitions affect the instanton map $\Gamma(z)$ and $\delta$, since these transformations can later be used to reduce the number of arbitrary components for both. We can understand their effect most naturally by looking at $X_{-1}(t)$ given in \eqref{eq:Xmin1}. The most general divisor-preserving coordinate redefinition takes the form $z^{i} \to z^{i}f(z)$, where $f(z)$ is any holomorphic function with $f(0) \neq 0$. In terms of the coordinates $t^i$ defined in \eqref{eq:Coordinates} this amounts to shifting $2\pi i t^i \to 2\pi i t^i + \log f(z)$. Applying this shift to $t^i N_i$ produces two terms, a constant term involving $\log[f(0)]$ and a holomorphic term involving $\log[f(z)/f(0)]$ which vanishes at $z=0$. Taking $f(0)$ to be real we can absorb the former into the phase operator $\delta$, and the latter into $\Gamma(z)$. To be precise, from \eqref{eq:Xmin1} we find the following shifts
\begin{equation}\label{eq:shift}
\begin{aligned}
\delta_{-1,-1} &\to \delta_{-1,-1} - \frac{1}{2\pi }  \log[ f(0)] \,  N_i \, , \\
\Gamma_{-1}(z) &\to \Gamma_{-1}(z)+ \frac{1}{2\pi i}  \log\Big[ \frac{f(z^{i})}{f(0)} \Big] \, N_{j}\, .
\end{aligned}
\end{equation}
Later we will expand both of these maps into a basis for $\mathfrak{sp}(2h^{2,1}+2)$ that is valued in the appropriate operator subspaces $\Lambda_{p,q}$. From these shifts we learn that we are free to set the components along the log-monodromy matrices $N_i$ to zero, effectively reducing the number of arbitrary coefficients that have to be dealt with. Let us also note that we did not yet exploit the full set of coordinate redefinitions: we can still rotate $f(0)$ by a complex phase, corresponding to a shift of the axion $x^i$ in $t^i=x^i+i y^i$. A shift $x^i \to x^i + c^i$ can then partially be absorbed by a basis transformation $e^{cN^i}$ for the Deligne splitting $\tilde{I}^{p,q}_{(n)}$, while it also rotates the complex phase of exponentially suppressed terms in the periods. The latter feature will prove to be useful in the explicit construction of the periods in one- and two-moduli settings in section \ref{sec:constructionperiods}, since it allows us to set the leading instanton coefficients to real values.

Finally, let us discuss the precise conditions that need to be imposed on $\Gamma(z)$ in order to realize the instanton terms required by \eqref{eq:instantonpresence}.  For Calabi-Yau threefolds we want that derivatives of the period vector together span the vector space $H^3(Y_3,\mathbb{C})$ as alluded to in section \ref{ssec:asympperiods}. In terms of the Hodge filtration $F^p$ this amounts to putting an equality sign in \eqref{eq:EdEsubset} when we take all possible linear combinations of the partial derivatives $E^{-1} \partial_i E$ into account on the left-hand side. Following \cite{CattaniFernandez2008} we can translate this statement into a more concrete condition involving the instanton map $\Gamma(z)$. The vector space $\tilde{F}^3_0 =\tilde I_{(n)}^{3,d}$ is one-dimensional, while the span of all $E^{-1} \partial_i E $ needs to be able to generate all lower lying spaces $I^{p,q}$ with $p<3$. The total dimension of these spaces is given by $2h^{2,1}+1$, so we find that
\begin{equation}\label{eq:rankGamma}
\dim \Big(\bigoplus_i  \img  (N_i+\partial_i \Gamma_{-1}) \Big)= 2h^{2,1}+1\, ,
\end{equation}
where we wrote out $E^{-1} \partial_i E $ in terms of $N_i$ and $\partial_i \Gamma_{-1}$ according to \eqref{eq:EdEdX} and \eqref{eq:Xmin1}. This condition can be understood intuitively by considering the Hodge-Deligne diamond \eqref{eq:HDDiamond}. It implies that either a log-monodromy matrix $N_i$ or the instanton map $\Gamma_{-1}$ should map into every space $\tilde I_{(n)}^{p,q}$ in the Deligne splitting apart from $\tilde I_{(n)}^{3,d}$. Since the log-monodromy matrices $N_i$ are $(-1,-1)$-maps and therefore only act vertically on \eqref{eq:HDDiamond}, this means we need $\Gamma_{-1}$ to generate the horizontally separated columns. In practice, we will use \eqref{eq:rankGamma} to determine which components of $\Gamma_{-1}$ are required to be non-vanishing. For the one-modulus setups discussed in section \ref{app:one-modulus} we find that $\Gamma_{-1}$ has only one functional degree of freedom, so \eqref{eq:rankGamma} dictates if this function must be non-vanishing. For the two-modulus setups studied in section \ref{sec:two-moduli} we find that $\Gamma_{-1}$ consists of several holomorphic functions, and \eqref{eq:rankGamma} will generically only indicate for some of these whether they must be non-vanishing.

\section{Models for one- and two-moduli periods}\label{sec:models}
Here we present general expressions for the periods near one- and two-moduli boundaries. We refer to sections \ref{app:one-modulus} and \ref{sec:two-moduli} for the construction of these periods to avoid distracting the reader by technical details. Crucially, we include the essential instanton terms for boundaries away from large complex structure in accordance with our discussion in section \ref{sec:instantonperiods}. This section is written such that it can be read without understanding the ingredients that go into this derivation. In particular, these periods can be used directly in studying four-dimensional supergravity theories, and to illustrate this point we readily compute the corresponding K\"ahler potentials, flux superpotentials and scalar potentials.

\subsection{Models for one-modulus periods} \label{sec:onemodels}
In this section we present general expressions for the periods near boundaries in one-dimensional moduli spaces, and refer to section \ref{app:one-modulus} for the details. Based on the classification reviewed in section \ref{ssec:classification} there are three possible types of boundaries for $h^{2,1}=1$, given by
\begin{equation}
\begin{aligned}
\mathrm{I}_1 &: \qquad  \text{conifold point}\, ,\\
\mathrm{II}_0 &: \qquad \text{Tyurin degeneration}\, ,\\
\mathrm{IV}_1 &: \qquad \text{large complex structure point}\, .
\end{aligned}
\end{equation}
As indicated, each of these types of boundaries has a natural geometrical interpretation in the complex structure moduli space of Calabi--Yau threefolds. The type $\mathrm{I}_1$ characterizes conifold points, which arise for instance in the moduli space of the mirror quintic, cf.~\cite{Candelas:1990rm}. Type $\mathrm{II}_0$ boundaries arise from so-called Tyurin degenerations \cite{Tyurin:2003}, and these periods have also been studied recently in the context of the swampland programme in \cite{Joshi:2019nzi}. Finally, $\mathrm{IV}_1$ boundaries correspond to large complex structure points, where the periods can be expressed in terms of the triple intersection numbers of the mirror Calabi--Yau manifold. By using \eqref{eq:columncount} we find that instanton terms play a crucial role in the asymptotic regime of $\mathrm{I}_1$ and $\mathrm{II}_0$ boundaries, so these provide us with an excellent setting to demonstrate how the formalism discussed in section \ref{sec:instanton_map} describes periods. In contrast, instanton terms are insignificant in the asymptotic regime of $\mathrm{IV}_1$ boundaries, as follows from \eqref{eq:instantonpresence}. Since these periods are already well-understood from the study of large complex structure points anyway, we do not include the periods at these boundaries in this work.

\subsubsection{Type $\text{I}_1$ boundaries}\label{ssec:I1}
We begin by writing down the periods for $\mathrm{I}_1$ boundaries. From the analysis of section \ref{ssec:constrconifold} we find that these periods can be expressed as
\begin{equation}\label{eq:I1periods}
\Pi = \begin{pmatrix}
1 +\frac{a^2}{8\pi} z^2\\
a z \\
i-\frac{i a^2}{8\pi} z^2 \\
\frac{i a}{2\pi} z \log[z]
\end{pmatrix},
\end{equation}
where $a\in \mathbb{R}$ is a model-dependent coefficient. These periods contain two instanton terms, i.e.~terms exponentially suppressed in the saxion $y$ in $t=x+iy=\log[z]/2\pi i$. The periods depend on the complex structure modulus $t$ solely through these exponentially suppressed terms, so instanton terms clearly cannot be ignored for these boundaries. In fact, one can verify that $\Pi, \partial_z \Pi, \partial_z^2 \Pi, \partial_z^3 \Pi$ together span a four-dimensional space only when we include the terms at order $z^2$, so including just the terms at order $z$ does not suffice. This matches nicely with \eqref{eq:columncount}, which indicates two instanton terms for these $\mathrm{I}_1$ boundaries.  

For illustrative purposes, we compute the K\"ahler potential \eqref{Kpot_Omega} from the above periods
\begin{align}\label{eq:kpI1}
e^{-K}=2-2a^2 e^{-4\pi y} y- \frac{a^4}{32\pi^2}e^{-8\pi y}\, ,
\end{align}
where we wrote $z=e^{2\pi i t}$ with $t=x+iy$. Let us now inspect this K\"ahler potential carefully. It depends exponentially on $y$, so by computing the K\"ahler metric one can straightforwardly verify that $\mathrm{I}_1$ boundaries are at finite distance. Also note that it does not depend on the axion $x$ even though these exponential terms are present, so close to the boundary a continuous shift symmetry $x\to x+c$ emerges for the K\"ahler metric.\footnote{Interestingly this differs from the usual K\"ahler potential one encounters through the prepotential \eqref{eq:I1prepotential}, where a cosine type term arises at order $|z|^2$. Compared to our formulation we have effectively removed this term by a K\"ahler transformation, so it does not make a difference at the level of the K\"ahler metric.}  Looking at the sign of the terms in the K\"ahler potential, we note that the subleading terms are fixed to be negative, which ensures the resulting K\"ahler metric is positive definite. From the perspective of asymptotic Hodge theory these signs follow from the polarization conditions \eqref{eq:pol} that the symplectic form satisfies.\footnote{To be precise, one finds that $a_0 \in P^{3,0}$, $a_1  \in P^{2,2}$ and $a_2 \in P^{0,3}$. The respective polarization conditions then imply that the coefficients of these terms satisfy $i\langle a_0\, , \ \bar{a}_0 \rangle >0$, $\langle a_1\, , \ N \bar{a}_1 \rangle < 0$ and $i \langle a_2\, , \  \bar{a}_2 \rangle <0$.}

Next we consider the flux superpotential \eqref{eq:superpotential}. From the above periods we obtain
\begin{equation}\label{eq:superpotentialI1}
W= ig_1-g_3-a e^{2\pi i t}\bigg(g_2 t +g_4 \bigg)-\frac{a^2 e^{4\pi i t} }{8\pi} (ig_1 +g_3)\, ,
\end{equation}
where we wrote out the fluxes as $G_3=(g_1,\ldots, g_4)$. In turn we find the leading polynomial scalar potential \eqref{eq:potential} to be
\begin{equation}\label{eq:potentialI1}
4 \cV^2 \Im \tau  V_{\rm lead}= \bar{G}_3 \, e^{-x N^T}\begin{pmatrix} 1 & 0 & 0 & 0  \\
0 & y-\frac{1}{2\pi}& 0& 0 \\
0 & 0&1& 0  \\
0 & 0 & 0 & \frac{1}{y-\frac{1}{2\pi}} \\
\end{pmatrix} e^{-x N} G_3 \, ,
\end{equation}
where the log-monodromy matrix $N$ is given in \eqref{eq:I1N}. We dropped exponentially suppressed terms in $y$, and left out the $\langle G_3 , \bar{G}_3 \rangle$ term for convenience. The $1/2\pi$ is an artefact of setting the phase operator equal to $\delta=-N/2\pi$ to simplify the periods, and could in principle be removed by a coordinate redefinition as discussed above \eqref{eq:shift}. It is interesting to point out that all fluxes appear at polynomial order in the scalar potential, while  $ig_1+g_3,g_2,g_4$ were exponentially suppressed in the superpotential \eqref{eq:superpotentialI1}. In order to obtain \eqref{eq:potentialI1} it is therefore crucial to include the terms linear in $e^{-2\pi y}$ in the superpotential, while the terms at order $e^{-4\pi y}$ lead to exponentially suppressed corrections.

\subsubsection{Type $\text{II}_0$ boundaries}
Next we consider the periods near $\mathrm{II}_0$ boundaries. From the analysis of section \ref{ssec:constrII0} we find that these periods can be written as
\begin{equation}
\Pi = \begin{pmatrix}
1+a z\\
i-iaz\\ 
\frac{\log[z]}{2 \pi i}+ \frac{az}{2\pi i} (\log[z]-2) \\
 \frac{\log[z]}{2 \pi } -\frac{az}{2\pi  } (\log[z]-2) 
\end{pmatrix},
\end{equation}
where $a \in \mathbb{R}$ is a model-dependent coefficient. Note that these periods do have polynomial terms in $t=\log[z]/2\pi i$, but we also have a restricted form for the periods at order $z=e^{2\pi it}$. One needs this exponentially suppressed term in $t$ in order to span a four-dimensional space with $\Pi , \partial_z \Pi, \partial^2_z \Pi, \partial^3_z \Pi$. This matches nicely with \eqref{eq:columncount}, which indicates one instanton term for $\mathrm{II}_0$ boundaries.  

For illustration, let us again compute the K\"ahler potential \eqref{Kpot_Omega} from the periods
\begin{align}\label{eq:kpII0}
e^{-K}=4  y + \frac{4  a^2(1+\pi y)}{\pi} e^{-4 \pi y} \, ,
\end{align}
where we wrote $z=e^{2\pi i t}$ with $t=x+iy$. Similar to $\mathrm{I}_1$ boundaries the K\"ahler potential does not depend on the axion $x$, both at leading and subleading order, so a continuous shift symmetry $x \to x+c$ emerges close to the boundary. Inspecting the sign of the terms in the K\"ahler potential, we note that both the leading polynomial term as the exponentially suppressed term are fixed to be positive. This ensures that the K\"ahler metric is positive definite, and these signs can again be traced back to the polarization conditions \eqref{eq:pol} of the symplectic form.\footnote{To be precise, one finds that $a_0 \in P^{3,1}$ and $(1+N/\pi i)a_1 \in P^{1,3}$, which implies that the coefficients satisfy $ \langle a_0 , N\bar{a}_0 \rangle >0$ and $\langle a_1, N \bar{a}_1 \rangle > 0$.} Finally, by computing the K\"ahler metric from \eqref{eq:kpI1} one finds that $\text{II}_0$ singularities are at infinite distance, as is expected from K\"ahler potentials that depend through polynomial terms on the saxion $y$ in the large field limit.

Next we consider the flux superpotential \eqref{eq:superpotential}. From the above periods we obtain
\begin{equation}
W=-g_3-ig_4+(g_1+ig_2)t +a e^{2\pi i t} \big(t-\frac{1}{\pi i}\big)(g_1-ig_2)-ae^{2\pi i t}(g_3-ig_4)\, ,
\end{equation}
where we wrote out the fluxes as $G_3=(g_1,\ldots, g_4)$. In turn we find the leading polynomial scalar potential \eqref{eq:potential} to be
\begin{equation}\label{eq:potentialII0}
4 \cV^2 \Im \tau  V_{\rm lead}=\bar{G}_3e^{-x N^T}\begin{pmatrix} 
y & 0& 0& 0 \\
0 & y&0& 0  \\
0 & 0 & \frac{1}{y} &0\\
0 & 0 & 0 & \frac{1}{y}  \\
\end{pmatrix} e^{-x N} G_3 \, ,
\end{equation}
where the log-monodromy matrix $N$ is given in \eqref{eq:II0N}. We again dropped exponentially suppressed terms in $y$, and left out the $\langle G_3 , \bar{G}_3 \rangle$ term for convenience. Interestingly all fluxes now appear at polynomial order in \eqref{eq:potentialII0}, while before the linear combinations $g_1-ig_2$ and $g_3-ig_4$ were exponentially suppressed in the superpotential. However, $g_1+ig_2$ and $g_3+ig_4$ do appear at polynomial order in the superpotential, so one finds that the instanton terms do not contribute at leading order, but instead result in exponential corrections to the leading polynomial scalar potential.

\subsection{Models for two-moduli periods}\label{sec:twomodels}
Having discussed the one-modulus periods, we now turn to periods in a two-moduli setting. We refer to section \ref{sec:two-moduli} for the construction of these periods, to avoid distracting the reader with technical details. Recall from section \ref{ssec:classification} that two-moduli boundaries are characterized by three types of limiting mixed Hodge structures, written as a 2-cube $\langle \mathrm{A}_1 | \mathrm{A}_{(2)} | \mathrm{A}_2 \rangle$. In addition to the intersection $y_1=y_2 =\infty$ with singularity type $\mathrm{A}_{(2)} $, we can consider the separate divisors $y_1 = \infty$ and $y_2=\infty$ characterized by the types $\mathrm{A}_1$ and $\mathrm{A}_2 $ respectively as well. Conveniently such 2-cubes have already been classified in \cite{Kerr2017}, where the possible combinations of singularity types for the boundaries were identified. We do not review the details of this classification by \cite{Kerr2017} in this work, but simply present the exhaustive set of 2-cubes that was obtained
\begin{equation}\label{eq:cubes}
\begin{aligned}
\text{$\mathrm{I}_2$ class} &: \quad \langle \mathrm{I}_1 | \mathrm{I}_2 | \mathrm{I}_1 \rangle\, , \ \langle \mathrm{I}_2 | \mathrm{I}_2 | \mathrm{I}_1 \rangle \, , \  \langle \mathrm{I}_2 | \mathrm{I}_2 | \mathrm{I}_2 \rangle    \, , \\
\text{Coni-LCS class} &: \quad \langle \mathrm{I}_1 | \mathrm{IV}_2 | \mathrm{IV}_1 \rangle \, , \   \langle \mathrm{I}_1 | \mathrm{IV}_2 | \mathrm{IV}_2 \rangle  \, , \\
\text{$\mathrm{II}_1$ class} &: \quad \langle \mathrm{II}_0 | \mathrm{II}_1 | \mathrm{I}_1 \rangle \, , \  \langle \mathrm{II}_1 | \mathrm{II}_1 | \mathrm{I}_1 \rangle \, , \   \langle \mathrm{II}_0 | \mathrm{II}_1 | \mathrm{II}_1 \rangle     \, , \ \langle \mathrm{II}_1 | \mathrm{II}_1 | \mathrm{II}_1 \rangle  \, ,\\
\text{LCS class} &: \quad \langle \mathrm{II}_1 | \mathrm{IV}_2 | \mathrm{III}_0 \rangle \, , \   \langle \mathrm{II}_1 | \mathrm{IV}_2 | \mathrm{IV}_2 \rangle  \, , \ \langle \mathrm{III}_0 | \mathrm{IV}_2 | \mathrm{III}_0 \rangle \, , \ \langle \mathrm{III}_0 | \mathrm{IV}_2 | \mathrm{IV}_1 \rangle\, , \\
& \qquad \langle \mathrm{III}_0 | \mathrm{IV}_2 | \mathrm{IV}_2 \rangle \, , \ \langle \mathrm{IV}_1 | \mathrm{IV}_2 | \mathrm{IV}_2 \rangle \, , \ \langle \mathrm{IV}_2 | \mathrm{IV}_2 | \mathrm{IV}_2 \rangle \, ,   \\
\end{aligned}
\end{equation}
where we chose to sort 2-cubes with similar characteristics together. For the first three classes we find that instanton corrections are needed in $\mathbf{\Pi}$ in order to recover the information in the nilpotent orbit $F^p_{\rm nil}$. We determine the general models for the corresponding period vectors in section \ref{sec:two-moduli}, and give a summary of the obtained results here. The fourth subset of 2-cubes consists of cases that can be realized for particular values of the coefficients $\cK_{ijk}$ describing a large complex structure region and hence specify the intersection numbers of a candidate mirror Calabi-Yau threefold. At these boundaries we do not have any predictive capabilities regarding the instanton series with our machinery and we recover the usual K\"ahler cone restrictions, so we will not discuss the periods for these two-moduli setups later.

\subsubsection{Class $\text{I}_2$ boundaries}\label{ssec:I2}
Let us begin with the class of $\mathrm{I}_2$ boundaries. From the analysis in section \ref{ssec:I2construction} we found that we can write the periods near these boundaries as
\begin{equation}\label{eq:I2periods}
\Pi = \begin{pmatrix}
1 -\frac{a^2}{8 \pi  k_{2}}  z_{1}^{2 k_{1}} z_{2}^{2 k_{2}}-\frac{b^2}{8 \pi  m_{1}}  z_{1}^{2 m_{1}}
   z_{2}^{2 m_{2}} \\
 a z_{1}^{k_{1}} z_{2}^{k_{2}} \\
 b z_{1}^{m_{1}} z_{2}^{m_{2}} \\
i+ \frac{i a^2}{8 \pi  k_{2}} z_{1}^{2 k_{1}} z_{2}^{2 k_{2}}+\frac{i b^2 }{8 \pi  m_{1}} z_{1}^{2 m_{1}}
   z_{2}^{2 m_{2}} \\
-\frac{a}{2\pi i}  z_{1}^{k_{1}} z_{2}^{k_{2}}  \big( n_1 \log[z_1]+\log [z_{2}]-1/k_1\big)+i b \delta_{1}
   z_{1}^{m_{1}} z_{2}^{m_{2}} \\
 -\frac{b}{2\pi i}  z_{1}^{m_{1}}
   z_{2}^{m_{2}} \big(  \log [z_{1}]+n_2 \log[z_2]-1/m_2 \big)+ i a \delta_{1} z_{1}^{k_{1}} z_{2}^{k_{2}}
\end{pmatrix}\, .
\end{equation}
Let us briefly discuss the parameters that appear in these periods, whose properties have been summarized in table \ref{table:I2parameters}. The numbers $n_1,n_2 \in \mathbb{Q}_{\geq 0}$ parametrize the monodromy transformations under $z_i \to e^{2\pi i} z_i$. The integers $k_1,k_2 \in \mathbb{N}$ and $m_1,m_2 \in \mathbb{N}$ specify the order in the instanton expansion. These orders are fixed to be the (smallest) integers such that $n_1=k_1/k_2$ and $n_2=m_2/m_1$ (with $m_1,k_2 >0$), which follows from the horizontality property \eqref{eq:horizontality} of the periods. We must furthermore require $n_1 n_2 \neq 1$ to ensure that the derivatives of the periods together span a six-dimensional space, i.e.~the full three-form cohomology $H^3(Y_3, \mathbb{C})$. Finally, we have real coefficients $\delta_1 \in \mathbb{R}$ and $a,b\in \mathbb{R}$. The coefficient $\delta_1$ coming from the phase operator is always real, while the instanton coefficients $a,b$ have been rotated to real values using shifts of the axions $x^i$ in $x^i+i y^i = \log[z^i]/2\pi i$ as explained below \eqref{eq:shift}. 

\begin{table}[h!]
\centering
\renewcommand*{\arraystretch}{2.0}
\begin{tabular}{| l | c | c | c | c |}
\hline parameters  & $ \langle \mathrm{I}_1 | \mathrm{I}_2 | \mathrm{I}_1 \rangle $ & $ \langle \mathrm{I}_2 | \mathrm{I}_2 | \mathrm{I}_1 \rangle $& $ \langle \mathrm{I}_2 | \mathrm{I}_2 | \mathrm{I}_2 \rangle $\\ \hline \hline 
log-monodromies $n_1, n_2$ & $n_1=n_2=0$ & $n_1 \in \mathbb{Q}_{>0}$, $n_2=0$ & $n_1,n_2 \in \mathbb{Q}_{>0}$, $n_1 n_2 \neq 1$  \\ \hline
instanton orders $k_1,k_2$ & $k_1=0,k_2=1$ & $k_1 = n_1 k_2$& $k_1 = n_1 k_2$ \\ \hline
instanton orders $m_1,m_2$ & $m_1=1,m_2=0$ & $m_1=1,m_2=0$ & $m_2 = n_2 m_1$  \\ \hline
instanton coefficients $a,b$ & \multicolumn{3}{c|}{$a,b \in \mathbb{R} -\{0\}$} \\ \hline
phase operator $\delta$ & \multicolumn{3}{c|}{$\delta_1 \in \mathbb{R}$} \\ \hline
\end{tabular}
\caption{\label{table:I2parameters} Summary for the properties of the parameters in the periods \eqref{eq:I2periods} for each of the possible boundaries of class $\mathrm{I}_2$. }
\end{table}

Let us now compute the K\"ahler potential \eqref{Kpot_Omega} from these periods. We find that
\begin{align}\label{eq:kpI2}
e^{-K} &= 2 -2 a^2 e^{-4\pi k_1 y_1-4\pi k_2 y_2} \Big( n_1 y_1+ y_2+\frac{1}{2\pi k_2} \Big) -2b^2 e^{-4\pi m_1 y_1-4\pi m_2 y_2} \Big(  y_1+n_2 y_2+\frac{1}{2\pi m_1} \Big) \nn \\
& \ \ \ +4 \delta_1 ab e^{-4\pi (k_1+m_1)y_1-4\pi (k_2+m_2)y_2}   \cos[2\pi (k_1-m_1)x_1+2\pi(k_2-m_2)x_2]  \, ,
\end{align}
where we only included terms up to square order in the two instanton expansions, i.e.~in $e^{-2\pi (k_1 y_1+ k_2 y_2)} $ and $e^{-2\pi (m_1 y_1+m_2 y_2)} $, and we used $2\pi i t_i=2\pi i(x_i+y_i) = \log z_i$ for convenience. Note that the sign of the first two non-constant terms is fixed to be negative similar to the one-modulus $\mathrm{I}_1$ boundaries, which again follows from the polarization conditions \eqref{eq:pol} that the symplectic form satisfies. The parameter $\delta_1$ of the phase operator controls the mixing between the two different instanton terms in the periods, i.e.~one coming from $a z_1^{k_1}z_2^{k_2}$ and the other from $ b z_{1}^{m_{1}} z_{2}^{m_{2}}$. This mixing term breaks the continuous shift symmetry for a particular linear combination of the axions $(k_1-m_1)x_1+(k_2-m_2)x_2$, while for the direction $(k_1-m_1)x_1=-(k_2-m_2)x_2$ we still find that a continuous shift symmetry emerges near the boundary for the K\"ahler potential. Finally, one can straightforwardly verify by computing the K\"ahler metric from \eqref{eq:kpI2} that class $\mathrm{I}_2$ boundaries are at finite distance for any large field limit in $y_1,y_2$ due to the exponential dependence. 

Next we consider the flux superpotential \eqref{eq:superpotential}. By using the above periods we find that
\begin{equation}
\begin{aligned}
W &=  ig_1-g_4-a \Big(g_2  \big( n_1 t_1+t_2 - \frac{1}{2 \pi  i k_1}\big)-i\delta_1 g_3 +g_5 \Big) e^{2\pi i (k_1 t_1+k_2 t_2)} \\
&\ \ \ -b \Big(g_3 \big( t_1 + n_2 t_2 -\frac{1}{2\pi i m_2} \big) -i\delta_1 g_2+g_6 \Big) e^{2\pi i (m_1 t_1+m_2 t_2)}\\
&\ \ \ +\frac{a^2}{8\pi k_2} (ig_1+g_4)e^{4\pi i (k_1 t_1+k_2 t_2)} +\frac{b^2}{8\pi m_1} (ig_1+g_4)e^{4\pi i (m_1 t_1+m_2 t_2)}   \, ,
\end{aligned}
\end{equation}
where we wrote out the fluxes as $G_3=(g_1,\ldots, g_6)$. In turn we find the leading polynomial scalar potential \eqref{eq:potential} to be
\begin{equation}
4 \cV^2 \Im \tau  V_{\rm lead}= \bar{G}_3 e^{-x^iN_i^T}\begin{pmatrix} 1 & 0 & 0 & 0 & 0 & 0 \\
0 & n_1 y_1+y_2 & \delta_1 & 0 & 0 & 0 \\
0 & \delta_1 & y_1+n_2 y_2 & 0 & 0 & 0 \\
0 & 0 & 0 & 1 & 0 & 0 \\
0 & 0 & 0 & 0 & \frac{y_1+n_2 y_2}{\Delta} & \frac{\delta_1}{\Delta}  \\
0 & 0 & 0 & 0 & \frac{\delta_1}{\Delta}  & \frac{n_1y_1+ y_2}{\Delta}  \\
\end{pmatrix} e^{-x^iN_i} G_3 \, ,
\end{equation}
where the log-monodromy matrices $N_i$ are given in \eqref{eq:I2N} and we wrote $\Delta=(n_1 y_1+y_2)(y_1+n_2 y_2)-\delta_1^2$. We dropped exponentially suppressed corrections in $y_1,y_2$ and left out the $\langle G_3 , \bar{G}_3 \rangle$ term. Note in particular that the linear combination of fluxes $ig_1+g_4$ as well as $g_2,g_3,g_5,g_6$ are exponentially suppressed in $y_1,y_2$ in the superpotential, while all fluxes appear at polynomial order in the scalar potential. We can trace these terms in the scalar potential back to the terms at orders $e^{-2\pi (k_1 y_1+ k_2 y_2)} $ and $e^{-2\pi (m_1 y_1+m_2 y_2)} $ in the superpotential, while the subleading corrections in the superpotential do produce exponential corrections in the scalar potential.

\subsubsection{Class $\text{II}_1$ boundaries}\label{ssec:II1}
We continue with the class of $\text{II}_1$ boundaries. Within this class, it is interesting to point out that the periods near the boundary $ \langle \mathrm{II}_0 | \mathrm{II}_1 | \mathrm{II}_1 \rangle $ cover a well-studied degeneration for the K3-fibered Calabi-Yau threefold in $\mathbb{P}_4^{1,1,2,2,6}[12]$\cite{Hosono:1993qy, Candelas_1994, Kachru:1995fv, Curio_2001,Lee:2019wij}. The precise match between the two sets of periods is included in appendix \ref{app:Seiberg-Witten}. As outlined in section \ref{ssec:II1construction}, the period vector for boundaries of class $\mathrm{II}_1$ can be written as 
\begin{equation}\label{eq:II1periods}
\begin{aligned}
\mathbf{\Pi}&= \begin{pmatrix}
1+ c z_1^{m_1} z_2^{m_2} +\frac{1}{4}  \big(a^2 n_1 z_2^2+ 2 a b z_1 z_2\frac{1-n_1 n_2^2}{1-n_2}+b^2 n_2 z_1^2 \big)\\
i -i c z_1^{m_1} z_2^{m_2} - \frac{i}{4} \big(a^2 n_1 z_2^2+ 2 a b z_1 z_2\frac{1-n_1n_2^2}{1-n_2}+b^2 n_2 z_1^2 \big) \\
b n_2 z_1 + a z_2\\
   \frac{\log [z_1] + n_2 \log[z_2] }{2\pi i }\Big(1+ c z_1^{m_1} z_2^{m_2} +\frac{1}{4}  \big(a^2 n_1 z_2^2+ 2 a b z_1 z_2\frac{1-n_1 n_2^2}{1-n_2}+b^2 n_2 z_1^2 \big) \Big)  +f(z)\\
\frac{\log [z_1] + n_2 \log[z_2] }{2\pi  } \Big(1- c z_1^{m_1} z_2^{m_2} -\frac{1}{4}  \big(a^2 n_1 z_2^2+ 2 a b z_1 z_2\frac{1-n_1 n_2^2}{1-n_2}+b^2 n_2 z_1^2 \big)  \Big)-i f(z)\\
i ( b n_2 z_1 + a z_2) \frac{n_1 \log[z_1]+ \log[z_2]}{2\pi}- \frac{1-n_1 n_2}{2\pi i}(b z_1 - a z_2)\end{pmatrix}\, ,
\end{aligned}
\end{equation}
where we wrote
\begin{equation}
 f(z)= \frac{i c\, z_1^{m_1} z_2^{m_2} }{m_1 \pi } + \frac{1-n_1 n_2}{8\pi i} \Big(  a^2 z_2^2  +2 ab\, z_1 z_2 \frac{1+n_2^2}{1-n_2} + b^2 n_2 z_1^2\Big)\, .
\end{equation}
The information about the parameters in these periods has been summarized in table \ref{table:II1parameters}. In the construction it was assumed that the coefficients $a,b$ or $a,c$ are non-vanishing, which ensures the presence of essential instanton terms needed in order to span the entire space $H^3(Y_3,\mathbb{C})$. Furthermore $a,b$ have been rotated to real values using the residual axion shift symmetry as discussed below \eqref{eq:shift}. The parameters $n_1,n_2$ control the form of the log-monodromy matrices \eqref{eq:II1N} under $z_i \to e^{2\pi i} z_i$, and hence determine which member of the $\text{II}_1$ boundary class we are looking at. Also note that there is an interplay between the parameter $n_2$ and the orders of the instanton expansion $m_1,m_2$ similar to the class $\mathrm{I}_2$ boundaries, owing to the horizontality property of the periods \eqref{eq:horizontality}.

\begin{table}[h!]
\centering
\renewcommand*{\arraystretch}{2.0}
\begin{tabular}{| l | c | c | c | c | c |}
\hline parameters  & $ \langle \mathrm{II}_0 | \mathrm{II}_1 | \mathrm{I}_1 \rangle $ & $ \langle \mathrm{II}_1 | \mathrm{II}_1 | \mathrm{I}_1 \rangle $ & $ \langle \mathrm{II}_0 | \mathrm{II}_1 | \mathrm{II}_1 \rangle $&$ \langle \mathrm{II}_1 | \mathrm{II}_1 | \mathrm{II}_1 \rangle $ \\ \hline \hline 
log-mon.  & $n_1=n_2=0$ &$n_1 \in \mathbb{Q}_{>0}$, $n_2=0$ & $n_1=0,n_2 \in \mathbb{Q}_{>0}$ & $n_1,n_2 \in \mathbb{Q}_{>0}$, $n_1 n_2 \neq 1$    \\ \hline
inst. orders & \multicolumn{2}{c|}{$m_1 = 1, m_2=0$  } & \multicolumn{2}{c|}{$m_2 = n_2 m_1$  } \\ \hline
inst. coeff.  & \multicolumn{4}{c|}{$a,b \in \mathbb{R}, c \in \mathbb{C} : \ \  a,b \neq 0 \parallel  a,c \neq 0 $}  \\ \hline
\end{tabular}
\caption{\label{table:II1parameters} Summary for the properties of the parameters in the periods \eqref{eq:II1periods} for each of the possible boundaries of class $\mathrm{II}_1$. }
\end{table}


Using these periods we calculate the K\"ahler potential \eqref{Kpot_Omega} for class $\text{II}_1$ boundaries to be
\begin{align}
e^{-K}  = \ & 4(y_1+n_2 y_2)  -2 a^2 e^{-4\pi y_2} \Big(  n_1 y_1+y_2 + \frac{1-n_1 n_2}{2\pi} \Big)  \nn \\
 &-2 n_2 b^2 e^{-4\pi y_1} \Big( n_2(n_1y_1+y_2) -\frac{1-n_1 n_2}{2\pi} \Big)  +4 |c|^2 e^{-4\pi y_1} \Big( y_1+n_2 y_2 + 1/m_1 \pi \Big)  \\
&-4 ab e^{-2\pi y_1-2\pi y_2}\Big(  n_2 ( n_1 y_1+y_2) - \frac{(1-n_2)(1-n_1 n_2)}{4\pi} \Big) \cos(2\pi(x_1-x_2)) +\cO(e^{-6\pi y})\, , \nn
\end{align}
where we used the coordinates $2 \pi t_i=2 \pi i (x_i + i y_i)= \log [z_i]$ for convenience, and dropped subleading corrections in the exponential expansion. For $n_2=0$ the coordinate dependence on $y_2$ enters only through exponentially suppressed terms as one would expect from the presence of an $\text{I}_1$ boundary associated with this coordinate. A noteworthy feature is also that the K\"ahler metric derived from the above potential does not require all instanton terms to become non-degenerate. The instanton term involving $a$ cures this degeneracy, while the one involving $b$ only does so for $n_2 \neq 0$. The instanton term involving $c$ does not suffice to fix the K\"ahler metric. This can be understood more precisely by looking at the derivatives $\partial_1 \Pi$ and $\partial_2 \Pi$ out of which the K\"ahler metric is constructed. For $a \neq 0$ or $b, n_2 \neq 0 $ these derivatives span a two-dimensional space, while if only $c \neq 0$ they are linearly dependent.  Finally, the signs of the first four leading terms in the K\"ahler potential are fixed by the polarization conditions \eqref{eq:pol}, similar to the examples we encountered previously. The remaining term breaks the continuous shift symmetry for the linear combination of axions $x_1-x_2$ at the level of the K\"ahler potential.

Next we consider the flux superpotential \eqref{eq:superpotential} and the corresponding scalar potential \eqref{eq:potential}. We find that the flux superpotential is given by
\begin{equation}\label{eq:superpotentialII1}
\begin{aligned}
W &=  ( g_1+ig_2) (t_1+n_{2}  t_2) - (g_4+i g_5)\\
& \ \ \ - a \, e^{2\pi i t_2}  \left(g_3 \Big( n_1 t_1 +t_2 - \frac{1-n_{1} n_{2}}{2\pi i} \Big)+    g_6 \right) \\
  & \ \ \ -  b  \, e^{2\pi i t_1} \left(g_3 \Big( n_2(n_1 t_1 +t_2) + \frac{1-n_{1} n_{2}}{2\pi i} \Big)  +
   n_2 g_6 \right) \\
  & \ \ \ +  c \, e^{2\pi i m_1 t_1} e^{2\pi i m_2 t_2} \left( \Big(t_1+n_2 t_2 +\frac{i}{m_1 \pi}\Big) (g_1-i g_2)-(g_4-ig_5) \right)+\mathcal{O}(e^{-4\pi y})  \,  ,
\end{aligned}
\end{equation}
where we wrote out $G_3=(g_1 , \ldots, g_6)$ and expanded up to first order in the instanton expansion. In turn, we find as leading polynomial scalar potential
\begin{equation}\label{eq:potentialII1}
4 \cV^2 \Im \tau  V_{\rm lead}= \bar{G} \begin{pmatrix}  y_1+ n_2y_2 & 0 & 0 & 0 & 0 & 0 \\
0 &  y_1+ n_2y_2 & 0 & 0 & 0 & 0 \\
0 & 0 & n_1 y_1+ y_2 & 0 & 0 & 0 \\
0 & 0 & 0 & \frac{1}{ y_1+n_2y_2} & 0 & 0 \\
0 & 0 & 0 & 0 &\frac{1}{ y_1+n_2y_2} & 0\\
0 & 0 & 0 & 0 & 0 & \frac{1}{ n_1y_1+ y_2} \\
\end{pmatrix} G \, ,
\end{equation}
where we absorbed the axion-dependence as $G=e^{-x^i N_i}G_3$ with $N_i$ the log-monodromy matrices given in \eqref{eq:II1N}. We again dropped exponentially suppressed corrections in $y_1,y_2$ and left out the $\langle G_3 , \bar{G}_3 \rangle$ term. Note that only the linear combinations of fluxes $g_1+ig_2$ and $g_4+i g_5$ appear at polynomial order in $t_i = \log[z_i]/2\pi i$ in the superpotential \eqref{eq:superpotentialII1}. In particular the fluxes $g_3,g_6$ only appear through exponential corrections in the superpotential, while they appear at polynomial order in the scalar potential. In the computation of \eqref{eq:potentialII1} these exponential factors cancel out against factors in the K\"ahler metric, resulting in polynomial terms for the scalar potential. In other words, we find that class $\mathrm{II}_1$ boundaries require us to include essential exponential corrections in the superpotential, even though these are at infinite distance. To be more precise, the terms at first order in $e^{-2\pi y_i}$ in the superpotential contribute to the leading polynomial scalar potential, while the other instanton terms lead to exponential corrections.

\subsubsection{Coni-LCS class boundaries}\label{sec:coniLCS}
Finally we come to the class of coni-LCS boundaries. While these boundaries are characterized by a $\text{IV}_2$ singularity type similar to large complex structure points, one has to include essential instanton terms in the periods. Recently the periods near such boundaries have been considered in the context of small flux superpotentials in \cite{Demirtas:2020ffz,Blumenhagen:2020ire} (see also \cite{Demirtas:2019sip} for the original study at large complex structure). The period vector that we construct in section \ref{ssec:IV2construction} is given by
\begin{align} \label{eq:coniLCSperiods}
\Pi= \begin{pmatrix}
1  \\
a z_1\\
\frac{ \log[z_2]}{2 \pi i} \\
-\frac{i \log[z_2]^3}{48 \pi^3}-\frac{ i a^2 n z_1^2 \log[z_2]}{4\pi }+   \frac{a^2}{4  \pi i} z_1^2+i \delta_2 +i \delta_1 a z_1\\
 - a z_1\frac{ \log[z_1] + n \log[z_2]}{2 \pi i} +i\delta_1\\
- \frac{\log[z_2]^2}{8 \pi^2}  -\frac{1}{2} a^2 n z_1^2\\
\end{pmatrix}.
\end{align}
Note that the modulus $t_1=\log[z_1]/2\pi i$ only appears in terms with exponential factors $ e^{2\pi i t_1}$, while $t_2 = \log[z_2]/2\pi i$ appears polynomially. The former we typically attribute to conifold points in the moduli space, while the latter is familiar from large complex structure points, hence the term coni-LCS boundary.

The information about the different parameters is summarized in table \ref{table:IV2parameters}. It is assumed that the coefficient $a$ is non-vanishing as this is required in order to span the entire three-form cohomology $H^3(Y_3,\mathbb{C})$ from derivatives of the period vector. Furthermore, we have used the residual axion shift symmetry to set $a$ to a real value as discussed below \eqref{eq:shift}. The parameter $n$ controls which member of the coni-LCS class we are considering. 

\begin{table}[h!]
\centering
\renewcommand*{\arraystretch}{2.0}
\begin{tabular}{| l | c | c | c | }
\hline parameters  & $ \langle \mathrm{I}_1 | \mathrm{IV}_2 | \mathrm{IV}_1 \rangle $ & $ \langle \mathrm{I}_1 | \mathrm{IV}_2 | \mathrm{IV}_2 \rangle $\\ \hline \hline 
log-monodromies $n_1,n_2$  & $n=0$ & $n \in \mathbb{Q}_{>0}$  \\ \hline
instanton coefficient $a$ & \multicolumn{2}{c|}{$a\in \mathbb{R}- \{0 \}$} \\ \hline
phase operator $\delta$ & \multicolumn{2}{c|}{$\delta_1,\delta_2 \in \mathbb{R}$} \\ \hline
\end{tabular}
\caption{\label{table:IV2parameters} Summary for the properties of the parameters in the periods \eqref{eq:coniLCSperiods} for each of the possible boundaries of the coni-LCS class. }
\end{table}

Using these periods we calculate the K\"ahler potential \eqref{Kpot_Omega} for coni-LCS class boundaries
\begin{equation}
\begin{aligned}
e^{-K} =\ & \frac{4y_2^3}{3} +2 \delta_2 +4a\delta_1 e^{-2\pi y_1} \cos[2\pi x_1] \\
&-2 a^2 e^{-4\pi y_1} \big( y_1+ny_2-(n y_2-1/4\pi ) \cos[4\pi x_1]\big)\, ,
\end{aligned}
\end{equation}
where we used the coordinates $2 \pi t_i=2 \pi i (x_i + i y_i)= \log [z_i]$ for convenience. The signs of the terms without parameters $\delta_i$ are fixed by the polarization conditions \eqref{eq:pol} similar to the previous examples. Note in particular that, as expected from the presence of a finite distance $\mathrm{I}_1$ divisor, the associated field $y_1$ only appears in exponentially suppressed terms. Furthermore, we can understand the role of the phase operator parameters $\delta_1,\delta_2$ by inspecting this K\"ahler potential. We find that $\delta_2$ gives rise to a constant term in the K\"ahler potential, similar to the Euler characteristic term at large complex structure. Interestingly, the parameter $\delta_1$ produces an axion-dependent term at order $e^{-2\pi y_1}$, which is leading compared to the usual term at order $e^{-4\pi y_2}$. See appendix \ref{app:coniLCS} for a more careful comparison with the standard large complex structure expressions.

We next consider the flux superpotential \eqref{eq:superpotential}. By inserting the above periods we find
\begin{equation}
\begin{aligned}
W= \ &- g_4+ig_2 \delta_1 +ig_1 \delta_2 - g_6 t_2+ \frac{1}{2} g_3 t_2^2 -\frac{1}{6} g_1 t_2^3\\
& ae^{2\pi i t_1} \Big( -g_2 (t_1+n t_2) -g_5+ig_1 \delta_1 \Big) -\frac{a^2 e^{4\pi i t_1}}{2} \Big(g_3 n+g_1(n t_2+\frac{1}{2\pi i}) \Big) \, .
\end{aligned}
\end{equation}
 We compute the corresponding scalar potential \eqref{eq:potential} to be
\begin{equation}\label{eq:potentialConiLCS}
4 \cV^2 \Im \tau \,  V_{\rm lead}=  \bar{G}_3 e^{-x^iN_i^T}\begin{pmatrix} \frac{y_2^3}{6} & 0 & 0 & 0 & 0 & 0 \\
0 & \frac{y_2}{2} & 0 & 0 & 0 & 0 \\
0 & 0 & y_1+n y_2 & 0 & 0 & 0 \\
0 & 0 & 0 & \frac{6}{y_2^3} & 0 & 0 \\
0 & 0 & 0 & 0 & \frac{2}{y_2} & 0\\
0 & 0 & 0 & 0 & 0 & \frac{1}{y_1+n_2 y_2} \\
\end{pmatrix} e^{-x^iN_i} G_3 \, ,
\end{equation}
where the log-monodromy matrices $N_i$ are given in \eqref{eq:IV2N}. We again dropped exponentially suppressed corrections in $y_1$ and left out the $\langle G_3 , \bar{G}_3 \rangle$ term for convenience. Note that the fluxes $g_1,g_2,g_5$ as well as the linear combination $g_4-i\delta_1 g_2-i\delta_2 g_1$ appear at polynomial order in the superpotential, while the other fluxes are exponentially suppressed. In computing \eqref{eq:potentialConiLCS} the terms at order $e^{-2\pi y_1}$ in the superpotential are crucial to obtain the polynomial terms for the fluxes $g_3,g_6$ in the scalar potential, while the other instanton terms lead to exponential corrections. This is similar to the corrections for the one-modulus $\mathrm{I}_1$ boundary, and can be traced back to the fact that the divisor $y^1=\infty$ is at finite distance.

\section{Construction of one- and two-moduli periods}\label{sec:constructionperiods}
Here we construct the asymptotic periods for all possible boundaries in one- and two-dimensional complex structure moduli spaces. We begin by writing down the nilpotent orbit data that characterizes these boundaries. For the one-modulus case this data has been constructed in \cite{GreenGriffithsKerr}, and for the two-moduli case we refer to our analysis in appendix \ref{app:boundarydata}. From the given nilpotent orbits we then construct the most general compatible periods following the procedure laid out in section \ref{sec:instanton_map}.

\subsection{Construction of one-modulus periods} \label{app:one-modulus}
In this section we explicitly construct general expressions for the periods near boundaries in one-dimensional moduli spaces. Recall from section \ref{sec:onemodels} that there are three possible singularity types for boundaries in complex structure moduli space when $h^{2,1}=1$: $\mathrm{I}_1$, $\mathrm{II}_0$ and $\mathrm{IV}_1$. Conveniently, we do not need to construct the boundary data from scratch as this was already done in \cite{GreenGriffithsKerr}, so we are simply going to record the results expressed in a different basis more suitable to us. With this information at hand, we write down the instanton map $\Gamma(z)$ as explained in section \ref{sec:instanton_map}, and use it to construct the periods including the necessary instanton terms. These results are nothing inherently new in the sense that the periods for the one-modulus cases can also be systematically constructed by using the so-called Meijer G-functions, see e.g.~\cite{Garcia-Etxebarria:2014wla,Joshi:2019nzi}. However, we find the exercise of re-deriving these periods useful as it serves to illustrate the method we are also going to employ to tackle the two-moduli cases, where there is no systematic construction for periods away from the large complex structure lamppost. Furthermore, it allows us to fix our notation for these expressions as they are also used for computations in section \ref{sec:onemodels}.

\subsubsection{Type $\text{I}_1$ boundaries}\label{ssec:constrconifold}
\begin{figure}[h!]
\centering
\begin{tikzpicture}[scale=1,cm={cos(45),sin(45),-sin(45),cos(45),(15,0)}]
  \draw[step = 1, gray, ultra thin] (0, 0) grid (3, 3);

  \draw[fill] (0, 3) circle[radius=0.05];
  \draw[fill] (1, 1) circle[radius=0.05];
  \draw[fill] (2, 2) circle[radius=0.05];
  \draw[fill] (3, 0) circle[radius=0.05];

\draw[->, red] (0.15,2.95) -- (1.9,2.1);
\draw[->, red] (1.1,0.9) -- (2.85,0.05);

\draw[->, green] (0.05,2.85) -- (0.9,1.1);
\draw[->, green] (2.1,1.9) -- (2.95,0.15);

\draw[->, blue] (0.1,2.9) -- (2.9,0.1);

\draw[->, black] (1.9,1.9) -- (1.1,1.1);
\end{tikzpicture}
\caption{\label{fig:I1} The Hodge-Deligne diamond that classifies $\mathrm{I}_1$ boundaries. We included colored arrows to denote the different components $\Gamma_{-1}, \Gamma_{-2}$ and $\Gamma_{-3}$ of the instanton map $\Gamma$ by red, green and blue respectively. We also used a black arrow to denote the action of the log-monodromy matrix $N$.}
\end{figure}
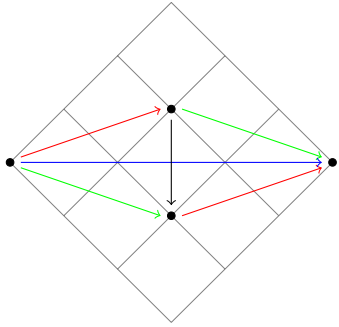

Let us begin by studying $\mathrm{I}_1$ boundaries. The Hodge-Deligne diamond representing these boundaries has been depicted in figure \ref{fig:I1}. The nilpotent orbit data consists of the $sl(2)$-split Deligne splitting $\tilde{I}^{p,q}$, the log-monodromy matrix $N$ and a phase operator $\delta$.  The vector spaces $\tilde{I}^{p,q}$ of the Deligne splitting are spanned by
\begin{equation}\label{eq:DsplitI1}
\begin{aligned}
\tilde{I}^{3,0}&: \quad \begin{pmatrix} 1, & 0,&i,&0 \end{pmatrix} , \\
\tilde{I}^{2,2}&: \quad \begin{pmatrix} 0, & 1,&0,&0 \end{pmatrix} ,  \\
\tilde{I}^{1,1}&: \quad \begin{pmatrix} 0, & 0,&0,&1 \end{pmatrix} ,    \\
\tilde{I}^{0,3}&: \quad  \begin{pmatrix} 1, & 0,&-i,&0 \end{pmatrix} , \\
\end{aligned}
\end{equation}
while the log-monodromy matrix and phase operator can be written as
\begin{align}\label{eq:I1N}
N=\begin{pmatrix}
0 & 0 & 0 & 0 \\
0 & 0 & 0 & 0 \\
0 & 0 & 0 & 0 \\
0 & -1 & 0 & 0
\end{pmatrix}\, , \qquad \delta = \delta_1 N\, .
\end{align}
Note that the phase operator is proportional to the log-monodromy matrix $N$. According to \eqref{eq:shift} we can therefore tune the parameter $\delta_1$ to simplify the periods later.

Now let us follow the procedure of section \ref{sec:instanton_map} to construct the most general periods compatible with this boundary data. First we construct the instanton map $\Gamma(z)$. Let us write down the most general Lie algebra-valued map in $\Lambda_{-}$ with holomorphic coefficients, which reads
\begin{align}
\Gamma(z)= \frac{1}{2}\begin{pmatrix}
 c(z) & i b(z) & -i c(z) & -i a(z) \\
 a(z) & 0 & -i a(z) & 0 \\
 -i c(z) & b(z) & -c(z) & -a(z) \\
 b(z) & 0 & -i b(z) & 0 \\
\end{pmatrix}\, ,
\end{align}
where $a(z),b(z)$ and $c(z)$ make up the charge $\Gamma_{-1}, \Gamma_{-2}$ and $\Gamma_{-3}$ components respectively, with $a(0)=b(0)=c(0)=0$. Note that we set the piece proportional to the log-monodromy matrix $N$ to zero by using \eqref{eq:shift}. The periods can then be written in terms of these coefficients as
\begin{equation}
\Pi = \begin{pmatrix}
 1+c(z)\\
 a(z) \\
 i-ic(z)\\
\frac{i a(z)}{2\pi} \log[z]+b[z]-i\delta_1\\
\end{pmatrix}\, .
\end{equation}
The holomorphic functions $a(z)$, $b(z)$ and $c(z)$ that appear in these periods must satisfy the recursion relations \eqref{eq:recursion}. We can write them out as differential constraints on the coefficients as
\begin{equation}\label{eq:I1diffs}
\begin{aligned}
z b(z)'   = \frac{1}{2\pi i} a(z)\, , \qquad c(z)' = \frac{i}{2} a(z)' b(z)\, .
\end{aligned}
\end{equation}
Since these coefficients are required to vanish at $z=0$, one finds that $b(z),c(z)$ are determined completely by $a(z)$, as can be verified by performing a holomorphic expansion in $z$. In order to obtain a more concrete model for the periods, let us include only the leading order term for $a(z)$ in this instanton expansion. We write as ansatz
\begin{equation}
a(z) = a z\, .
\end{equation}
Plugging this ansatz into the differential equations \eqref{eq:I1diffs} we can solve for the other two functions
\begin{equation}
b(z) = \frac{a}{2\pi i } z\, , \qquad c(z) = \frac{a^2}{8\pi} z^2\, .
\end{equation}
We then obtain the following expression for the asymptotic periods near $\mathrm{I}_1$ boundaries
\begin{align}
\Pi= \begin{pmatrix}
1 + \frac{ a^2}{8 \pi} z^2 \\
a z \\
i - \frac{i  a^2}{8 \pi} z^2 \\
\frac{ia}{2\pi} z \log[z]
\end{pmatrix}\, ,
\end{align}
where we set $\delta_1=-1/2\pi$.

\subsubsection{Type $\text{II}_0$ boundaries}\label{ssec:constrII0}
Next we consider $\mathrm{II}_0$ boundaries. The Hodge-Deligne diamond representing these boundaries has been depicted in figure \ref{fig:II0}. Again let us begin by writing down the nilpotent orbit data. The spaces $\tilde{I}^{p,q}$ of the $sl(2)$-split Deligne splitting are spanned by
\begin{equation}
\begin{aligned}
\tilde{I}^{3,1}&: \quad \begin{pmatrix}
1, & i, &  0, & 0 
\end{pmatrix}, \\
\tilde{I}^{2,0} &: \quad \begin{pmatrix}
0, &  0, &1, & i
\end{pmatrix} ,\\
\tilde{I}^{1,3}&: \quad \begin{pmatrix}
1, & -i, &  0, & 0 
\end{pmatrix}, \\
\tilde{I}^{0,2} &: \quad \begin{pmatrix}
0, &  0, &1, & -i
\end{pmatrix},
\end{aligned}
\end{equation} 
while the log-monodromy matrix can be written as
\begin{align}\label{eq:II0N}
N=\begin{pmatrix}
0 & 0 & 0 & 0 \\
0 & 0 & 0 & 0 \\
1 & 0 & 0 & 0 \\
0 & 1 & 0 & 0
\end{pmatrix},
\end{align}
and the phase operator $\delta = \delta_1 N$ has been set to zero by a coordinate shift \eqref{eq:shift}.

\begin{figure}[h!]
\centering
\begin{tikzpicture}[scale=1,cm={cos(45),sin(45),-sin(45),cos(45),(15,0)}]
  \draw[step = 1, gray, ultra thin] (0, 0) grid (3, 3);

  \draw[fill] (0, 2) circle[radius=0.05];
  \draw[fill] (1, 3) circle[radius=0.05];
  \draw[fill] (2, 0) circle[radius=0.05];
  \draw[fill] (3, 1) circle[radius=0.05];

\draw[->, red] (0.1,2) -- (2.9,1);

\draw[->, green] (1.1,2.9) -- (2.9,1.1);
\draw[->, green] (0.1,1.9) -- (1.9,0.1);

\draw[->, blue] (1,2.9) -- (2,0.1);

\draw[->] (0.9,2.9) -- (0.1,2.1);
\draw[->] (2.9,0.9) -- (2.1,0.1);
\end{tikzpicture}
\caption{\label{fig:II0} The Hodge-Deligne diamond that classifies $\mathrm{II}_0$ boundaries. We included colored arrows to denote the different components $\Gamma_{-1}, \Gamma_{-2}$ and $\Gamma_{-3}$ of the instanton map $\Gamma$ by red, green and blue respectively, and black arrows for the log-monodromy matrix $N$.}
\end{figure}

Using the procedure of section \ref{sec:instanton_map} we now want to write down the most general periods compatible with this boundary data. We again begin by constructing the instanton map $\Gamma(z)$. In this case the most general Lie algebra-valued map in $\Lambda_-$ with holomorphic coefficients reads
\begin{align}
\Gamma=\frac{1}{2}\begin{pmatrix}
 b(z) & -i b(z) & i a(z) & a(z) \\
 -i b(z) & -b(z) & a(z) & -i a(z) \\
 c(z) & -i c(z) & -b(z) & i b(z) \\
 -i c(z) & -c(z) & i b(z) & b(z) \\
\end{pmatrix}
\end{align}
where $a(z),b(z)$ and $c(z)$ make up the charge $\Gamma_{-1}, \Gamma_{-2}$ and $\Gamma_{-3}$ components respectively, with $a(0)=b(0)=c(0)=0$. Note that we have again used coordinate transformations \eqref{eq:shift} to set the piece proportional to $N$ to zero. We can then write the periods in terms of these coefficients as
\begin{equation}\label{eq:II0genperiods}
\Pi = \begin{pmatrix}
1+b(z) \\
i-ib(z) \\
\big(1+b(z) \big) \frac{\log[z]}{2\pi i} +c(z) \\
i \big(1-b(z) \big) \frac{\log[z]}{2\pi i} +c(z)
\end{pmatrix}\, .
\end{equation}
The functions $a(z)$, $b(z)$ and $c(z)$ that appear in the periods must satisfy the recursion relations \eqref{eq:recursion}. These can be written out as differential constraints
\begin{equation}\label{eq:II0diffs}
\begin{aligned}
z b'(z) = \frac{1}{2\pi} a(z)\, , \qquad z c'(z) = \frac{i}{\pi} b(z)\, .
\end{aligned}
\end{equation}
Note that while $a(z)$ did not appear in the periods directly, it does determine $b(z)$ and $c(z)$ uniquely similar to $\mathrm{I}_1$ boundaries. In order to obtain more concrete expressions for the periods near the boundary, let us include only the leading order term $a(z)$ in the holomorphic expansion. We write as ansatz
\begin{equation}
a(z) = 2\pi a z\, ,
\end{equation}
where we included a factor of $2\pi$ for later convenience. By using \eqref{eq:II0diffs} the other two functions are then found to be
\begin{equation}
b(z) = a z\, , \qquad c(z) = \frac{i}{\pi} a z\, .
\end{equation}
By plugging these expressions into \eqref{eq:II0genperiods} we find as asymptotic periods for $\mathrm{II}_0$ boundaries
\begin{align}
\Pi= \begin{pmatrix}
1+a z\\
i-iaz\\ 
\frac{\log[z]}{2 \pi i}+ \frac{az}{2\pi i} (\log[z]-2) \\
 \frac{\log[z]}{2 \pi } - \frac{az}{2\pi } (\log[z]-2) 
\end{pmatrix}\, .
\end{align}

\subsection{Construction of two-moduli periods}\label{sec:two-moduli}
In this section we derive general expressions for the periods near all possible two-moduli boundaries. Recall from section \ref{sec:twomodels} that there are three classes of boundaries we focus on: $\mathrm{I}_2, \mathrm{II}_1$ and coni-LCS class. The boundary data characterizing these classes has been constructed in appendix \ref{app:boundarydata}. We use the techniques discussed in section \ref{sec:instanton_map} to construct the most general periods compatible with these sets of data. Before we begin, let us already note that we now find that \eqref{eq:dXdX} imposes non-trivial constraints on the coefficients of the component $\Gamma_{-1}(z)$ of the instanton map. In contrast to the one-modulus case, this means that one cannot consider any choice of holomorphic functions for these coefficients, but there will be some differential equations that have to be satisfied. For this reason we choose to make a simplified leading order ansatz for $\Gamma_{-1}(z)$, which allows us to illustrate the qualitative features of the models more easily.

\subsubsection{Class $\text{I}_{2}$ boundaries}\label{ssec:I2construction}
Let us begin by considering the class of $\mathrm{I}_2$ boundaries. The Hodge-Deligne diamond representing these boundaries has been depicted in figure \ref{fig:I2}. The nilpotent orbit data has been constructed in appendix \ref{app:I2data}, and is again given by the $sl(2)$-split Deligne splitting $\tilde{I}_{(2)}^{p,q}$ together with the log-monodromy matrices $N_i$ and the phase operator $\delta$. The Deligne splitting is spanned by
\begin{equation}\label{eq:DsplitI2}
\begin{aligned}
\tilde{I}_{(2)}^{3,0}&: \quad \begin{pmatrix} 1, & 0,&0,&i,&0,&0 \end{pmatrix} , \\
\tilde{I}_{(2)}^{2,2}&: \quad \begin{pmatrix} 0, & 1,&0,&0,&0,&0 \end{pmatrix} , \  \begin{pmatrix} 0, & 0,&1,&0,&0,&0 \end{pmatrix} ,  \\
\tilde{I}_{(2)}^{1,1}&: \quad \begin{pmatrix} 0, & 0,&0,&0,&1,&0 \end{pmatrix} ,   \  \begin{pmatrix} 0, & 0,&0,&0,&0,&1 \end{pmatrix} ,  \\
\tilde{I}_{(2)}^{0,3}&: \quad \begin{pmatrix} 1, & 0,&0,&-i,&0,&0\end{pmatrix} , \\
\end{aligned}
\end{equation}
while the log-monodromy matrices are written as
\begin{equation}\label{eq:I2N}
\begin{aligned}
N_1 &= -\begin{pmatrix}
 0 & 0 & 0 & 0 & 0 & 0 \\
 0 & 0 & 0 & 0 & 0 & 0 \\
 0 & 0 & 0 & 0 & 0 & 0 \\
 0 & 0 & 0 & 0 & 0 & 0 \\
 0 & 1 & 0 & 0 & 0 & 0 \\
 0 & 0 & n_1 & 0 & 0 & 0 \\
\end{pmatrix}, \quad &N_2 &=- \begin{pmatrix}
 0 & 0 & 0 & 0 & 0 & 0 \\
 0 & 0 & 0 & 0 & 0 & 0 \\
 0 & 0 & 0 & 0 & 0 & 0 \\
 0 & 0 & 0 & 0 & 0 & 0 \\
 0 & n_2 & 0 & 0 & 0 & 0 \\
 0 & 0 & 1 & 0 & 0 & 0 
\end{pmatrix}, 
\end{aligned}
\end{equation}
and the phase operator is given by
\begin{equation}
\delta = \begin{pmatrix}
 0 & 0 & 0 & 0 & 0 & 0 \\
 0 & 0 & 0 & 0 & 0 & 0 \\
 0 & 0 & 0 & 0 & 0 & 0 \\
 0 & 0 & 0 & 0 & 0 & 0 \\
 0 & 0 & \delta_1 & 0 & 0 & 0 \\
 0 & \delta_1 & 0 & 0 & 0 & 0 \\
\end{pmatrix}\, .
\end{equation}

\begin{figure}[h!]
\centering
\begin{minipage}{0.4\textwidth}
\begin{tikzpicture}[scale=1,cm={cos(45),sin(45),-sin(45),cos(45),(15,0)}]
  \draw[step = 1, gray, ultra thin] (0, 0) grid (3, 3);

  \draw[fill] (0, 3) circle[radius=0.05];
  \draw[fill] (0.9, 1.1) circle[radius=0.05];
  \draw[fill] (1.9, 2.1) circle[radius=0.05];
  \draw[fill] (1.1, 0.9) circle[radius=0.05];
  \draw[fill] (2.1, 1.9) circle[radius=0.05];
  \draw[fill] (3, 0) circle[radius=0.05];

\draw[->, red] (0.15,2.95) -- (1.9,2.1);
\draw[->, red] (0.9,1.1) -- (2.9,0.1);

\draw[->, green] (0.1,2.9)  -- (2.1,1.9);
\draw[->, green] (1.1,0.9) -- (2.85,0.05);

\draw[->, blue] (1.9,2.1) -- (1.1,0.9);
\draw[->, blue] (2.1,1.9) -- (0.9,1.1);
\end{tikzpicture}
\end{minipage}
\begin{minipage}{0.4\textwidth}
\begin{tikzpicture}[scale=1,cm={cos(45),sin(45),-sin(45),cos(45),(15,0)}]
  \draw[step = 1, gray, ultra thin] (0, 0) grid (3, 3);

  \draw[fill] (0, 3) circle[radius=0.05];
  \draw[fill] (0.9, 1.1) circle[radius=0.05];
  \draw[fill] (1.9, 2.1) circle[radius=0.05];
  \draw[fill] (1.1, 0.9) circle[radius=0.05];
  \draw[fill] (2.1, 1.9) circle[radius=0.05];
  \draw[fill] (3, 0) circle[radius=0.05];

\draw[->, purple] (1.9,2.1) --(0.9,1.1);
\draw[->, black] (2.1,1.9) --(1.1,0.9);
\end{tikzpicture}
\end{minipage}

\centering

\caption{\label{fig:I2} The Hodge-Deligne diamond that classifies $\mathrm{I}_2$ boundaries. Note that we split up $\tilde{I}_{(2)}^{1,1},\tilde{I}_{(2)}^{2,2}$ according to the vectors that span the spaces given in \eqref{eq:DsplitI2}, where left vertices correspond to the first two vectors and the right vertices to the last two vectors. On the left we included colored arrows to denote the different components of $\Gamma_{-1}(z)$  of the instanton map $\Gamma(z)$, where we labeled $a(z)$, $b(z)$ and $c(z)$ by red, green and blue respectively. On the right the purple and black arrow denote the action of $N_1$ and $N_2$ respectively (setting $n_1,n_2=0$ for simplicity).}
\end{figure}
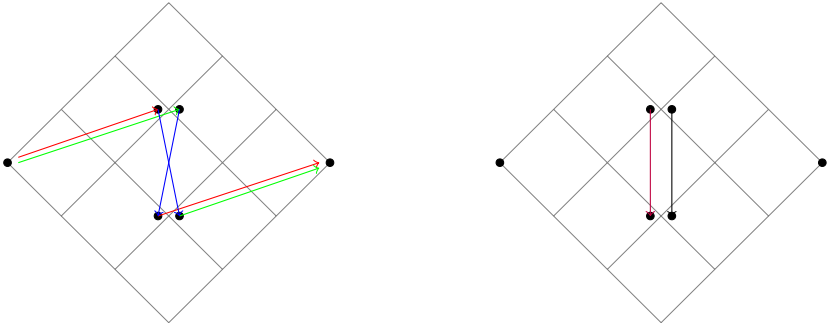
Now we want to write down the most general periods compatible with this boundary data. As explained in section \ref{sec:instanton_map} we begin by considering the most general instanton map $\Gamma(z)$. For the above data the most general Lie algebra-valued map in $\Lambda_-$ with holomorphic coefficients is given by
\begin{equation}\label{eq:I2Gamma}
\Gamma(z_1,z_2) =\frac{1}{2} \begin{pmatrix}
 f(z) & -i d(z) & -i e(z) & -i f(z) & -i
   a(z) & -i b(z) \\
 a(z) & 0 & 0 & -i a(z) & 0 & 0 \\
 b(z) & 0 & 0 & -i b(z) & 0 & 0 \\
 -i f(z) & -d(z) & -e(z) & -f(z) &
   -a(z) &  -b(z) \\
 -d(z) & 0 & -c(z) & i d(z) & 0 & 0 \\
 -e(z) & -c(z) & 0 & i e(z) & 0 & 0 \\
\end{pmatrix},
\end{equation}
where $a,b,c$ make up the charge component $\Gamma_{-1}$, $d,e$ correspond to $\Gamma_{-2}$ and $f$ to $\Gamma_{-3}$. Note that we set the coefficients proportional to the log-monodromy matrices $N_1,N_2$ to zero by using coordinate redefinitions \eqref{eq:shift}, where it is important that $n_1 n_2 \neq 1$. For illustration we depicted the action of the $\Gamma_{-1}$ coefficients on the Deligne splitting in figure \ref{fig:I2}. We can then use \eqref{eq:PiGamma} to write the periods in terms of these coefficients as
\begin{equation}\label{eq:I2periodsfunctions}
\Pi = \begin{pmatrix}
1+f(z)+ \frac{i}{12}   a(z) b(z) c(z) \\
 a(z) \\
 b(z) \\
i-i f(z)  +\frac{1}{12}   a(z) b(z) c(z) \\
- a(z) \frac{ n_1 \log[z_1] + \log
   [z_2]}{2 \pi i}+i\delta_1
   b(z)-d(z)-\frac{1}{4} b(z) c(z)\\
- b(z) \frac{  \log[z_1] + n_2 \log
  [z_2]}{2 \pi i}+i \delta_1
   a(z)-e(z)- \frac{1}{4} a(z) c(z)\\
\end{pmatrix}.
\end{equation}
The holomorphic functions appearing in these periods are constrained by several sets of differential equations. Recall from section \ref{sec:instanton_map} that we must first impose \eqref{eq:dXdX} on the coefficients $a,b,c$ of $\Gamma_{-1}$. Subsequently the coefficients $d,e,f$ of $\Gamma_{-2},\Gamma_{-3}$ are fixed uniquely by $a,b,c$ through \eqref{eq:recursion}. Let us write out these equations explicitly in terms of the holomorphic coefficients. We find that \eqref{eq:dXdX} imposes
\begin{equation}\label{eq:dXdXI2}
\begin{aligned}
 z_{1} a^{(1,0)}- n_1 z_{2} a^{(0,1)}&=i \pi  z_{1} z_{2} \big(b^{(0,1)}c^{(1,0)}-b^{(1,0)} c^{(0,1)}\big)\, , \\
  z_{2} b^{(0,1)}-n_2 z_{1} b^{(1,0)}&=i\pi  z_{1} z_{2} \big(a^{(1,0)}c^{(0,1)}-a^{(0,1)}
   c^{(1,0)}\big)\, .
\end{aligned}
\end{equation}
Inspecting these differential equations carefully, we note that the right-hand side only contains mixed terms in $z_1,z_2$ after expanding the holomorphic functions around $z_1=z_2=0$. Therefore we obtain the following relations on their coefficients
\begin{equation}
a_{k0}=0\, ,\qquad b_{0l} = 0\, , \qquad n_1 a_{0l} =0\, , \qquad  n_2 b_{k0} = 0\, .
\end{equation}
For the remainder of the coefficients we have the relations
\begin{equation}
\begin{aligned}
(n_1 l - k)a_{kl} &= \pi i \sum_{m,n} (k-m)n \, b_{k-m,l-n} c_{mn} -\pi i \sum_{m,n} (l-n)m \,  b_{k-m,l-n} c_{mn} \, ,\\
(n_2 k - l)b_{kl} &= \pi i \sum_{m,n} (k-m)n \, a_{k-m,l-n} c_{mn} - \pi i \sum_{m,n} (l-n)m \,  a_{k-m,l-n} c_{mn} \, .
\end{aligned}
\end{equation}
Note that coefficients $a_{kl}$ with $k=n_1 l$ and $b_{kl}$ with $l=n_2 k$ do not appear on the left-hand side of this equation, so they are unfixed by these differential constraints.

Next let us write down the differential equations that fix $d,e,f$ uniquely in terms of $a,b,c$. We find that \eqref{eq:recursion} imposes the following set of constraints
\begin{equation}\label{eq:I2recursion}
\begin{aligned}
z_1d^{(1,0)} &=- \frac{n_1}{2\pi i}a+\frac{1}{4} z_1 \big( b^{(1,0)}c-bc^{(1,0)}\big)\, , \\
z_1e^{(1,0)} &=- \frac{1}{2\pi i} b+\frac{1}{4} z_1 \big( a^{(1,0)} c-ac^{(1,0)} \big)\, , \\
z_2d^{(0,1)} &=- \frac{1}{2\pi i} a+\frac{1}{4} z_2 \big( b^{(0,1)}c-bc^{(0,1)} \big)\, , \\
z_2e^{(0,1)} &=- \frac{n_2}{2\pi i} b+\frac{1}{4} z_2 \big( a^{(0,1)} c-ac^{(0,1)} \big)\, , \\
i f^{(1,0)}&=-\frac{1}{24}  a^{(1,0)} b
   c+\frac{1}{2}   a^{(1,0)}
   d-\frac{1}{24}   a
   b^{(1,0)} c+\frac{1}{12} 
   ab c^{(1,0)}+\frac{1}{2}  
   b^{(1,0)} e\, , \\
if^{(0,1)}   &=-\frac{1}{24}  a^{(0,1)} b
   c+\frac{1}{2}   a^{(0,1)}
   d-\frac{1}{24}  a
   b^{(0,1)} c+\frac{1}{12} 
   a b c^{(0,1)}+\frac{1}{2}  
   b^{(0,1)} e\, . \\
\end{aligned}
\end{equation}
Now we want to find out what \eqref{eq:rankGamma} imposes on the functions $a,b,c$ that make up the $\Gamma_{-1}$ component of the instanton map. The dimension of the image of this set of matrices must be equal to 5. By inspecting \eqref{eq:I2Gamma} we find that the function $c$ is irrelevant, since they span the same part of the vector space as $N_1$ and $N_2$. On the other hand we can satisfy \eqref{eq:rankGamma} by turning on $a,b$. This can also be seen from figure \ref{fig:I2}, because in order to span $\tilde{I}_{(2)}^{2,2}$ and $\tilde{I}_{(2)}^{0,3}$ we need the components of $a,b$. Let us therefore take the following ansatz for the $\Gamma_{-1}$ coefficients
\begin{equation}
a(z)=a \, z_1^{k_1} z_2^{k_2}\, , \quad b(z)=b \, z_1^{m_1} z_2^{m_2}\, , \quad c(z)=0\, , 
\end{equation}
where $a,b \in \mathbb{C}$. Some comments are in order here. When $(k_1,k_2)\neq (m_1,m_2)$ we can use shifts of the axions $x^i$ to set $a,b \in \mathbb{R}$, while for $(k_1,k_2)= (m_1,m_2)$ this is generally only possible for one of the two. Also note that vanishing of $c(z)$ was not required by \eqref{eq:rankGamma}. Nevertheless $c$ only appear in products with $a(z),b(z)$ in the periods \eqref{eq:I2periodsfunctions}, so it would lead only to subleading corrections. Finally, we only wrote down one leading term for $a(z)$ and $b(z)$. We are however expanding with respect to two coordinates $z_1,z_2$, so there could in principle be two different leading terms for the two expansions. We will see shortly that \eqref{eq:dXdXI2} fixes the orders $k_1,k_2$ and $m_1,m_2$ for the leading terms, justifying the above expansion.  

Let us now solve the differential constraints that the functions of $\Gamma(z)$ must satisfy. We begin with \eqref{eq:dXdXI2}, which reduces to the following two conditions on our ansatz
\begin{equation}\label{eq:orders}
 n_1=k_1/k_2\, , \qquad n_2 = m_2/m_1\, .
\end{equation}
Thus the orders of the leading terms in the instanton expansion are fixed as the pairs of coprime integers $k_1,k_2 \in \mathbb{Z}$ and $m_1,m_2 \in \mathbb{Z}$ such that \eqref{eq:orders} holds. We can then continue and solve \eqref{eq:I2recursion}, which yields
\begin{equation}
\begin{aligned}
d(z) &=\frac{ia}{2\pi k_2} z_1^{k_1} z_2^{k_2} \, , \qquad e(z)=\frac{ib}{2\pi m_1} z_1^{m_1} z_2^{m_2} \, , \\
 f(z) &= \frac{a^2}{8\pi k_2} z_1^{2k_1} z_2^{2k_2} +\frac{b^2}{8\pi m_1} z_1^{2m_1} z_2^{2m_2}  \, .
\end{aligned}
\end{equation}
By inserting these expressions into \eqref{eq:I2periodsfunctions} we obtain the periods
\begin{equation}
\Pi = \begin{pmatrix}
1 +\frac{a^2 z_{1}^{2 k_{1}} z_{2}^{2 k_{2}}}{8 \pi  k_{1}}+\frac{b^2 z_{1}^{2 m_{1}}
   z_{2}^{2 m_{2}}}{8 \pi  m_{2}} \\
 a z_{1}^{k_{1}} z_{2}^{k_{2}} \\
 b z_{1}^{m_{1}} z_{2}^{m_{2}} \\
i- \frac{i a^2 z_{1}^{2 k_{1}} z_{2}^{2 k_{2}}}{8 \pi  k_{1}}-\frac{i b^2 z_{1}^{2 m_{1}}
   z_{2}^{2 m_{2}}}{8 \pi  m_{2}} \\
-a  z_{1}^{k_{1}} z_{2}^{k_{2}}  \frac{n_1 \log[z_1]+\log [z_{2}]-1/k_1}{2 \pi i}+i b \delta_{1}
   z_{1}^{m_{1}} z_{2}^{m_{2}} \\
- b  z_{1}^{m_{1}}
   z_{2}^{m_{2}} \frac{ \log [z_{1}]+ n_2\log[z_2]-1/m_2}{2 \pi i }+ i a \delta_{1} z_{1}^{k_{1}} z_{2}^{k_{2}}
   \end{pmatrix}.
\end{equation}

\subsubsection{Class $\text{II}_1$ boundaries}
\label{ssec:II1construction}
Next we study class $\mathrm{II}_1$ boundaries. The Hodge-Deligne diamond representing such boundaries has been depicted in figure \ref{fig:II1}. The relevant boundary data has been constructed in appendix \ref{app:II1data}, and is again given by the $sl(2)$-split Deligne splitting $\tilde{I}_{(2)}^{p,q}$ together with the log-monodromy matrices $N_i$ and the phase operator. The Deligne splitting is spanned by 
\begin{equation}\label{eq:DsplitII1}
\begin{aligned}
\tilde{I}_{(2)}^{3,1}&: \quad \begin{pmatrix} 1, & i,&0,&0,&0,&0 \end{pmatrix} ,\qquad  &\tilde{I}_{(2)}^{2,0}&: \quad\begin{pmatrix} 0, & 0,&0,&1,&i,&0 \end{pmatrix} , \\
\tilde{I}_{(2)}^{2,2}&: \quad \begin{pmatrix} 0, & 0,&1,&0,&0,&0 \end{pmatrix} , \qquad &\tilde{I}_{(2)}^{1,1}&: \quad \begin{pmatrix} 0, & 0,&0,&0,&0,&1 \end{pmatrix} ,  \\
\tilde{I}_{(2)}^{1,3}&:\quad \begin{pmatrix} 1, & -i,&0,&0,&0,&0\end{pmatrix} , \qquad &\tilde{I}_{(2)}^{0,2}&:\quad \begin{pmatrix} 0, & 0,&0,&1,&-i,&0\end{pmatrix} , 
\end{aligned}
\end{equation}
while the log-monodromy matrices are given by
\begin{equation}\label{eq:II1N}
\begin{aligned}
N_1 &= \begin{pmatrix}
 0 & 0 & 0 & 0 & 0 & 0 \\
 0 & 0 & 0 & 0 & 0 & 0 \\
 0 & 0 & 0 & 0 & 0 & 0 \\
 1 & 0 & 0 & 0 & 0 & 0 \\
 0 & 1 & 0 & 0 & 0 & 0 \\
 0 & 0 & -n_1 & 0 & 0 & 0 \\
\end{pmatrix}, \qquad N_2 &= \begin{pmatrix}
 0 & 0 & 0 & 0 & 0 & 0 \\
 0 & 0 & 0 & 0 & 0 & 0 \\
 0 & 0 & 0 & 0 & 0 & 0 \\
 n_2 & 0 & 0 & 0 & 0 & 0 \\
 0 & n_2 & 0 & 0 & 0 & 0 \\
 0 & 0 & -1 & 0 & 0 & 0 
\end{pmatrix}, 
\end{aligned}
\end{equation}
where $n_1, n_2 \in \mathbb{Q}$ and $n_1,n_2 \geq 0$. The phase operator $\delta$ has been set to zero by using coordinate transformations \eqref{eq:shift}.

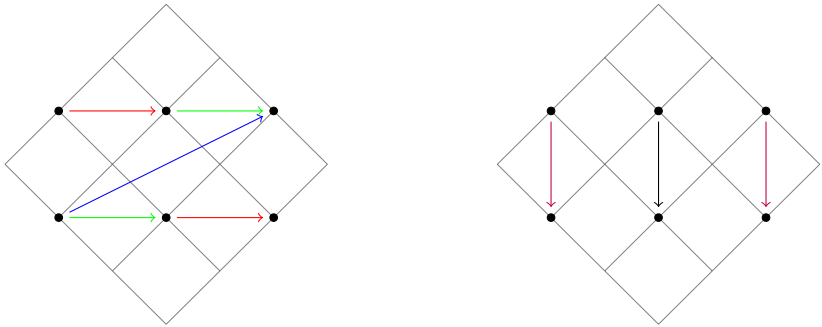
\begin{figure}[h!]
\centering
\begin{minipage}{0.4\textwidth}
\begin{tikzpicture}[scale=1,cm={cos(45),sin(45),-sin(45),cos(45),(15,0)}]
  \draw[step = 1, gray, ultra thin] (0, 0) grid (3, 3);

  \draw[fill] (0, 2) circle[radius=0.05];
  \draw[fill] (1, 3) circle[radius=0.05];
  \draw[fill] (2, 2) circle[radius=0.05];
  \draw[fill] (1, 1) circle[radius=0.05];
  \draw[fill] (2, 0) circle[radius=0.05];
  \draw[fill] (3, 1) circle[radius=0.05];

\draw[->, red] (1.1,2.9) -- (1.9,2.1);
\draw[->, red] (1.1,0.9) -- (1.9,0.1);

\draw[->, blue] (0.15,1.95) -- (2.85,1.05);

\draw[->, green]  (0.1,1.9) -- (0.9,1.1);
\draw[->, green]  (2.1,1.9) -- (2.9,1.1);
\end{tikzpicture}
\end{minipage}
\begin{minipage}{0.4\textwidth}
\begin{tikzpicture}[scale=1,cm={cos(45),sin(45),-sin(45),cos(45),(15,0)}]
  \draw[step = 1, gray, ultra thin] (0, 0) grid (3, 3);

  \draw[fill] (0, 2) circle[radius=0.05];
  \draw[fill] (1, 3) circle[radius=0.05];
  \draw[fill] (2, 2) circle[radius=0.05];
  \draw[fill] (1, 1) circle[radius=0.05];
  \draw[fill] (2, 0) circle[radius=0.05];
  \draw[fill] (3, 1) circle[radius=0.05];

\draw[->, purple] (0.9,2.9) -- (0.1,2.1);
\draw[->, purple] (2.9,0.9) -- (2.1,0.1);

\draw[->] (1.9,1.9) -- (1.1,1.1);

\end{tikzpicture}
\end{minipage}
\caption{\label{fig:II1} The Hodge-Deligne diamond that classifies $\mathrm{II}_1$ boundaries. On the left included colored arrows to denote the different components of $\Gamma_{-1}(z)$  of the instanton map $\Gamma(z)$, where we labeled $a(z)$, $b(z)$ and $c(z)$ by red, green and blue respectively. On the right the purple and black arrows denote the action of $N_1$ and $N_2$ respectively (setting $n_1,n_2=0$ for simplicity).}
\end{figure}

Now we want to write down the most general expressions for the periods compatible with this boundary data. Following section \ref{sec:instanton_map} we begin by writing down the most general instanton map $\Gamma(z)$. It is the most general Lie algebra-valued map in $\Lambda_-$ with respect to \eqref{eq:DsplitII1} that has holomorphic coefficients
\begin{equation}
\Gamma(z) =\frac{1}{2} \begin{pmatrix}
 e(z) & -i e(z) & b(z) & c(z) & -i
   c(z) & 0 \\
 -i e(z) & -e(z) & -i b(z) & -i c(z) &
   -c(z) & 0 \\
 a(z) & -i a(z) & 0 & 0 & 0 & 0 \\
 f(z) & -i f(z) & -d(z) & -e(z) & i
   e(z) & -a(z) \\
 -i f(z) & -f(z) & i d(z) & i e(z) &
   e(z) & i a(z) \\
 -d(z) & i d(z) & 0 & -b(z) & i b(z) & 0 \\
\end{pmatrix},
\end{equation}
where $a(z),b(z),c(z)$ make up the $\Gamma_{-1}$ component of the instanton map, $d(z),e(z)$ the  $\Gamma_{-2}$ component and $f(z)$ the  $\Gamma_{-3}$ component. Note that we used coordinate redefinitions \eqref{eq:shift} to set the pieces along $N_1$ and $N_2$ to zero. The period vector then reads
\begin{equation}
\Pi= \begin{pmatrix}
1+\frac{1}{4} a(z) b(z) + e(z) \\
i - \frac{i}{4} a(z) b(z) -i e(z) \\
a(z)\\
\frac{\log [z_1] + n_2 \log[z_2] }{2\pi i } \big(1+\frac{1}{4}a(z)b(z)+e(z) \big)  +f(z)  \\
\frac{\log [z_1] + n_2 \log[z_2] }{2\pi  } \big(1-\frac{1}{4}a(z)b(z)-e(z) \big)-if(z)  \\
- a(z)\frac{ n_{1} \log [z_{1}] + \log[z_2] }{2\pi i }  -d(z) \\
\end{pmatrix}\, .
\end{equation}
Note in particular that the function $c(z)$ does not explicitly appear in the period vector. This can be attributed to the fact that we only read off how $\exp[\Gamma(z)]$ acts on $\mathbf{a}_0$, not the full matrix. Nevertheless, $c(z)$ enters indirectly in the periods through the recursion relations \eqref{eq:recursion}, as we will see shortly.

From \eqref{eq:dXdX} we obtain two differential constraints on the $\Gamma_{-1}$ coefficients
\begin{equation}
\begin{aligned}
z_{2} n_1 a^{(0,1)}-z_{2} b^{(0,1)}(z) &=  z_{1} a^{(1,0)}(z)-n_{2} z_{1}
   b^{(1,0)}(z)\, ,\\
z_{2} c^{(0,1)}(z) -n_{2} z_{1}
   c^{(1,0)}(z)&= 2 i \pi  z_{1} z_{2}  \big(a^{(0,1)}(z) b^{(1,0)}(z) -a^{(1,0)}(z) b^{(0,1)}(z) \big)\, .  \label{eq:II1DiffConstraints}
\end{aligned}
\end{equation}
By expanding the holomorphic functions around $z_1=z_2=0$ we then obtain the relations
\begin{equation}
\begin{aligned}
k(a_{kl}-n_2 b_{kl})&= l(n_1  a_{kl} -  b_{kl})  \, ,\\
l\, c_{kl}-n_2 k \, c_{kl} &=  2\pi i \sum_{m,n} (k-m)n \, a_{k-m,l-n} b_{mn}  - 2\pi i \sum_{m,n} (l-n)m \,  b_{k-m,l-n} a_{mn} \, .
\end{aligned}
\end{equation}
In particular, for $k=0$ or $l=0$ we find that
\begin{equation}
n_1 a_{0l}=b_{0l}\, ,\quad  a_{k0} =n_2 b_{k0}\, , \quad c_{0l}=0 \, , \quad n_2  c_{k0}= 0\, .
\end{equation}
Next we consider the component-wise constraints given in \eqref{eq:recursion}. These result in the following set of relations
\begin{equation} \label{eq:II1DiffConstraints2}
\begin{aligned}
2 i \pi  z_{1} d^{(1,0)}(z) &= b(z)-n_1 a(z)\, , \\
2 i \pi  z_{2} d^{(0,1)}(z) &= n_{2} b(z)- a(z)\, , \\
4 i \pi  z_{1} e^{(1,0)}(z) &= 2c(z)+i \pi  z_{1} \big(  a^{(1,0)}(z) b(z)-a(z)
   b^{(1,0)}(z) \big)\, , \\
4 i \pi  z_{2} e^{(0,1)}(z) &= 2 n_{2} c(z) +i \pi  z_{2}\big(  a^{(0,1)}(z) b(z)-a(z)
   b^{(0,1)}(z) \big) \, , \\
 i \pi  z_{1}
   f^{(1,0)}(z) &= -  e(z)+ i \pi  z_{1} a^{(1,0)}(z) d(z)\, , \\
 i \pi  z_{2}
   f^{(0,1)}(z) &= - n_2 e(z)+ i \pi  z_{2} a^{(0,1)}(z) d(z)\, .
\end{aligned}
\end{equation}
Similarly these relations can be cast into expressions for the series coefficients $e_{kl},f_{kl},g_{kl}$, but for brevity we do not write them down here. 

We then make a leading order ansatz for the components of $\Gamma_{-1}$. Turning on as many linear terms in $z_1,z_2$ as possible without violating \eqref{eq:II1DiffConstraints}, we write for these coefficients
\beq
\begin{aligned}
a(z) &= n_2 b\,  z_1+a\,  z_2\, ,  \qquad b(z)= b\, z_1 +n_1 a \, z_2 \, , \\
 c(z)&= 2\pi i m_1 c\,  z^{m_1}_1 z^{m_2}_2 +\pi i \, a b \frac{1-n_1 n_2}{1-n_2} z_1 z_2\,,
\end{aligned}
\eeq
where $a,b,c \in \mathbb{C}$, and we have non-negative integers $m_1, m_2$ with $m_1 > 0$ that satisfy
\begin{equation}
n_2 = m_2 /m_1 \, .
\end{equation}
When $n_1 n_2 \neq 1$ we can use axion shifts as discussed below \eqref{eq:shift} to set $a,b \in \mathbb{R}$, while for $n_1 n_2 = 1$ this is generally only possible for one of the two. The rank condition \eqref{eq:rankGamma} tells us that we must always require $a\neq 0$, and additionally $b \neq 0$ or $c \neq 0$ should hold. Note that the second term in $c(z)$ is ill-defined for $n_2 = 1$, meaning we should set $b=0$. In this special case the term at order $z_1 z_2$  in $c(z)$ then already follows from the first term by $m_1=m_2 = 1$. 

We can now solve the recursive differential equations  \eqref{eq:II1DiffConstraints2} for the components of $\Gamma_{-q}$ with $q< -1$, from which we obtain
\begin{equation}\label{eq:II1def}
\begin{aligned}
d(z)&=   \frac{1-n_1 n_2}{2\pi i} (b \, z_1-a \, z_2)\, ,\qquad  e(z) =c \, z_1^{m_1} z_2^{m_2} +ab\frac{(1+n_2)(1-n_1 n_2)}{4(1-n_2)} z_1 z_2\, ,\\
 f(z)&= \frac{i c\, z_1^{m_1} z_2^{m_2}}{\pi m_1 } + \frac{1-n_1 n_2}{8\pi i} \Big(  a^2 z_2^2 + b^2 n_2 z_1^2 +2 ab\, z_1 z_2 \frac{1+n_2^2}{1-n_2} \Big)\, .
 \end{aligned}
\end{equation}
Putting all this together we get as asymptotic period vector 
\begin{equation}
\mathbf{\Pi}= \begin{pmatrix}
1+ c z_1^{m_1} z_2^{m_2} +\frac{1}{4}  \big(a^2 n_1 z_2^2+ 2 a b z_1 z_2\frac{1-n_1 n_2^2}{1-n_2}+b^2 n_2 z_1^2 \big)\\
i -i c z_1^{m_1} z_2^{m_2} - \frac{i}{4} \big(a^2 n_1 z_2^2+ 2 a b z_1 z_2\frac{1-n_1n_2^2}{1-n_2}+b^2 n_2 z_1^2 \big) \\
b n_2 z_1 + a z_2\\
   \frac{\log [z_1] + n_2 \log[z_2] }{2\pi i }\Big(1+ c z_1^{m_1} z_2^{m_2} +\frac{1}{4}  \big(a^2 n_1 z_2^2+ 2 a b z_1 z_2\frac{1-n_1 n_2^2}{1-n_2}+b^2 n_2 z_1^2 \big) \Big)  +f(z)\\
\frac{\log [z_1] + n_2 \log[z_2] }{2\pi  } \Big(1- c z_1^{m_1} z_2^{m_2} -\frac{1}{4}  \big(a^2 n_1 z_2^2+ 2 a b z_1 z_2\frac{1-n_1 n_2^2}{1-n_2}+b^2 n_2 z_1^2 \big)  \Big)-i f(z)\\
i ( b n_2 z_1 + a z_2) \frac{n_1 \log[z_1]+ \log[z_2]}{2\pi}- \frac{1-n_1 n_2}{2\pi i}(b z_1 - a z_2)\end{pmatrix}\, ,
\end{equation}
with $f(z)$ given in \eqref{eq:II1def}.

\subsubsection{Coni-LCS class boundaries}
\label{ssec:IV2construction}
\begin{figure}[h!]
\centering
\begin{minipage}{0.4\textwidth}
\begin{tikzpicture}[scale=1,cm={cos(45),sin(45),-sin(45),cos(45),(15,0)}]
  \draw[step = 1, gray, ultra thin] (0, 0) grid (3, 3);

  \draw[fill] (3, 3) circle[radius=0.05];
  \draw[fill] (0.9, 1.1) circle[radius=0.05];
  \draw[fill] (1.9, 2.1) circle[radius=0.05];
  \draw[fill] (1.1, 0.9) circle[radius=0.05];
  \draw[fill] (2.1, 1.9) circle[radius=0.05];
  \draw[fill] (0, 0) circle[radius=0.05];

\draw[->, red] (2.95,2.95) -- (1.95,2.15);
\draw[->, red] (0.85,1.05) -- (0.05,0.05);

\draw[->, blue] (2.1,1.9) -- (1.1,0.9);

\draw[->, green] (1.9,2.1) -- (1.1,0.9);
\draw[->, green] (2.1,1.9) -- (0.9,1.1);

\end{tikzpicture}
\end{minipage}
\begin{minipage}{0.4\textwidth}
\begin{tikzpicture}[scale=1,cm={cos(45),sin(45),-sin(45),cos(45),(15,0)}]
  \draw[step = 1, gray, ultra thin] (0, 0) grid (3, 3);

  \draw[fill] (3, 3) circle[radius=0.05];
  \draw[fill] (0.9, 1.1) circle[radius=0.05];
  \draw[fill] (1.9, 2.1) circle[radius=0.05];
  \draw[fill] (1.1, 0.9) circle[radius=0.05];
  \draw[fill] (2.1, 1.9) circle[radius=0.05];
  \draw[fill] (0, 0) circle[radius=0.05];

\draw[->] (2.95,2.95) -- (2.15,1.95);
\draw[->] (1.05,0.85) -- (0.05,0.05);

\draw[->] (2.1,1.9) -- (1.1,0.9);

\draw[->, purple] (1.9,2.1) -- (0.9,1.1);

\end{tikzpicture}
\end{minipage}
\caption{\label{fig:IV2} The Hodge-Deligne diamond that classifies $\mathrm{IV}_2$ boundaries of coni-LCS type. On the left we included colored arrows to denote the different components of $\Gamma_{-1}(z)$  of the instanton map $\Gamma(z)$, where we labeled $a(z)$, $b(z)$ and $c(z)$ by red, green and blue respectively. On the right we included purple and black arrows to indicate the action of the log-monodromy matrices $N_1,N_2$ respectively, where we considered $n=0$ for simplicity. Note that we split up $\tilde{I}_{(2)}^{1,1},\tilde{I}_{(2)}^{2,2}$ according to the vectors that span the spaces given in \eqref{eq:DsplitIV2}, where left vertices correspond to the first two vectors and the right vertices to the last two.}
\end{figure}
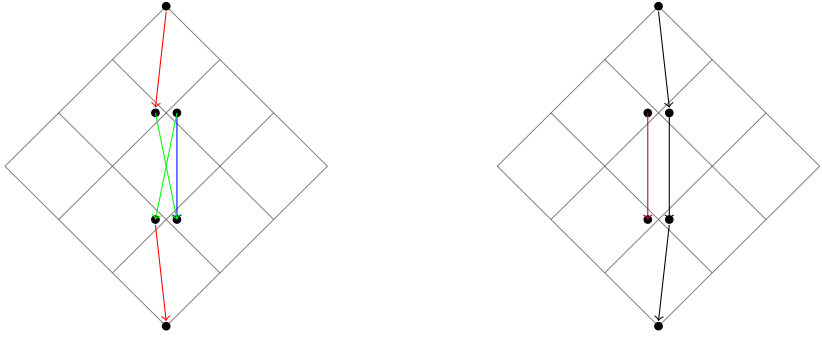

Finally we study coni-LCS class boundaries. Such boundaries are characterized by a $\mathrm{IV}_2$ Hodge-Deligne diamond, which has been depicted in figure \ref{fig:IV2}. The nilpotent orbit data has been constructed in section \ref{app:coniLCSdata}, and is again given by the Deligne splitting $\tilde{I}_{(2)}^{p,q}$ together with the log-monodromy matrices $N_i$ and the phase operator $\delta$. Recall that the Deligne splitting is spanned by
\begin{equation}\label{eq:DsplitIV2}
\begin{aligned}
\tilde{I}_{(2)}^{3,3}&: \quad \begin{pmatrix} 1, & 0,&0,&0,&0,&0 \end{pmatrix} , \\
\tilde{I}_{(2)}^{2,2}&: \quad \begin{pmatrix} 0, & 1,&0,&0,&0,&0 \end{pmatrix} , \ \  \begin{pmatrix} 0, & 0,&1,&0,&0,&0 \end{pmatrix} , \\
\tilde{I}_{(2)}^{1,1}&: \quad \begin{pmatrix} 0, & 0,&0,&0,&1,&0 \end{pmatrix} , \ \  \begin{pmatrix} 0, & 0,&0,&0,&0,&1 \end{pmatrix} , \\
\tilde{I}_{(2)}^{0,0}&: \quad \begin{pmatrix} 0, & 0,&0,&1,&0,&0 \end{pmatrix} ,
\end{aligned}
\end{equation}
while the log-monodromy matrices are
\begin{align}\label{eq:IV2N}
N_1= \begin{pmatrix}
0 & 0 & 0 & 0 & 0 & 0 \\
0 & 0 & 0 & 0 & 0 & 0 \\
0 & 0 & 0 & 0 & 0 & 0 \\
0 & 0 & 0 & 0 & 0 & 0 \\
0 & 1 & 0 & 0 & 0 & 0 \\
0 & 0 & 0 & 0 & 0 & 0 \\
\end{pmatrix} \, ,  \quad 
N_2=\begin{pmatrix}
0 & 0 & 0 & 0 & 0 & 0 \\
0 & 0 & 0 & 0 & 0 & 0 \\
1 & 0 & 0 & 0 & 0 & 0 \\
0 & 0 & 0 & 0 & 0 & -1 \\
0 & -n & 0 & 0 & 0 & 0 \\
0 & 0 & 1 & 0 & 0 & 0 \\
\end{pmatrix},
\end{align}
and the phase operator $\delta$ is given by
\begin{equation}
\delta =  \begin{pmatrix} 0 & 0 & 0 & 0 & 0 & 0 \\
 0 & 0 & 0 & 0 & 0 & 0 \\
0 & 0 & 0 & 0 & 0 & 0 \\
 \delta_2 & \delta_1 & 0& 0 & 0 & 0 \\
 \delta_1 & \frac{1}{2\pi} & 0 & 0 & 0 & 0 \\
 0 & 0 &0 & 0 & 0 & 0 
\end{pmatrix},
\end{equation}
where we chose to set $\delta_3=1/2\pi$ and $\delta_4=0$.

We now want to construct the most general periods compatible with this boundary data. Following section \ref{sec:instanton_map} we begin by considering the most general instanton map $\Gamma(z)$. For the above boundary data the most general Lie algebra-valued map in $\Lambda_-$ is given by
\begin{align}
\Gamma= \begin{pmatrix}
0 & 0 & 0 & 0 & 0 & 0 \\
a(z) & 0 & 0 & 0 & 0 & 0 \\
0 & 0 & 0 & 0 & 0 & 0 \\
f(z) & e(z) & d(z) & 0 & -a(z) & 0 \\
e(z) & 0 & b(z) & 0 & 0 & 0 \\
d(z) & b(z) & c(z) & 0 & 0 & 0   
\end{pmatrix},
\end{align}
where the holomorphic coefficients $a(z), b(z), c(z)$ specify the $\Gamma_{-1}$ component, while $d(z),e(z)$ and $f(z)$ correspond to $\Gamma_{-2}$ and $\Gamma_{-3}$ respectively. Note that we used coordinate shifts \eqref{eq:shift} to set coefficients associated with $N_1,N_2$ to zero. Next we can use \eqref{eq:PiGamma} to write the periods in terms of the holomorphic coefficients as
\begin{align}\label{eq:periodsIV2general}
\Pi = \begin{pmatrix}
1 \\ 
a(z) \\
\frac{ \log[z_2]}{2 \pi i} \\
i \delta_2 + i \delta_1 a(z) +f(z) + \frac{i (\frac{1}{2} a(z) c(z) + d(z)) \log[z_2]}{2 \pi } - \frac{i \log[z_2]^3}{48 \pi^3} \\
i \delta_1 +e(z)- \frac{ a(z)(\log[z_1]+n \log[z_2])}{2 \pi i} \\
\frac{1}{2} a(z) c(z) + d(z) - \frac{\log[z_2]^2}{8 \pi^2} \\
\end{pmatrix}.
\end{align}
The coefficients of the instanton map $\Gamma$ now must satisfy several differential equations. The $\Gamma_{-1}$ coefficients obey \eqref{eq:dXdX}, while in turn the $\Gamma_{-2},\Gamma_{-3}$ coefficients are fixed uniquely by \eqref{eq:recursion}. We can write out \eqref{eq:dXdX} as
\begin{equation}\label{eq:dXdXIV2}
\begin{aligned}
z_2 a^{(0,1)}(z)&=  z_1 (n a^{(1,0)}(z) b^{(1,0)}(z))\, , \\
 c^{(1,0)}(z) &= 2 \pi i z_2 (a^{(1,0)} (z)b^{(0,1)}(z)-a^{(0,1)} (z)b^{(1,0)}(z)) \, ,  
\end{aligned}
\end{equation}
while \eqref{eq:recursion} imposes
\begin{equation}\label{eq:recursionIV2}
\begin{aligned}
 2 d^{(1,0)}(z) &= a^{(1,0)}(z) b(z) - a(z) b^{(1,0)}(z) \, , \\
 2 \pi i z_2  d^{(0,1)}(z)&=c(z) +  \pi i z_2 (a^{(0,1)} (z)b(z) - a(z) b^{(0,1)}(z)) \, , \\
 2 \pi i z_1 e^{(1,0)}(z)&=  a(z) \, ,\\
 2 \pi i z_2 e^{(0,1)}(z)&= b(z) + n a(z) \, ,\\
 f^{(1,0)}(z) &= a^{(1,0)}(z) e(z) \, ,\\
 \pi f^{(0,1)}(z)&= \pi a^{(0,1)}(z) e(z) -id(z)\, .
\end{aligned}
\end{equation}
We now want to construct asymptotic models for the periods by performing a leading order expansion for the $\Gamma_{-1}$ coefficients. First we consider the rank condition \eqref{eq:rankGamma}, which gives us an indication which coefficients have to be turned on. In terms of the Hodge-Deligne diamond it implies that there should be an ingoing arrow due to $\Gamma_{-1}$ or $N_i$ for every vertex apart from $\tilde{I}_{(2)}^{3,3}$. Looking at figure \ref{fig:IV2} this means that $a(z)$ must be turned on. In order to solve \eqref{eq:dXdX} we find that we must also turn on $b(z)$. Let us therefore make as ansatz
\begin{align}
a(z)= a z_1 \, , \qquad b(z)=b z_1 \, , \qquad c(z)=0\, ,
\end{align}
where $a \in \mathbb{R}$ has been rotated to a real value using the axion shift symmetry described below \eqref{eq:shift}. We then find that \eqref{eq:dXdXIV2} reduces on our ansatz to
\begin{equation}
b = -n a\, .
\end{equation}
Consequently we can solve \eqref{eq:recursionIV2} for the coefficients of $\Gamma_{-2},\Gamma_{-3}$ as
\begin{align}
d(z)=0\, , \qquad e(z)= \frac{a}{2 \pi i} \, , \qquad f(z)= \frac{a^2}{4 \pi i } z_1^{2 } \, .
\end{align}
Inserting these leading order behaviors into \eqref{eq:periodsIV2general} we find as asymptotic model for the periods
\begin{align}\label{eq:IV2Period}
\Pi= \begin{pmatrix}
1  \\
a z_1\\
\frac{ \log[z_2]}{2 \pi i} \\
-\frac{i \log[z_2]^3}{48 \pi^3}-\frac{ i a^2 n z_1^2 \log[z_2]}{4\pi }+   \frac{a^2}{4  \pi i} z_1^2+i \delta_2 +i \delta_1 a z_1\\
 - a z_1\frac{ \log[z_1] + n \log[z_2]}{2 \pi i} +i\delta_1\\
- \frac{\log[z_2]^2}{8 \pi^2}  -\frac{1}{2} a^2 n z_1^2\\
\end{pmatrix}.
\end{align}

\section{Conclusions}
With the aim to identify universal properties of string theory compactifications in the asymptotic 
regimes of the moduli space, we have initiated the general study of asymptotic period vectors of Calabi-Yau manifolds. 
We have focused on a detailed study of Calabi-Yau threefold periods, which are obtained by considering the integrals of the 
distinguished $(3,0)$-form over a basis of three-cycles. 
In order to obtain abstract models for these periods we employed the powerful techniques of 
asymptotic Hodge theory. This mathematical machinery allowed us to formulate consistency constraints on the 
asymptotic Hodge decomposition and translate these as conditions on the asymptotic period vectors. The first condition is a
completeness requirement that ensures that the complete complex middle cohomology of the threefold 
is split into $(p,q)$-eigenspaces and all spaces can be spanned by considering forms obtained from the derivatives of the period vector. 
The second condition is that the split has to ensure the positivity of the Hodge norm. The final condition is due to the existence 
of the monodromy symmetry, which becomes particularly constraining near the boundaries to which it is associated. We have then used 
these general principles together with the classification of boundaries in complex structure moduli space to draw general conclusions 
about the asymptotic periods and construct general one- and two-moduli models.

As a first finding we have shown that the asymptotic periods near any co-dimension $h^{2,1}$ boundary component away from large complex structure have to contain 
non-perturbative corrections. More concretely, we considered intersections of $h^{2,1}$ boundary loci $z^i=0$ and argued by 
using completeness and the action of the monodromy matrices. The starting point for the argument is the nilpotent orbit, which was shown in \cite{Schmid} to  
be sufficient to encode a well-behaved asymptotic Hodge decomposition. The information contained in this orbit can be integrated into 
the period vector and we have shown that this general fact suffices to argue for the presence of non-perturbative terms.  To determine the concrete 
expressions for these non-perturbative terms we have introduced the instanton map $\Gamma$ and argued following \cite{CattaniFernandez2000} that its leading terms 
are constrained by the matching with the nilpotent orbit formulation. Taken together with the classification of all one-moduli and two-moduli boundaries 
we were then able to determine general models for the associated asymptotic periods. In this construction we consider a leading order ansatz for the instanton map, which helps us to illustrate their characteristic features. These models still depend on a number of free parameters, which 
would have to be determined by considering concrete Calabi-Yau threefold examples. Nevertheless, we can use positivity properties and completeness 
relations to determine general constraints on these free coefficients. In other words, we restrict the free parameters in such a way that the resulting asymptotic 
periods define a well-behaved asymptotic Hodge decomposition. Let us note that explicit expressions for period vectors 
have been computed for many Calabi-Yau threefold examples by solving Picard-Fuchs equations, performing analytic continuations, and solving the 
constraints from the variation of Hodge structure. We have checked that indeed the explicit periods in e.g.~\cite{Kachru:1995fv} nicely fit into the general models  
determined in this work. It would be interesting to make the connection with these established techniques more concrete. In particular, one could try to set up a combined approach and hope to get further constraints on the periods.

Given the models for the one- and two-moduli periods we have computed some basic physical quantities. Firstly, we have determined 
the K\"ahler potential relevant both in the $\cN=2$ Type IIB vector moduli space and in certain Type IIB orientifold compactifications. Secondly, we have then 
determined the flux superpotential and extracted the leading terms of the scalar potential. In accordance with the general expectation, see the recent analysis 
in the more general fourfold case \cite{Grimm:2019ixq}, we have shown that the leading scalar potential admits a polynomial behavior. We have shown that in some cases, such as 
in class II$_1$ boundaries, these polynomial terms can stem from exponentially suppressed corrections in the superpotential and K\"ahler potential. In general and in 
particular in higher-dimensional moduli spaces, it is therefore crucial to keep non-perturbative terms in $K$ and $W$. Alternatively, one can first take derivatives 
of $K,W$ with the general periods and then use the nilpotent orbit as an approximation. As already stressed in \cite{Bastian:2020egp} these processes do not commute.
Despite the presence of non-perturbative contributions in the asymptotic K\"ahler potential near most boundaries, we have found that it always is independent 
of at least one real coordinate direction. The presence of this axion-like field was linked with infinite distances in moduli space in \cite{Grimm:2018ohb} and recently 
played a central role in the swampland program \cite{Lanza:2020qmt,Lanza:2021qsu}. Here we find that the existence of such axions is tied to the presence of a non-trivial 
log-monodromy matrix. Hence, an axionic direction exist in our models at all infinite distance boundaries which are of type II, III, IV as has been conjectured, but also 
in the models that admit finite distance boundaries that are of type I. 

The directions in which we can continue the programme initiated in this work are twofold. To begin with, there are numerous interesting applications of the models presented in this work. An immediate next step \cite{toappear} is to use the one- and two-moduli 
periods in order to test some of the asymptotic swampland conjectures including the essential non-perturbative corrections. This will allow us, for example, 
to compute subleading corrections to the weak gravity bounds found in \cite{Bastian:2020egp}. Furthermore, one can also study moduli stabilization in detail \cite{Bastian:2021hpc,inprogress}. Eventually these results can be employed to construct phenomenological 
models, for example, when trying to implement axion inflation. The other direction is to apply the methods used in this work to more general settings.
Firstly, it is desirable to construct models for the 
asymptotic period vector for moduli spaces of higher dimension. While technically more involved, we see no obstacle that this can be done systematically for 
an arbitrary number of moduli. One might hope to find general closed expressions that determine the asymptotic periods in terms of the 
enhancement graph introduced in \cite{Grimm:2019bey}. This would give us a powerful tool for instance to test recent conjectures about tadpole problems \cite{Bena:2020xrh,Bena:2021wyr}. Secondly, an immediate generalization would be to determine periods in Calabi-Yau fourfolds and study F-theory 
models. Here again the technology will remain the same, even though the models will get technically more involved. On a more fundamental level it will eventually 
be necessary to constrain the free parameters in the periods for the constructed general models. While this will require to go beyond the mathematical tools introduced here 
it appears to be central to many applications and deserves much attention in the future.

\subsubsection*{Acknowledgments}
It is a pleasure to thank Stefano Lanza, Chongchuo Li, Jeroen Monnee, Eran Palti, Erik Plauschinn and Irene Valenzuela for very useful discussions and correspondence. 
This research is partly supported by the Dutch Research Council (NWO) via a Start-Up grant and a VICI grant. 

\appendix

\section{Construction of nilpotent orbit data for two-moduli periods}\label{app:boundarydata}
In this section we construct the nilpotent orbits of the two-moduli periods. Following figure \ref{fig:flowchart}, we begin by writing down an $sl(2)$-splitting each boundary. Recall from section \ref{sec:sl2splitting} that the $sl(2)^n$-boundary data is encoded in a set of commuting $sl(2)$-triples $(N_i^{\pm},N_i^0)$ and the $sl(2)$-split Deligne splitting $\tilde{I}^{p,q}_{(2)}$ characterizing the intersection. By using the building blocks given in table \ref{table:buildingblocks} we can straightforwardly obtain simple expressions for this $sl(2)^n$-data. The main task in this appendix is then to complete this $sl(2)^n$-boundary data into the most general compatible nilpotent orbit, which comes in two parts. First we construct the most general log-monodromy matrices $N_i$ out of the lowering operators $N_i^-$. Secondly we determine the most general phase operator $\delta$ that rotates the $sl(2)$-split $\tilde{I}^{p,q}_{(2)}$ into a generic Deligne splitting. This construction follows the approach laid out in section \ref{sec:nilpotent_orbit}.

\subsection{Class $\text{I}_2$ boundaries}\label{app:I2data}
Let us begin by considering boundaries of class $\mathrm{I}_2$, which consists of the 2-cubes $\langle \mathrm{I}_1|\mathrm{I}_2|\mathrm{I}_1 \rangle$, $\langle \mathrm{I}_2|\mathrm{I}_2|\mathrm{I}_1 \rangle$ and $\langle \mathrm{I}_2|\mathrm{I}_2|\mathrm{I}_2 \rangle$. For $\langle \mathrm{I}_1|\mathrm{I}_2|\mathrm{I}_1 \rangle$ we consider the enhancement chain $\mathrm{I}_1 \to \mathrm{I}_2$, while we consider $\mathrm{I}_2 \to \mathrm{I}_2$ for $\langle \mathrm{I}_2|\mathrm{I}_2|\mathrm{I}_1 \rangle$ and $\langle \mathrm{I}_2|\mathrm{I}_2|\mathrm{I}_2 \rangle$. Both these enhancement chains have the singularity type $\mathrm{I}_2$ for $y^1=y^2=\infty$ in common. We span the vector spaces $\tilde{I}_{(2)}^{p,q}$ of this $sl(2)$-split Deligne splitting by
\begin{equation}
\begin{aligned}
\tilde{I}_{(2)}^{3,0}&: \quad \begin{pmatrix} 1, & 0,&0,&i,&0,&0 \end{pmatrix} , \\
\tilde{I}_{(2)}^{2,2}&: \quad \begin{pmatrix} 0, & 1,&0,&0,&0,&0 \end{pmatrix} , \  \begin{pmatrix} 0, & 0,&1,&0,&0,&0 \end{pmatrix} ,  \\
\tilde{I}_{(2)}^{1,1}&: \quad \begin{pmatrix} 0, & 0,&0,&0,&1,&0 \end{pmatrix} ,   \  \begin{pmatrix} 0, & 0,&0,&0,&0,&1 \end{pmatrix} ,  \\
\tilde{I}_{(2)}^{0,3}&: \quad \begin{pmatrix} 1, & 0,&0,&-i,&0,&0\end{pmatrix} . \\
\end{aligned}
\end{equation}

\subsubsection*{Enhancement step $\text{I}_1 \to \text{I}_2$}
Here we construct the nilpotent orbit data for the 2-cube $\langle \mathrm{I}_1 | \mathrm{I}_2 | \mathrm{I}_1 \rangle$. Let us begin by writing down the commuting $sl(2)$-triples as
\begin{equation}
\begin{aligned}
 N_1 &= N_1^- =  \begin{pmatrix}
 0 & 0 & 0 & 0 & 0 & 0 \\
 0 & 0 & 0 & 0 & 0 & 0 \\
 0 & 0 & 0 & 0 & 0 & 0 \\
 0 & 0 & 0 & 0 & 0 & 0 \\
 0 & 0 & 0 & 0 & 0 & 0 \\
 0 & 0 & -1 & 0 & 0 & 0 \\
 \end{pmatrix}, \qquad &Y_1 &= \begin{pmatrix}
 0 & 0 & 0 & 0 & 0 & 0 \\
 0 & 0 & 0 & 0 & 0 & 0 \\
 0 & 0 & 1 & 0 & 0 & 0 \\
 0 & 0 & 0 & 0 & 0 & 0 \\
 0 & 0 & 0 & 0 & 0 & 0 \\
 0 & 0 & 0 & 0 & 0 & -1 \\
\end{pmatrix}\, , \\
 N_2^- &= \begin{pmatrix}
 0 & 0 & 0 & 0 & 0 & 0 \\
 0 & 0 & 0 & 0 & 0 & 0 \\
 0 & 0 & 0 & 0 & 0 & 0 \\
 0 & 0 & 0 & 0 & 0 & 0 \\
 0 & -1 & 0 & 0 & 0 & 0 \\
 0 & 0 & 0 & 0 & 0 & 0 \\
 \end{pmatrix}, \qquad &Y_2 &= \begin{pmatrix}
 0 & 0 & 0 & 0 & 0 & 0 \\
 0 & 1 & 0 & 0 & 0 & 0 \\
 0 & 0 & 0 & 0 & 0 & 0 \\
 0 & 0 & 0 & 0 & 0 & 0 \\
 0 & 0 & 0 & 0 & -1 & 0 \\
 0 & 0 & 0 & 0 & 0 & 0 \\
\end{pmatrix}\, ,
\end{aligned}
\end{equation}
where we did not include $N_i^+$ since the raising operators are irrelevant for our discussion. Let us point out that the sign of the coefficients in the lowering operators is fixed by \eqref{eq:pol}, which requires $\eta N_1^-$ and $\eta N_2^-$ to have negative eigenvalues.

Next we want to write down the most general log-monodromy matrix $N_2$ compatible with the above boundary data. From \eqref{eq:NtoNmin} we find that we must identify matrices with eigenvalue $\ell\leq -2$ under the adjoint action of $Y_1$. The only map that satisfies this property is proportional to $N_1$, so we find that
\begin{equation}
N_2 = N_2^- + n N_1 = \begin{pmatrix}
 0 & 0 & 0 & 0 & 0 & 0 \\
 0 & 0 & 0 & 0 & 0 & 0 \\
 0 & 0 & 0 & 0 & 0 & 0 \\
 0 & 0 & 0 & 0 & 0 & 0 \\
 0 & -1 & 0 & 0 & 0 & 0 \\
 0 & 0 & -n & 0 & 0 & 0 \\
 \end{pmatrix}.
\end{equation}
Similar to the lowering operators we find that \eqref{eq:pol} requires $N_2$ to have two negative eigenvalues, so we must impose $n\geq 0$. For $n=0$ we find that $N_2$ produces a $\mathrm{I}_1$ singularity type for the $y^2=\infty$ divisor, while for $n>0$ it produces a $\mathrm{I}_2$ singularity. Thus for the 2-cube $\langle \mathrm{I}_1 | \mathrm{I}_2 | \mathrm{I}_1 \rangle$ one can simply take $N_1^-$ and $N_2^-$ as log-monodromy matrices.

\subsubsection*{Enhancement step $\text{I}_2 \to \text{I}_2$}
Here we construct the nilpotent orbit data for the 2-cubes $\langle \mathrm{I}_2|\mathrm{I}_2|\mathrm{I}_1 \rangle$ and $\langle \mathrm{I}_2|\mathrm{I}_2|\mathrm{I}_2 \rangle$. Let us begin by writing down the commuting $sl(2)$-triples as
\begin{equation}
 N_1 =N_1^-= -\begin{pmatrix}
 0 & 0 & 0 & 0 & 0 & 0 \\
 0 & 0 & 0 & 0 & 0 & 0 \\
 0 & 0 & 0 & 0 & 0 & 0 \\
 0 & 0 & 0 & 0 & 0 & 0 \\
 0 & 1 & 0 & 0 & 0 & 0 \\
 0 & 0 & 1 & 0 & 0 & 0 \\
 \end{pmatrix}, \qquad Y_1 = \begin{pmatrix}
 0 & 0 & 0 & 0 & 0 & 0 \\
 0 & 1 & 0 & 0 & 0 & 0 \\
 0 & 0 & 1 & 0 & 0 & 0 \\
 0 & 0 & 0 & 0 & 0 & 0 \\
 0 & 0 & 0 & 0 & -1 & 0 \\
 0 & 0 & 0 & 0 & 0 & -1 \\
\end{pmatrix},
\end{equation}
while the second $sl(2)$-triple is trivial, i.e.~$N_2^-=Y_2=0$. The sign of the coefficients in $N_1$ is fixed by the requirement of the polarization conditions \eqref{eq:pol} that $\eta N_1$ has two negative eigenvalues.

Next we want to determine the most general log-monodromy matrix $N_2$ compatible with the above boundary data. There are three matrices with eigenvalue $\ell=-2$ under the adjoint action of $Y_1$ that are infinitesimal isometries of the symplectic pairing $\langle \cdot, \cdot \rangle$. These matrices map from $\tilde{I}_{(2)}^{2,2}$ to $\tilde{I}_{(2)}^{1,1}$, and the most general linear combination is given by

\begin{equation}
N_2
=- \begin{pmatrix}
 0 & 0 & 0 & 0 & 0 & 0 \\
 0 & 0 & 0 & 0 & 0 & 0 \\
 0 & 0 & 0 & 0 & 0 & 0 \\
 0 & 0 & 0 & 0 & 0 & 0 \\
 0 & n_1 & n_3 & 0 & 0 & 0 \\
 0 & n_3 & n_2 & 0 & 0 & 0 \\
\end{pmatrix}\, .
\label{NilpI1}
\end{equation}
We now want to simplify this expression by considering a change of basis. First we want to diagonalize the $2\times 2$ block that appears in $N_2$ by a symplectic basis transformation (while keeping $N_1$ the same). This yields
\begin{equation}
N_2
= -\begin{pmatrix}
 0 & 0 & 0 & 0 & 0 & 0 \\
 0 & 0 & 0 & 0 & 0 & 0 \\
 0 & 0 & 0 & 0 & 0 & 0 \\
 0 & 0 & 0 & 0 & 0 & 0 \\
 0 & \lambda_1 & 0 & 0 & 0 & 0 \\
 0 & 0 & \lambda_2 & 0 & 0 & 0 \\
\end{pmatrix}\, ,
\end{equation}
where the $2\times 2$ matrix needs to have at least one non-zero eigenvalue, which we take to be $\lambda_1 \neq 0$. This allows us to rescale by the symplectic basis transformation $M=\text{diag}(1,\sqrt{\lambda_1},1,1,1/\sqrt{\lambda_1},1)$, after which our log-monodromy matrices become
\begin{equation}
 N_1 =-\begin{pmatrix}
 0 & 0 & 0 & 0 & 0 & 0 \\
 0 & 0 & 0 & 0 & 0 & 0 \\
 0 & 0 & 0 & 0 & 0 & 0 \\
 0 & 0 & 0 & 0 & 0 & 0 \\
 0 & n_1 & 0 & 0 & 0 & 0 \\
 0 & 0 & 1 & 0 & 0 & 0 \\
 \end{pmatrix}, \qquad N_2
= -\begin{pmatrix}
 0 & 0 & 0 & 0 & 0 & 0 \\
 0 & 0 & 0 & 0 & 0 & 0 \\
 0 & 0 & 0 & 0 & 0 & 0 \\
 0 & 0 & 0 & 0 & 0 & 0 \\
 0 & 1 & 0 & 0 & 0 & 0 \\
 0 & 0 & n_2 & 0 & 0 & 0 \\
\end{pmatrix}\, ,
\end{equation}
by relabeling $n_1=1/\lambda_1$ and $n_2=\lambda_2$. From the polarization conditions \eqref{eq:pol} we find that $\eta N_2$ needs to have one positive and one non-negative eigenvalue, so $n_2 \geq 0$. For $n_2=0$ we are dealing with a $\mathrm{I}_1$ divisor at $y^2=\infty$, while for $n_2>0$ it is a $\mathrm{I}_2$ divisor. Note that the log-monodromy matrices for the $\langle \mathrm{I}_1 | \mathrm{I}_2 | \mathrm{I}_1\rangle$ boundary in the $\mathrm{I}_1 \to \mathrm{I}_2$ enhancement step take the same form with $n_1=n_2=0$. This conveniently allows us to use the same form for the log-monodromy matrices for all boundaries of $\mathrm{I}_2$ class.

\subsubsection*{Construction of the phase operator}
Based on the above data let us write down the phase operator $\delta$ according to \eqref{eq:delta}. The only possible matrices that are valued in $\Lambda_{-p,-q}$ with $p,q>0$ map from $\tilde{I}_{(2)}^{2,2}$ to $\tilde{I}_{(2)}^{1,1}$, effectively reducing $\delta$ to a $2\times 2$ sub-block. We therefore find that the most general $\delta$ takes the form
\begin{equation}
\delta = \begin{pmatrix}
 0 & 0 & 0 & 0 & 0 & 0 \\
 0 & 0 & 0 & 0 & 0 & 0 \\
 0 & 0 & 0 & 0 & 0 & 0 \\
 0 & 0 & 0 & 0 & 0 & 0 \\
 0 & \delta_2 & \delta_1 & 0 & 0 & 0 \\
 0 & \delta_1 & \delta_3 & 0 & 0 & 0 \\
\end{pmatrix}\, ,
\end{equation}
where the off-diagonal components of the sub-block are set to be equal by requiring $\delta^T \eta+\eta \delta = 0$. One can reduce $\delta$ further by using coordinate shifts following \eqref{eq:shift}, allowing one to set $\delta_2=\delta_3=0$.

\subsection{Class $\mathrm{II}_1$ boundaries} \label{app:II1data}
Next we consider boundaries of class $\mathrm{II}_1$, which consists of the 2-cubes $\langle \mathrm{II}_0 | \mathrm{II}_1 | \mathrm{I}_1 \rangle$, $\langle \mathrm{II}_0 | \mathrm{II}_1 | \mathrm{II}_1 \rangle$,  $\langle \mathrm{II}_1 | \mathrm{II}_1 | \mathrm{I}_1 \rangle$ and  $\langle \mathrm{II}_1 | \mathrm{II}_1 | \mathrm{II}_1 \rangle$. For the first two boundaries we reconstruct the nilpotent orbit data starting from the enhancement chain $\mathrm{II}_0 \to \mathrm{II}_1$, while for the latter two boundaries we consider the enhancement chain $\mathrm{II}_1 \to \mathrm{II}_1$. In either case the enhancement chains have the singularity type $\mathrm{II}_1$ at $y^1=y^2=\infty$ in common. Let us therefore begin by writing down the vectors that span the spaces $\tilde{I}_{(2)}^{p,q}$ of this $sl(2)$-split Deligne splitting as
\begin{equation}\label{eq:appII1}
\begin{aligned}
\tilde{I}_{(2)}^{3,1}&: \quad \begin{pmatrix} 1, & i,&0,&0,&0,&0 \end{pmatrix} ,\qquad  &\tilde{I}_{(2)}^{2,0}&: \quad\begin{pmatrix} 0, & 0,&0,&1,&i,&0 \end{pmatrix} , \\
\tilde{I}_{(2)}^{2,2}&: \quad \begin{pmatrix} 0, & 0,&1,&0,&0,&0 \end{pmatrix} , \qquad &\tilde{I}_{(2)}^{1,1}&: \quad \begin{pmatrix} 0, & 0,&0,&0,&0,&1 \end{pmatrix} ,  \\
\tilde{I}_{(2)}^{1,3}&:\quad \begin{pmatrix} 1, & -i,&0,&0,&0,&0\end{pmatrix} , \qquad &\tilde{I}_{(2)}^{0,2}&:\quad \begin{pmatrix} 0, & 0,&0,&1,&-i,&0\end{pmatrix} .
\end{aligned}
\end{equation}

\subsubsection*{Enhancement step $\text{II}_0 \to \text{II}_1$}
Here we construct the nilpotent orbit data for the 2-cubes $\langle \mathrm{II}_0 | \mathrm{II}_1 | \mathrm{I}_1 \rangle$ and $\langle \mathrm{II}_0 | \mathrm{II}_1 | \mathrm{II}_1 \rangle$. Let us begin by writing down the $sl(2)$-triples as
\begin{equation}
 N_1 = N_1^- =  \begin{pmatrix}
 0 & 0 & 0 & 0 & 0 & 0 \\
 0 & 0 & 0 & 0 & 0 & 0 \\
 0 & 0 & 0 & 0 & 0 & 0 \\
 1 & 0 & 0 & 0 & 0 & 0 \\
 0 & 1 & 0 & 0 & 0 & 0 \\
 0 & 0 & 0 & 0 & 0 & 0 \\
 \end{pmatrix}, \qquad Y_1 = \begin{pmatrix}
 1 & 0 & 0 & 0 & 0 & 0 \\
 0 & 1 & 0 & 0 & 0 & 0 \\
 0 & 0 & 0 & 0 & 0 & 0 \\
 0 & 0 & 0 & -1 & 0 & 0 \\
 0 & 0 & 0 & 0 & -1 & 0 \\
 0 & 0 & 0 & 0 & 0 & 0 \\
\end{pmatrix},
\end{equation}
and
\begin{equation}
 N_2^- = \begin{pmatrix}
 0 & 0 & 0 & 0 & 0 & 0 \\
 0 & 0 & 0 & 0 & 0 & 0 \\
 0 & 0 & 0 & 0 & 0 & 0 \\
 0 & 0 & 0 & 0 & 0 & 0 \\
 0 & 0 & 0 & 0 & 0 & 0 \\
 0 & 0 & -1 & 0 & 0 & 0 \\
 \end{pmatrix}, \qquad Y_2 = \begin{pmatrix}
 0 & 0 & 0 & 0 & 0 & 0 \\
 0 & 0 & 0 & 0 & 0 & 0 \\
 0 & 0 & 1 & 0 & 0 & 0 \\
 0 & 0 & 0 & 0 & 0 & 0 \\
 0 & 0 & 0 & 0 & 0 & 0 \\
 0 & 0 & 0 & 0 & 0 & -1 \\
\end{pmatrix},
\end{equation}
where the signs of the coefficients in $N_1^-,N_2^-$ are fixed by the polarization conditions \eqref{eq:pol}. To be precise,  $\eta N_1^-$ must have two positive eigenvalues and $\eta N_2^-$ one negative eigenvalue.

Next we determine the most general log-monodromy matrix $N_2$ compatible with the above boundary data. There are three real maps with eigenvalue $\ell\leq -2$ under the adjoint action with $Y_1$, which means we find as log-monodromy matrix
 \begin{equation}
 N_2 = \begin{pmatrix}
 0 & 0 & 0 & 0 & 0 & 0 \\
 0 & 0 & 0 & 0 & 0 & 0 \\
 0 & 0 & 0 & 0 & 0 & 0 \\
 n_2 & n_4 & 0 & 0 & 0 & 0 \\
 n_4 & n_3 & 0 & 0 & 0 & 0 \\
 0 & 0 & -1 & 0 & 0 & 0 \\
 \end{pmatrix} \,.
\end{equation}
Additionally recall that  $N_2$ must be a $(-1,-1)$-map with respect to the Deligne splitting \eqref{eq:appII1}, which requires us to put $n_2=n_3$ and $n_4=0$. Polarization conditions \eqref{eq:pol} then tell us that $\eta N_2$ should have two non-negative eigenvalues and one negative eigenvalue, which sets $n_2 \geq 0$. For $n_2=0$ we encounter a $\mathrm{I}_1$ divisor at $y^2=\infty$, while for $n_2>0$ we encounter a $\mathrm{II}_1$ divisor.

\subsubsection*{Enhancement step $\text{II}_1 \to \text{II}_1$}
Here we construct the nilpotent orbit data for the 2-cubes $\langle \mathrm{II}_1 | \mathrm{II}_1 | \mathrm{I}_1 \rangle$ and $\langle \mathrm{II}_1 | \mathrm{II}_1 | \mathrm{II}_1 \rangle$. Let us begin by writing down the $sl(2)$-triples as 
\begin{equation}
 N_1 = N_1^- =  \begin{pmatrix}
 0 & 0 & 0 & 0 & 0 & 0 \\
 0 & 0 & 0 & 0 & 0 & 0 \\
 0 & 0 & 0 & 0 & 0 & 0 \\
 1 & 0 & 0 & 0 & 0 & 0 \\
 0 & 1 & 0 & 0 & 0 & 0 \\
 0 & 0 & -1 & 0 & 0 & 0 \\
 \end{pmatrix}, \qquad Y_1 = \begin{pmatrix}
 1 & 0 & 0 & 0 & 0 & 0 \\
 0 & 1 & 0 & 0 & 0 & 0 \\
 0 & 0 & 1 & 0 & 0 & 0 \\
 0 & 0 & 0 & -1 & 0 & 0 \\
 0 & 0 & 0 & 0 & -1 & 0 \\
 0 & 0 & 0 & 0 & 0 & -1 \\
\end{pmatrix},
\end{equation}
while the second $\mathfrak{sl}(2,\mathbb{R})$-triple is trivial, i.e.~$N_2^-=Y_2=0$. The signs of $N_1$ are fixed by the polarization conditions \eqref{eq:pol}, which requires  $\eta N_1^-$ to have two positive and one negative eigenvalue.

We now want to construct the most general log-monodromy matrix $N_2$ compatible with the above boundary data. The most general map with eigenvalue $\ell \leq -2$ under the adjoint action with $Y_1$ is given by
\begin{equation}
 N_2 = \begin{pmatrix}
 0 & 0 & 0 & 0 & 0 & 0 \\
 0 & 0 & 0 & 0 & 0 & 0 \\
 0 & 0 & 0 & 0 & 0 & 0 \\
  n_1 & n_4 & n_5 & 0 & 0 & 0 \\
 n_4 & n_2 & n_6 & 0 & 0 & 0 \\
 n_5 & n_6 & -n_3 & 0 & 0 & 0 \\
 \end{pmatrix}  \,.
\end{equation}
where we required the $3\times 3$ block to be symmetric to ensure that $N_2^T \eta + \eta N_2 = 0$. Additionally we must require that $N_2$ is a $(-1,-1)$-map with respect to the Deligne splitting \eqref{eq:appII1}. This requires us to set $n_1=n_2$ and $n_4=n_5=n_6=0$. Polarization conditions \eqref{eq:pol} then require us to put $n_1>0$ and $n_3>0$.

Let us now try to bring these log-monodromy matrices into a similar form as we found for the enhancement chain $\mathrm{II}_0 \to \mathrm{II}_1$. We can apply a symplectic basis transformation $M = \text{diag}(1,1,\sqrt{n_3},1,1,1/\sqrt{n_3})$, which yields
\begin{equation}
N_1 = \begin{pmatrix}
 0 & 0 & 0 & 0 & 0 & 0 \\
 0 & 0 & 0 & 0 & 0 & 0 \\
 0 & 0 & 0 & 0 & 0 & 0 \\
 1 & 0 & 0 & 0 & 0 & 0 \\
 0 & 1 & 0 & 0 & 0 & 0 \\
 0 & 0 & -n_1 & 0 & 0 & 0 \\
 \end{pmatrix}, \qquad N_2 = \begin{pmatrix}
 0 & 0 & 0 & 0 & 0 & 0 \\
 0 & 0 & 0 & 0 & 0 & 0 \\
 0 & 0 & 0 & 0 & 0 & 0 \\
 n_2 & 0 & 0 & 0 & 0 & 0 \\
 0 & n_2 & 0 & 0 & 0 & 0 \\
 0 & 0 & -1 & 0 & 0 & 0 \\
\end{pmatrix} \,,
\end{equation}
where we relabeled $n_1=1/n_3$. 
\subsubsection*{Construction of the phase operator}
Based on the above data let us write down the most general phase operator $\delta$ according to \eqref{eq:delta} and \eqref{eq:deltacom}. For the above Deligne splitting there are two real maps $\delta_{-p,-q}$ with $p,q>0$ that satisfy $\delta_{-p,-q}^T \eta+\eta \delta_{-p,-q}=0$, one mapping $\tilde{I}_{(2)}^{3,1},\tilde{I}_{(2)}^{1,3}$ to $\tilde{I}_{(2)}^{2,0},\tilde{I}_{(2)}^{0,2}$ and another from $\tilde{I}_{(2)}^{2,2}$ to $\tilde{I}_{(2)}^{1,1}$. Taking $\delta_1,\delta_2  \in \mathbb{R}$ as proportionality constants for these two maps we find
\begin{equation}
\delta = \begin{pmatrix}
 0 & 0 & 0 & 0 & 0 & 0 \\
 0 & 0 & 0 & 0 & 0 & 0 \\
 0 & 0 & 0 & 0 & 0 & 0 \\
 \delta_1 & 0 & 0 & 0 & 0 & 0 \\
 0 & \delta_1 & 0 & 0 & 0 & 0 \\
 0 & 0 & \delta_2& 0 & 0 & 0 \\
\end{pmatrix}\, .
\end{equation}
Note that this matrix is precisely of the same form as the log-monodromies $N_1,N_2$, so we can use coordinate shifts \eqref{eq:shift} to set $\delta_1=\delta_2=0$.

\subsection{Coni-LCS class boundaries}\label{app:coniLCSdata}
Finally we consider coni-LCS class boundaries, which consists of the 2-cubes $\langle \mathrm{I}_1 | \mathrm{IV}_2 | \mathrm{IV}_1 \rangle$ and $\langle \mathrm{I}_1 | \mathrm{IV}_2 | \mathrm{IV}_2 \rangle$. These boundaries share a $\mathrm{IV}_2$ singularity type at the intersection $y^1=y^2=\infty$. Let us begin by writing down a basis for the vector spaces $\tilde{I}_{(2)}^{p,q}$ of this $sl(2)$-split Deligne splitting as
\begin{equation}
\begin{aligned}
\tilde{I}_{(2)}^{3,3}&: \quad \begin{pmatrix} 1, & 0,&0,&0,&0,&0 \end{pmatrix} , \\
\tilde{I}_{(2)}^{2,2}&: \quad \begin{pmatrix} 0, & 1,&0,&0,&0,&0 \end{pmatrix} , \ \  \begin{pmatrix} 0, & 0,&1,&0,&0,&0 \end{pmatrix} , \\
\tilde{I}_{(2)}^{1,1}&: \quad \begin{pmatrix} 0, & 0,&0,&0,&1,&0 \end{pmatrix} , \ \  \begin{pmatrix} 0, & 0,&0,&0,&0,&1 \end{pmatrix} , \\
\tilde{I}_{(2)}^{0,0}&: \quad \begin{pmatrix} 0, & 0,&0,&1,&0,&0 \end{pmatrix} .
\end{aligned}
\end{equation}

\subsubsection*{Enhancement step $\text{I}_1 \to \text{IV}_2$}
Here we construct the log-monodromy matrices for the 2-cubes $\langle \mathrm{I}_1 | \mathrm{IV}_2 | \mathrm{IV}_1 \rangle$ and $\langle \mathrm{I}_1 | \mathrm{IV}_2 | \mathrm{IV}_2 \rangle$. Let us begin by writing down the $sl(2)$-triples as
\begin{equation}
\begin{aligned}
N_1 = N_1^- &= \begin{pmatrix}
 0 & 0 & 0 & 0 & 0 & 0 \\
 0 & 0 & 0 & 0 & 0 & 0 \\
 0 & 0 & 0 & 0 & 0 & 0 \\
 0 & 0 & 0 & 0 & 0 & 0 \\
 0 & -1 & 0 & 0 & 0 & 0 \\
 0 & 0 & 0 & 0 & 0 & 0 
\end{pmatrix},  \qquad &Y_1 &= \begin{pmatrix}
 0 & 0 & 0 & 0 & 0 & 0 \\
 0 & 1 & 0 & 0 & 0 & 0 \\
 0 & 0 & 0 & 0 & 0 & 0 \\
 0 & 0 & 0 & 0 & 0 & 0 \\
 0 & 0 & 0 & 0 & -1 & 0 \\
 0 & 0 & 0 & 0 & 0 & 0 
\end{pmatrix} ,\\
\quad N_2^- &= \begin{pmatrix}
 0 & 0 & 0 & 0 & 0 & 0 \\
 0 & 0 & 0 & 0 & 0 & 0 \\
 1 & 0 & 0 & 0 & 0 & 0 \\
 0 & 0 & 0 & 0 & 0 & -1 \\
 0 & 0 & 0 & 0 & 0 & 0 \\
 0 & 0 & 1 & 0 & 0 & 0 
\end{pmatrix}, \qquad &Y_2 &= \begin{pmatrix}
 3 & 0 & 0 & 0 & 0 & 0 \\
 0 & 0 & 0 & 0 & 0 & 0 \\
 0 & 0 & 1 & 0 & 0 & 0 \\
 0 & 0 & 0 & -3 & 0 & 0 \\
 0 & 0 & 0 & 0 & 0 & 0 \\
 0 & 0 & 0 & 0 & 0 & -1 
\end{pmatrix}.
\end{aligned}
\end{equation}
The signs of the coefficients of the lowering operators $N^-_1,N^-_2$ are fixed by the polarization condition \eqref{eq:pol}. We must require $\eta N^-_1$ and $\eta (N^-_2)^3$ both to have one negative eigenvalue. In turn the condition that $N_2$ is an infinitesimal isomorphic of the symplectic product $(N^-_2)^T \eta + \eta N_2^-=0$ fixes the sign of the other two coefficients of $N_2$.

Next we construct the most general log-monodromy matrix $N_2$ compatible with the above boundary data. There is only one map with weight $\ell\leq -2$ under the adjoint action of $Y_1$, which is $N_1$. By using \eqref{eq:NtoNmin} we therefore find
\begin{equation}
N_2 = N_2^- + n N_1  = \begin{pmatrix}
 0 & 0 & 0 & 0 & 0 & 0 \\
 0 & 0 & 0 & 0 & 0 & 0 \\
 1 & 0 & 0 & 0 & 0 & 0 \\
 0 & 0 & 0 & 0 & 0 & -1 \\
 0 & -n & 0 & 0 & 0 & 0 \\
 0 & 0 & 1 & 0 & 0 & 0 
 \end{pmatrix},
\label{NilpIV2}
\end{equation}
where polarization conditions require $n\geq 0$. For $n=0$ we encounter a $\mathrm{IV}_1$ divisor at $y^2=\infty$, while for $n>0$ it is a $\mathrm{IV}_2$ divisor.

\subsubsection*{Construction of the phase operator}
Based on the above data let us construct the most general phase operator $\delta$ according to \eqref{eq:delta} and \eqref{eq:deltacom}. For the given Deligne splitting there are four real maps $\delta_{-p,-q}$ with $p,q\geq 0$ we can write down that are infinitesimal isometries of $\langle \cdot, \cdot \rangle$. Taking their linear combination gives us
\begin{equation}
\delta =  \begin{pmatrix} 0 & 0 & 0 & 0 & 0 & 0 \\
 0 & 0 & 0 & 0 & 0 & 0 \\
 \delta_4 & 0 & 0 & 0 & 0 & 0 \\
 \delta_2 & \delta_1 & 0& 0 & 0 & -\delta_4 \\
 \delta_1 & \delta_3 & 0 & 0 & 0 & 0 \\
 0 & 0 & \delta_4 & 0 & 0 & 0 
\end{pmatrix}.
\end{equation}
The components related to the coefficients $\delta_3,\delta_4$ are proportional to $N_1$ and $N_2$, so these can be tuned by using \eqref{eq:shift}.

\section{Embedding periods for geometrical examples}\label{app:embedding}
In this appendix we show how the periods constructed in our work relate to some familiar geometrical examples. We rewrite the periods of the one-modulus $\mathrm{I}_1$ boundary and two-modulus coni-LCS class boundaries in terms of the prepotential formulation of the conifold and coni-LCS periods. Furthermore we show how the periods found for the 2-cube $\langle \mathrm{II}_0 | \mathrm{II}_1 | \mathrm{II}_1 \rangle$ cover the periods for the Calabi-Yau threefold in $\mathbb{P}_4^{1,1,2,2,6}[12]$ near a particular degeneration.

\subsection{Conifold point}
We begin by rewriting the periods \eqref{eq:I1periods} of $\mathrm{I}_1$ boundaries in terms of the prepotential formulation in e.g.~\cite{PhysRevLett.62.1956,Candelas:1989js,Strominger_1995}. In this frame the periods can be written as $\Pi=(X^0,X^1,\cF_0,\cF_1)$, where the $\cF_i = \partial_{X^i}\cF$ are obtained by taking derivatives of the prepotential $\cF(X^i)$. In order to bring our periods into this frame, one has to perform K\"ahler transformations and basis changes. We typically set $X^0=1$, so let us first rescale the periods by an overall factor $\Pi \to e^f \Pi$ with $f=1-\frac{a^2 z^2}{8\pi}$ to set the first entry equal to one. Next we want to set the second entry equal to the special coordinate $X^1 =z$, so we also apply a symplectic basis transformation $M=\text{diag}(1,1/a,1,a)$. Consequently the transformed period vector reads
\begin{equation}
\Pi = \begin{pmatrix}
1 \\ 
z \\
i-\frac{ia^2}{4\pi}z^2 \\
\frac{ia^2}{2\pi} z\log[z]
\end{pmatrix},
\end{equation}
up to corrections in $z^3$. One can straightforwardly verify that these periods indeed match with the prepotential
\begin{equation}\label{eq:I1prepotential}
\cF= \frac{i}{2}(X^0)^2 +\frac{i a^2}{4\pi} (X^1)^2 \log\big[X^1/X^0\big] - \frac{i a^2}{8\pi} (X^1)^2 \, ,
\end{equation}
where afterwards we can set $X^0=1$ and $X^1=z=e^{2\pi i t}$. Note in particular that $a^2>0$ now fixes the sign of the second term in this prepotential.

\subsection{Coni-LCS point}\label{app:coniLCS}
Next we rewrite the periods \eqref{eq:coniLCSperiods} near coni-LCS boundaries in the prepotential formulation, see e.g.~\cite{Demirtas:2020ffz,Blumenhagen:2020ire} for recent constructions using different methods. Again we want to set one period proportional to the conifold modulus as $X^1=z_1$, so let perform a symplectic basis transformation $M = \text{diag}(1,1/a,1,1,a,1)$ analogous to the $\mathrm{I}_1$ boundaries. The periods then read
\begin{equation}
\Pi = \begin{pmatrix}
1  \\
 z_1\\
\frac{ \log[z_2]}{2 \pi i} \\
-\frac{i \log[z_2]^3}{48 \pi^3}-\frac{ i a^2 n z_1^2 \log[z_2]}{4\pi }+   \frac{a^2}{4  \pi i} z_1^2+i \delta_2 +i \delta_1 a z_1\\
 - a^2 z_1\frac{ \log[z_1] + n \log[z_2]}{2 \pi i} +i\delta_1 a\\
- \frac{\log[z_2]^2}{8 \pi^2}  -\frac{1}{2} a^2 n z_1^2\\
\end{pmatrix}.
\end{equation} 
Equivalently these periods can be obtained from the prepotential
\begin{equation}
\cF = \frac{1}{6} \frac{\cK_{ijk} X^i X^j X^k}{X^0} - \frac{1}{2}A_{ij} X^i X^j+B_i X^0 X^i+C(X^0)^2 + D (X^1)^2 \log[X^1/X^0] \, ,   
\end{equation}
where we set $X^0=1$, $X^1=z_1 $ and $X^2=\frac{ \log[z_2]}{2 \pi i}$ afterwards. The coefficients in the periods are then related by
\begin{equation}
 \quad \cK_{112}=-a^2n\, , \quad \cK_{222}=1\, , \quad A_{11} =\frac{ia^2}{4\pi}\, , \quad B_1= i\delta_1 a \, , \quad C=\frac{i\delta_2}{2} \, , \quad D=\frac{ia^2}{4\pi}\, ,
\end{equation}
and all other coefficients vanish. Note that the sign of the imaginary piece of $D$ is again fixed by $a^2>0$, similar to the $\mathrm{I}_1$ boundary.

\subsection{Degeneration for the Calabi-Yau threefold in $\mathbb{P}_4^{1,1,2,2,6}[12]$}\label{app:Seiberg-Witten}
In this appendix we illustrate how the period vector near the Seiberg-Witten point of the K3-fibered Calabi-Yau threefold in $\mathbb{P}_4^{1,1,2,2,6}[12]$ can be embedded into our models. This geometry has been studied in detail in the literature, see e.g.~\cite{Hosono:1993qy, Candelas_1994, Kachru:1995fv, Curio_2001,Lee:2019wij}, and we will follow the analysis of the periods of \cite{Lee:2019wij} here. In our models this Seiberg-Witten point corresponds to the $\langle \mathrm{II}_0 | \mathrm{II}_1 | \mathrm{II}_1 \rangle$ boundary, whose periods have been constructed section \ref{ssec:II1construction}. The period vector as computed from the relevant Picard-Fuchs equations takes the form
\begin{align}
\Pi_{\rm P12}=\frac{1}{\pi} \begin{pmatrix}
1+ \frac{5}{36} z_1 \\
z_1 \\
- \sqrt{z_1}  \\
\frac{i}{\pi} (\log[z_2]-6 \log[2]+7) \sqrt{z_1} \\
\frac{i}{2 \pi }  (5 + 2 \log[z_1]+\log[z_2] )(1+\frac{5}{36} z_1) \\
\frac{i}{2 \pi }(1+2 \log[z_1] + \log[z_2] )z_1
\end{pmatrix}.
\end{align}
One observation that we can immediately make is that the period vector depends on square roots of the coordinates, which results in monodromy transformations that are only quasi-unipotent. We remedy this by an appropriate coordinate transformation further below. As generic solutions to the Picard-Fuchs equations, these periods are not in a symplectic basis. To find the appropriate basis transformation one uses the fact that such a basis can naturally be found at the LCS point and then by analytic continuation one can compute the transition matrix. The latter is also given in \cite{Lee:2019wij}. Furthermore, we require another basis transformation to bring the periods into the symplectic basis we use in this work. The combined transition matrix is given by 
\begin{align}
M_{\rm P12}=\frac{\sqrt{2}}{X} \begin{pmatrix}
1 &- \frac{5}{36}+ X^2 & 0 & 0 & 0 & 0 \\
i &  - i  (\frac{5}{36 }+X^2) & 0 & 0 & 0 & 0 \\
0 & 0 & X & 0 & 0 & 0 \\
\frac{5 i}{8 \pi } & \frac{5i (36 X^2-1)}{288 \pi }&0 &0 & \frac{-1}{4}& \frac{1}{144}(5-36 X^2) \\
-\frac{5}{8 \pi} & \frac{5(36X^2 +1)}{288 \pi} & 0 & 0 & - \frac{i}{4} & \frac{i}{144}(5 + 36 X^2) \\
0 & 0 & \frac{3 i X}{2 \pi}(\log[4] -1)& -\frac{X}{2} & 0 & 0
\end{pmatrix}
\end{align}
where $X=\frac{\Gamma(3/4)^4}{\sqrt{3} \pi^2}$. In addition, we perform the divisor preserving coordinate redefinition
\begin{align}
 z_1 \to z_1^2 \exp \Big( \frac{5}{16}z_2+\frac{131}{2048}z_2^2 \Big) \, ,  \qquad z_2 \to z_2 \exp \Big(-\frac{5}{8}z_2-\frac{131}{1024}z_2^2 \Big) \,,
\end{align}
that, among other things, makes the monodromies unipotent. After these transformations the period vector takes the form
\begin{align}
\Pi_{\rm P12}=\begin{pmatrix}
1 + X^2 z_1^2    \\
i - i  X^2 z_1^2   \\
-X z_1 \\
- \frac{i}{8 \pi} (4 \log[z_1] + \log[z_2]) + \frac{iX^2}{8 \pi} (4- 4 \log[z_1]-\log[z_2])  z_1^2   \\
 \frac{1}{8 \pi} (4 \log[z_1] + \log[z_2]) + \frac{X^2}{8 \pi} (4- 4 \log[z_1]-\log[z_2])  z_1^2 \\
-\frac{iX}{2 \pi} (4+\log[z_2] )z_1  
\end{pmatrix} \, .\label{eq:P12Period}
\end{align}
Our model \eqref{eq:II1periods} reproduces the periods given in \eqref{eq:P12Period} upon identifying
\begin{equation}
n_1=0\, , \qquad n_2 = 1/4 \, , \qquad b = -4 X \, , \qquad a=c =0\, .
\end{equation}

\bibliographystyle{jhep}
\bibliography{references}

\end{document}